\documentclass[aps, twocolumn, prd, preprintnumbers, amsmath, amssymb, amsfonts, superscriptaddress, eqsecnum, nofootinbib]{revtex4-1}
\pdfoutput=1

\usepackage[linktocpage,pagebackref=false,hidelinks]{hyperref}
\usepackage[pagebackref=false,hidelinks]{hyperref}
\usepackage[normalem]{ulem}
\usepackage{diagbox}
\usepackage{setspace,latexsym}
\usepackage{color}
\usepackage{epsfig}
\usepackage{graphicx}
\usepackage[export]{adjustbox}
\usepackage[dvipsnames]{xcolor}
\usepackage{float}
\usepackage{tabularx}
\usepackage{placeins}
\usepackage{multirow}

\newcommand{\beq}{\begin{equation}}
\newcommand{\eeq}{\end{equation}}
\newcommand{\bea}{\begin{eqnarray}}
\newcommand{\eea}{\end{eqnarray}}
\newcommand{\nn}{\nonumber}

\newcommand{\keV}{\mathrm{keV}}
\newcommand{\GeV}{\mathrm{GeV}}
\newcommand{\MeV}{\mathrm{MeV}}
\newcommand{\TeV}{\mathrm{TeV}}

\newcommand{\PS}{{\rm PS}}

\newcommand{\SM}{\mathrm{SM}}

\newcommand{\Mpl}{M_\mathrm{Pl}}

\newcommand{\mb}{\mathrm{mb}}
\newcommand{\fm}{\mathrm{fm}}

\newcommand{\kicm}{k^{\rm in}_{{\rm cm}}}
\newcommand{\kfcm}{k^{\rm out}_{{\rm cm}}}

\newcommand{\kcm}{k_{\rm cm}}
\newcommand{\Ecm}{E_{\rm cm}}

\newcommand{\Tlab}{T_{\rm lab}}

\newcommand{\lab}{{\rm lab}}

\newcommand{\He}{$^4{\rm He}$}

\newcommand{\np}{{$n\leftrightarrow p$} }
\newcommand{\llangle}{\langle\!\langle}
\newcommand{\rrangle}{\rangle\!\rangle}

\newcommand{\Yp}{{\rm Y}_{\rm p}}

\renewcommand{\sec}{\mathrm{s}}

\def\bal#1\eal{\begin{align}#1\end{align}}

\begin{document}

\title{New Bounds on Heavy QCD Axions from Big Bang Nucleosynthesis}

\author{Tae Hyun Jung}
\email{thjung0720@gmail.com}
\affiliation{Particle Theory and Cosmology Group, Center for Theoretical Physics of the Universe, Institute for Basic Science (IBS), Daejeon, 34126, Korea}

\author{Takemichi Okui}
\email{tokui@fsu.edu}
\affiliation{Department of Physics, Florida State University, Tallahassee, FL 32306, USA}
\affiliation{High Energy Accelerator Research Organization (KEK), Tsukuba 305-0801, Japan}

\author{Kohsaku Tobioka}
\email{ktobioka@fsu.edu}
\affiliation{Department of Physics, Florida State University, Tallahassee, FL 32306, USA}
\affiliation{High Energy Accelerator Research Organization (KEK), Tsukuba 305-0801, Japan}

\author{Jiabao Wang}
\email{jwang21@fsu.edu}
\affiliation{Department of Physics, Florida State University, Tallahassee, FL 32306, USA}

\preprint{CTPU-PTC-25-30}

\begin{abstract}
We study Big Bang Nucleosynthesis (BBN) constraints on heavy QCD axions. BBN offers a powerful probe of new physics that modifies the neutron-to-proton ratio during the process, thanks to the precisely measured primordial Helium-4 abundance. A heavy QCD axion provides an attractive target for this probe, because not only is it a well-motivated hypothetical particle by the strong CP problem, but also it dominantly decays to hadrons if kinematically allowed. A range of its lifetime is thus excluded where the hadronic decays would significantly alter the neutron-to-proton ratio. We compute axion-induced modification of the neutron-to-proton ratio, and obtain robust upper bounds on the axion lifetimes, as low as 0.017~s for the axion mass higher than 300~MeV\@. Remarkably, this is stronger than projected future CMB bounds via $N_{\rm eff}$. Our bounds are largely insensitive to uncertainties in hadronic cross sections and the axion's branching fractions into various hadrons, as well as to the precise value of the initial axion abundance. We also incorporate, for the first time, several key improvements, such as scattering processes by energetic $K_L$ and secondary hadrons, that can also be important for studying general hadronic injections during BBN, not limited to those from axion decays. 
\end{abstract}

\maketitle

\setcounter{tocdepth}{1}
\tableofcontents

\section{Introduction}
Big Bang Nucleosynthesis (BBN) offers a powerful probe of new physics that modifies the neutron-to-proton ratio during the process, 
thanks to the well understood underlying Standard Model (SM) processes and precisely measured primordial ${}^4{\rm He}$ abundance.
The BBN probe is especially sensitive to new physics that injects extra hadrons into the plasma~\cite{Reno:1987qw} (with a first comprehensive quantitative analysis by \cite{Kohri:2001jx}). 
These extra hadrons modify the neutron-to-proton ratio via the strong interactions (e.g., $\pi^- p \to \pi^0 n$), 
whose cross sections are roughly 16 orders of magnitude larger than the weak interaction processes that govern the ratio in the standard BBN (e.g., $e^- p \to \nu_e n$) around the freeze-out temperature of the neutron-to-proton ratio ($T \sim 1\>\MeV$).
(Roughly speaking, the strong cross sections are on the order of $\mb\sim {\rm GeV}^{-2}$, while the weak cross sections are on the order of $G_F^2 \,{\rm MeV}^2 \sim 10^{-16}\>{\rm GeV}^{-2}$.)
Therefore, even an exponentially suppressed amount of hadronic injection can still significantly impact the neutron-to-proton ratio. 
An example of such hadronic injection is a hadronically decaying new particle 
whose lifetime is much shorter than the freeze-out time of the neutron-to-proton ratio ($\sim 1\>\sec$).
Indeed, as we will find in this paper, BBN can be sensitive to lifetimes of ${\cal O}(0.01)\>\sec$ for heavy QCD axions.
BBN constraints on hadronic injection have also been studied for a variety of new physics scenarios such as primordial black holes~\cite{Kohri:1999ex},
the gravitino~\cite{Kawasaki:2000en, Jedamzik:2004er, Kawasaki:2004yh, Kawasaki:2004qu, Kohri:2005wn, Jedamzik:2006xz, Kawasaki:2008qe, Cyburt:2009pg, Cyburt:2010vz, Cyburt:2013fda, Kawasaki:2017bqm, Hasegawa:2019jsa, Angel:2025dkw}, 
Higgs-portal scalars~\cite{Fradette:2017sdd}, 
dark photons~\cite{Fradette:2014sza, Berger:2016vxi},
heavy neutral leptons~\cite{Boyarsky:2020dzc, Chen:2024cla}, dark matter annihilation~\cite{Jedamzik:2009uy,Henning:2012rm,Omar:2025jue},
and as an attempt to solve the Lithium problem~\cite{Pospelov:2010cw}.

A heavy QCD axion provides an especially attractive target for the BBN probe, which we study for the first time. 
Firstly, it is a well-motivated hypothetical particle that can solve the strong CP problem without suffering from the so-called quality problem.
Secondly, it dominantly decays to hadrons if kinematically allowed.
This is because a QCD axion---heavy or not---by definition dominantly couples to the SM sector via the coupling:
\begin{align}
\mathcal{L} \supset \dfrac{\alpha_s}{8\pi} \dfrac{a}{f_a} G\tilde{G} \,,\label{eq:aGGdual}
\end{align}
where $a$ and $G$ denote the axion field and gluon field strength, respectively, and $f_a$ is a mass scale called the axion decay constant.
Hence, 
if its mass $m_a$ is above $\sim 300\;\MeV$ to kinematically allow a hadronic decay channel,  
the axion will decay dominantly to hadrons via the coupling~Eq.\,\eqref{eq:aGGdual}.
Since this coupling is what makes $a$ to solve the strong CP problem via the Peccei-Quinn (PQ) mechanism~\cite{Peccei:1977hh, Peccei:1977ur, Weinberg:1977ma, Wilczek:1977pj, Kim:1979if, Shifman:1979if, Dine:1981rt, Zhitnitsky:1980tq}, the dominance of hadronic decays is well-motivated.

For a further motivation for considering \emph{heavy} QCD axions in particular, let us pay attention to the ``quality'' of $U(1)_{\rm PQ}$, 
an approximate global symmetry under which $a$ shifts. 
The PQ mechanism requires that the location of the minimum of the axion potential be dictated by the $U(1)_{\rm PQ}$ breaking due to Eq.\,\eqref{eq:aGGdual} (the condition for ``good quality'').  
Since quantum gravity effects might completely violate global symmetries at the Planck scale ($M_{\rm Pl} \simeq 1.2\times 10^{19}\>{\rm GeV}$), the question arises as to whether the good-quality condition is well respected by $U(1)_{\rm PQ}$-violating operators suppressed by $M_{\rm Pl}$ (the quality problem).
It is straightforward to see that a high quality $U(1)_{\rm PQ}$ prefers a high value of $m_a$ for any given $f_a$. 
To see the idea, let $\Phi$ be the $U(1)_{\rm PQ}$-breaking scalar field with $\langle\Phi\rangle = f_a$.
A $U(1)_{\rm PQ}$ violating operator of the form $\Phi^n / M_{\rm Pl}^{n-4}$ with $n > 4$ with a generic $\mathcal{O}(1)$ complex coefficient would not observably shift the location of the minimum if $f_a^n / M_{\rm Pl}^{n-4} \lesssim \theta_{\rm max} f_a^2 m_a^2$, where $\theta_{\rm max} \sim 10^{-10}$ is the current upper bound on the QCD vacuum angle.
This shows that the PQ quality is maintained for a sufficiently high $m_a$ for any given $f_a$.

To motivate the region of the parameter space we will explore in this work, we rewrite the above condition as
\begin{align}
f_a &\lesssim \theta_{\rm max}^\frac{1}{n-2} M_{\rm Pl} \Bigl( \frac{m_a}{M_{\rm Pl}} \Bigr)^{\!\frac{2}{n-2}} 
\nonumber\\
&\sim \begin{cases}
10^7\>{\rm GeV} \Bigl( \dfrac{m_a}{\rm GeV} \Bigr)^{\!\frac12} &\text{if $n=6$} \\
10^{11}\>{\rm GeV} \Bigl( \dfrac{m_a}{\rm GeV} \Bigr)^{\!\frac13} &\text{if $n=8$} 
\end{cases}
\label{Eq:PQ_quality}
\end{align}
Such a region of the parameter space is not compatible with the standard relation $m_a f_a \simeq m_\pi f_\pi$ unless we take $f_a$ to be so low that it is already excluded by existing axion searches~\cite{Goudzovski:2022vbt}.
Since the standard relation assumes that the dominant $U(1)_{\rm PQ}$ breaking is given by the coupling of Eq.\,\eqref{eq:aGGdual},   
a heavy QCD axion requires an additional source of $U(1)_{\rm PQ}$ breaking.
This must not reintroduce the quality problem, but, unlike the original quality problem, this time it is unrelated to quantum gravity and hence can be firmly addressed by model building within conventional QFT\@.
Three types of models are known depending on 
whether additional contributions to the axion mass originate from (i) an additional confinement gauge group unified with $SU(3)_c$ into a larger gauge group in the UV~\cite{Dimopoulos:1979pp, Tye:1981zy, Agrawal:2017ksf, Rubakov:1997vp, Valenti:2022tsc}, (ii) the $SU(3)_c$ instanton contribution with somehow large strong gauge coupling~\cite{Dimopoulos:1979pp, Holdom:1982ex, Holdom:1985vx, Dine:1986bg, Flynn:1987rs, Choi:1988sy, Choi:1998ep, Gherghetta:2020keg}, or (iii) a mirror QCD~\cite{Berezhiani:2000gh, Fukuda:2015ana, Hook:2019qoh, Kelly:2020dda}.
Our bounds are independent of the origin of the additional axion mass as long as the axion dominantly decays to hadrons. We assume that the additional mass is generated at a sufficiently high energy scale that we can treat $m_a$ as constant during the cosmological evolution relevant to our analysis. 

As we will find in this paper, BBN constrains the heavy QCD axion lifetime to be $\lesssim 0.02\>\sec$.
This bound occupies a unique place in comparison with other experimental constraints when viewed as bounds on the heavy QCD axion lifetime. 
Collider experiments can probe prompt and displaced decays, where the longest lifetimes they can probe are set by the detector geometries to be around $\sim 10^{-9}\>\sec$.
For the mass range of our interest, $m_a\gtrsim 300\,\MeV$,%
\footnote{For lower masses, bounds from kaon decays become competitive. See the recent review~\cite{Goudzovski:2022vbt} for a compilation of various bounds and a list of relevant references. 
There are also bounds from supernova for lower masses and intermediate values of the decay constant $f_a\lesssim 10^8\>\GeV$~\cite{Chang:2018rso, Ertas:2020xcc}.}
competitive bounds in this range of lifetimes are from the LHC~\cite{LHCb:2015nkv, LHCb:2016awg, Bauer:2021mvw, ATLAS:2022abz, CMS:2017dcz, CidVidal:2018blh, LHCb:2025gbn} 
and $B$-factories~\cite{Aloni:2018vki, Chakraborty:2021wda, Bertholet:2021hjl, BaBar:2021ich}.
Proton beam dump experiments and similar setups can probe much longer lifetimes up to around $\sim 10^{-6}\>\sec$~\cite{Bergsma:1985qz, Blumlein:1990ay, ArgoNeuT:2022mrm, Afik:2023mhj, NA62:2025yzs} due to much larger distances to the detector. 
BBN probes even longer lifetimes, ranging from about $10^{-2}\>\sec$ to a few minutes. The CMB can also probe lifetimes longer than about 0.1~s via $N_{\rm eff}$~\cite{Dunsky:2022uoq}, 
which is especially relevant when the hadronic decays are kinematically forbidden. 
As we will show, once a phase space for hadronic decays opens, our bounds from the BBN neutron-to-proton ratio become stronger than the CMB's and constrain lifetimes of $\gtrsim 0.017\>\sec$.  

Since a well-motivated axion decays to hadrons if kinematically allowed, it is essential to include hadrons in studies of BBN bounds. Consequently, our bound is an order-of-magnitude stronger than the ones for the axion-like particles that assume diphoton decay to be dominant even at high masses\,\cite{Depta:2020wmr,Balazs:2022tjl,Cadamuro:2011fd,Millea:2015qra}. 


\section{Overview \label{sec:overview}}
Before discussing technical details, we would like to provide an overview of our study and results.
Readers mainly interested in the results may proceed directly to Sec.$\,$\ref{sec:result} after reading this section.
In addition, we also highlight key improvements we made in the methodology of studying BBN constraints on general hadronically decaying long-lived particles, not necessarily limited to heavy QCD axions. 

\subsection{Overall framework}
\label{sec:overall}
We begin by imagining a post-inflation reheating temperature that is sufficiently high to allow axions to be in thermal and chemical equilibrium with the SM particles.
As the temperature drops, but while still relativistic, the axions decouple from the rest, and their number density freezes out.
This sets the initial condition for our BBN analysis, 
with the initial axion abundance given by $Y_a = n_a/s \sim 1/g_*^{\rm FO}$, 
where $g_*^{\rm FO}$ is the effective number of relativistic degrees of freedom at the axion freezeout temperature $T_{\rm FO}$.
In~\cite{Dunsky:2022uoq}, $T_{\rm FO}$ and $Y_a$ are estimated for different values of $m_a$ and $f_a$ across the QCD deconfinement-confinement transition.
We basically follow their method of estimation, which will be discussed in detail in Sec.$\,$\ref{sec:thermal}\@. 

In Sec.$\,$\ref{sec:decay} the axion decay width and branching fractions will be calculated in two ways depending on whether the quark-gluon or hadronic description should be used.
For $m_a<2\,\GeV$ where the hadronic picture is appropriate, a 
data-driven method\,\cite{Aloni:2018vki, Cheng:2021kjg, Bisht:2024hbs, Bai:2025fvl, Balkin:2025enj} with the data in~\cite{Bisht:2024hbs} will be used to capture nonperturbative physics necessary for our purpose, in particular the branching fractions of the axion decay into various hadrons.
For $m_a > 2\,\GeV$, perturbative QCD (pQCD) will be employed to calculate the axion decay rate at NNLO following~\cite{Chetyrkin:1998mw} (with more technical details in Appendix~\ref{app:a_decay}), and the branching fractions are computed by the hadronic shower programs (\texttt{PYTHIA 8.306}\,\cite{Bierlich:2022pfr} and \texttt{Herwig v7.3.0}\,\cite{Corcella:2000bw, Bahr:2008pv, Bellm:2015jjp}). 
We find large discrepancies in the predictions of these two programs, especially when $m_a$ is close to $2\>{\rm GeV}$. 
Neither program can be trusted there, because they both fail to respect basic properties such as the parity of the axion.
While this indicates a serious need for improvement in these programs, for our purpose, we fortunately find that those order-one uncertainties in the axion branching fractions do not affect our final bounds on the axion lifetime beyond a few percent.

We aim to obtain an upper bound of ${\cal O}(0.01)\,\sec$ for the axion lifetime, where the relevant experimental observable being probed is the primordial {\He} abundance. 
In the standard BBN (see Ref.\,\cite{Pitrou:2018cgg} for a review), the process starts with the decoupling of neutrinos at $T \sim 2\,\MeV$ ($t\sim0.2\,\sec$), and the freeze-out of neutron-proton conversion around $T_n \sim 1\>\MeV$ ($t_n\sim 0.7\,\sec$), where the neutron fraction is defined as $X_n \equiv n_n / n_b$ with $n_n$ and $n_b$ being the number densities of neutrons and baryons, respectively.
Later, once $T$ falls below the so-called deuterium bottleneck temperature $ T_{\rm D} \sim 70\,\keV$ ($t_{\rm D}\sim 200\,\sec$),  these neutrons (after slight reduction in number due to beta decay) are all consumed to produce deuterium and then tritium, but at the end nearly all of the neutrons end up in {\He}.
Thus, the primordial {\He} abundance is determined by $X_n$ at $T_{\rm D}$. 

\begin{table*}[t!]
\centering
\begin{tabular}{c|c|c|c|c|c}
                         & $n \to p$                  & $p \to n$              & ~$Q/\MeV$~  & ~Secondary~ &  $\sigma$ \\ 
\hline\hline
\multirow{2}{*}{$\pi^-$} &                        & $p \pi^- \to n \pi^0$      & $3.30$ & & Fig.\,\ref{Fig:ppim_pi0}\\
                         &                            & $p \pi^- \to n \gamma$ & $138$  & & Fig.\,\ref{Fig:ppim_photon}\\ 
\hline
\multirow{2}{*}{$\pi^+$} & $n \pi^+ \to p \pi^0$      &                        & $5.89$ & & Fig.\,\ref{Fig:npip_pi0}\\
                         & $n \pi^+ \to p \gamma$     &                        & $141$  & & Fig.\,\ref{Fig:npip_photon}\\ 
\hline\hline
\multirow{3}{*}{$K^-$}   & & $\quad p K^- \to \pi Y$ $(Y \to n)\quad$& $\gtrsim 100$ & $\pi^\pm$ & Fig.\,\ref{Fig:pKm_Y}\\
                         &                 & $p K^- \to n \bar{K}^0$           & $-5.23$ & ${K_S}^\ddagger$, ${K_L}^{\ddagger}$  & Fig.\,\ref{Fig:pKm_K0}\\%
                         & $\quad n K^- \to \pi Y$ $(Y \to p)\quad $&& $\gtrsim 100$ & $\pi^-$ & Fig.\,\ref{Fig:nKm_Y}\\ 
\hline
$K^+$                    & $n K^+ \to p K^0$          &                        & $-2.64$ & ${K_S}^\ddagger$, ${K_L}^{\ddagger}$ & Fig.\,\ref{Fig:pKL_Kp}\\ %
\hline
\multirow{6}{*}{$K_L$}   &           & $p K_L \to \pi Y$ $(Y \to n)$ & $\gtrsim 100$ & $\pi^\pm$  & Figs.\,\ref{Fig:pKL_Y}, \ref{Fig:KLp_averaged}\\
                         &                            & $p K_L \to n K^+$      & $2.64$ &   $K^+$  & Figs.\,\ref{Fig:pKL_Kp}, \ref{Fig:KLp_averaged}\\
                         & $n K_L \to \pi Y$ $(Y \to p)$    &        & $\gtrsim 100$ & $\pi^\pm$  & Figs.\,\ref{Fig:nKL_Y}, \ref{Fig:KLn_averaged}\\
                         & $n K_L \to p K^-$          &                        & $5.23$ &   $K^-$  & Figs.\,\ref{Fig:pKm_K0}, \ref{Fig:KLn_averaged}\\ 
\cline{2-6}
                         & \multicolumn{2}{c|}{$p K_L \to p K_S$}    & $0$ & $K_S(\to \pi^\pm)$     & Fig.\,\ref{fig:NKL_El_Rg} \\
                         & \multicolumn{2}{c|}{$n K_L \to n K_S$}    & $0$ & $K_S(\to \pi^\pm)$     & Fig.\,\ref{fig:NKL_El_Rg} \\
\hline\hline
$\bar n$                 &              & $p \bar n \to {\rm mesons}$ & $1600$ &  ${\rm mesons}^\ddagger$ &  Fig.\,\ref{Fig:NNbar} \\ 
\hline
$\bar p$                 & $n \bar p \to {\rm mesons}$  &              & $1600$ &  ${\rm mesons}^\ddagger$ &  Fig.\,\ref{Fig:NNbar}
\end{tabular}
\caption{Summary of all $n \leftrightarrow p$ conversion processes considered in our analysis.  $Y$ denotes a hyperon: $\Sigma^\pm$, $\Sigma^0$, or $\Lambda$.
$Q$ is the total mass of the initial state minus that of the final state. 
Processes with negative $Q$ are kinematically forbidden unless the initial state has sufficient kinetic energy.
We keep track of the time evolution of $n$, $p$, and all hadrons listed in the leftmost column. 
The $\ddagger$ symbol on a secondary hadron indicates that its contribution is neglected in the Boltzmann equation for that hadron, for the reasons explained in~Sec.$\,$\ref{sec:improvements}.}
\label{Tab:reactions}
\end{table*}

Conventionally, the {\He} abundance is given as its mass fraction to the total baryon energy density: $\Yp \equiv \dfrac{\rho(^4{\rm He})}{\rho_b}$. 
We adopt the PDG-recommended value~\cite{ParticleDataGroup:2024cfk}:
\bal
\Yp  = 0.245 \pm 0.003 \,,
\label{Eq:YpValue}
\eal
which is based on~\cite{Srianand:2010un, Valerdi:2019beb,  Fernandez:2019hds, Kurichin:2021ppm, Hsyu:2020uqb, Valerdi_2021, Aver:2021rwi}.%
\footnote{The EMPRESS collaboration reports $\Yp = 0.2370^{+0.0034}_{-0.0033}$~\cite{Matsumoto:2022tlr},  
whose central value is about $2\sigma$ smaller than the one above, but fortunately, their uncertainties are the same. 
Since our constraints will be based on a fractional change $\delta\Yp / \Yp$ (or see Eq.\,\eqref{Eq:criterion_deltaXn}), this difference in the central values does not matter.}
As nearly all the neutrons become {\He} below $T_{\rm D}$, it is a very good approximation to take $\Yp \simeq 2 X_n(T_{\rm D})$ and just focus on calculating how the heavy QCD axion changes $X_n(T_{\rm D})$.
We thus obtain a 2$\sigma$ bound by
\bal
\left. \frac{\delta X_n}{X_n^\SM} \right|_{T=T_{\rm D}} \!\!\!\! < R_{\Yp} \equiv 2.45\%
\label{Eq:criterion_deltaXn}
\eal
with $\delta X_n \equiv X_n - X_n^\SM$, and
we define $T_{\rm D}$ by $\Yp=2X^{\rm SM}_n(T_{\rm D})$, where $\Yp$ is given in Eq.\,\eqref{Eq:YpValue} and we calculate $X_n^{\rm SM}(T)$ by removing the axion from our Boltzmann equations, although  
our constraints from Eq.\,\eqref{Eq:criterion_deltaXn} will be insensitive to the precise definition of $T_{\rm D}$. 
We would also like to point out that the standard BBN predicts the central value of $\Yp = 0.247$, slightly higher than the observed central value of Eq.\,\eqref{Eq:YpValue}, so our condition Eq.\,\eqref{Eq:criterion_deltaXn} with $X_n^{\rm SM}$ matched to Eq.\,\eqref{Eq:YpValue} provides a conservative bound because it turns out that $\delta X_n > 0$.

Next, we would like to summarize what hadrons and reactions are relevant to $n \leftrightarrow p$ conversion.
As we will explain shortly, the lifetime of a hadron dictates whether it should be included in the analysis. 
In Table~\ref{Tab:reactions}, we list all hadrons and reactions included in our analysis.
Relevant hadrons are limited to $\pi^\pm$, $K^\pm$, $K_L$, nucleons $N = p, n$, and anti-nucleons $\bar{N} = \bar{p}, \bar{n}$. 
Other hadrons, such as hyperons, decay too rapidly to participate in $n \leftrightarrow p$ conversion, even though their cross sections may be larger.  

To identify the hadrons important for $n \leftrightarrow p$ conversion, consider the disappearance terms in the Boltzmann equation for an injected hadron $h$, 
\bal
\dot n_h \supset -\bigl(3H + \Gamma_h + \langle\sigma v\rangle_{\rm dis}\, n_b \bigr) n_h  \,,\label{eq:nhdot-disappearance}
\eal
where $H$ is the Hubble expansion rate, $\Gamma_h$ isthe hadron's decay width, and $\langle \sigma v \rangle_{\rm dis}$ is the sum of the thermally averaged cross sections for all baryon-$h$ scattering processes that consume the $h$, which in particular include $n \leftrightarrow p$ conversion processes.
First, consider an $h$ other than $N$ or $\bar{N}$.
On the right-hand side, since $H$ is $\mathcal{O}(1)\;\sec^{-1}$, the $3H$ term is always negligible in comparison with the two other terms. 
Therefore, in order for the $h$ to participate in $n \leftrightarrow p$ conversion before it decays, the $\Gamma_h$ term must be smaller than or comparable to the scattering rate:%
\bal
\Gamma_h 
\lesssim
\langle \sigma v \rangle_{\rm dis} \, n_b(T)
\sim 
10^{-17} \>\GeV \> 
\frac{\langle \sigma v \rangle_{\rm dis}}{10\>{\rm mb}} \, 
\Bigl( \frac{T}{\MeV} \Bigr)^{\!3} .
\eal
Among mesons, this condition may be satisfied only by the following three:
\bal
&\text{$\pi^\pm\;$:} 
~
\tau_{\pi^\pm}=2.603\times 10^{-8}\,\sec,
~
\Gamma_{\pi^\pm}=2.53\times 10^{-17}\,\GeV, 
\nn \\
&\text{$K^\pm$:} 
~
\tau_{K^\pm \!}=1.238\times 10^{-8}\,\sec,
~
\Gamma_{K^\pm \!}=5.32\times 10^{-17}\,\GeV, 
\nn \\
&\text{$K_L\,$:}
~
\tau_{K_L \!}=5.116\times 10^{-8}\,\sec, 
~
\Gamma_{K_L \!}=1.29\times 10^{-17}\,\GeV
.
\nn
\eal
Among baryons, all unflavored heavy baryons (e.g., $\Delta$) are clearly too short-lived to satisfy the above condition. Hyperons, though much longer-lived, also fail to satisfy the condition.%
\footnote{The decay widths of hyperons are on the order of $10^{-15}\,\GeV$, while their scattering cross sections are at most $\sim 100\,\mb$, thereby failing to meet the condition.}
Finally, needless to say, nucleons and anti-nucleons ($p$, $n$, $\bar{p}$, and $\bar{n}$) are important in $n\leftrightarrow p$ conversion.
Here, once produced from axion decays, the anti-nucleons will immediately annihilate with  existing nucleons, e.g., 
$n\bar{p} \to \pi^- \pi^0 + \cdots$ and 
$p\bar{n} \to \pi^+ \pi^0 + \cdots$.

\begin{figure*}[t!]
    \centering
    \includegraphics[width=1\linewidth]{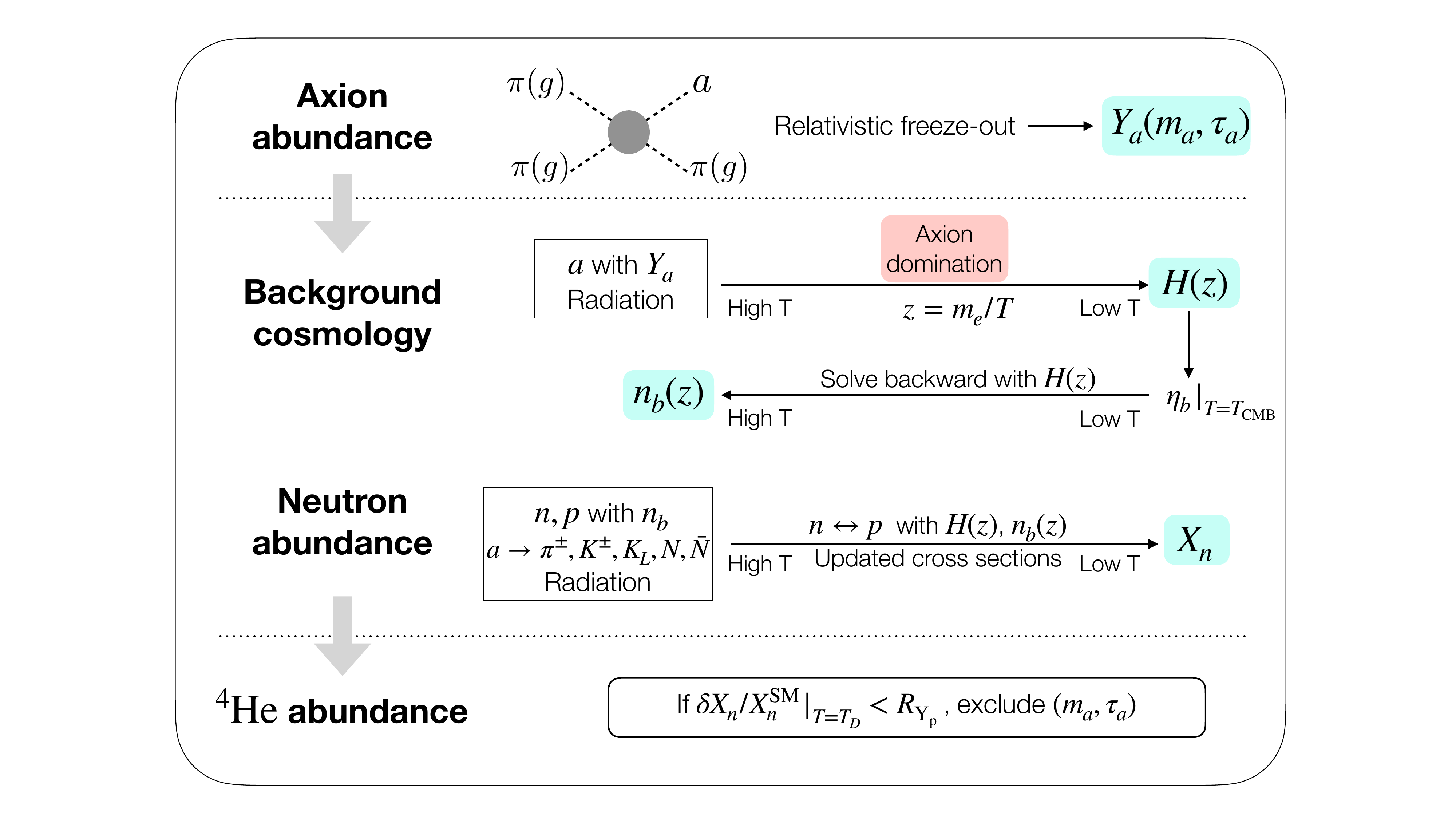}
    \vspace{-20pt}
    \caption{Schematic depiction of the thermal history of the universe as well as steps taken in our analysis as summarized in Sec.$\,$\ref{sec:overall}.
    We adopt the 2$\sigma$ bound of {\He} abundance $R_{\Yp} = 2.45\%$. }
    \label{fig:skematic}
\end{figure*}

Let us comment on the impact of the axions on the background cosmology. 
The energy density in the axions alters the evolution of the Hubble rate. 
For axion masses greater than a few~$\GeV$, there is even a short period of a matter-dominated era by non-relativistic axions, which ends as the axions decay into the radiation component, restarting the radiation-dominated era.
This alters the evolution of the Hubble rate, which in turn shifts the freeze-out temperature of the neutron-to-proton ratio.
Our Boltzmann equations account for this effect.

Another effect on the background cosmology is that axion decays after neutrino decoupling effectively dilute the neutrino population relative to the plasma, as the latter heats up by the axion decay products.  
As discussed in~\cite{Dunsky:2022uoq}, this changes the relative evolution of the neutrino temperature and results in a value of $N_{\rm eff}$ smaller than the SM value. 
We find that our BBN bounds are stronger than projected future CMB bounds via $N_{\rm eff}$. 

Heating up the plasma also dilutes the baryon-to-photon ratio, $\eta_b\equiv n_b/n_\gamma$,  whose value is tightly constrained by CMB in the late Universe. 
Therefore, we fix the the value of $\eta_b$ after the dilution to be $\eta_b = 6.115 \times 10^{-10}$ as measured by Planck~\cite{Planck:2018vyg}. Once the background cosmology and $n_b$ evolution are obtained, we solve Boltzmann equations for the hadrons and examine if $\delta X_n$ is compatible with the {\He} measurement. 

In Fig.\,\ref{fig:skematic}, we schematically depict the thermal history of the universe as well as the steps taken in our analysis described in this subsection.

\subsection{Improvements in the treatment of hadrons}
\label{sec:improvements}
We also introduce important improvements in the quantitative treatment of the reactions listed in Table~\ref{Tab:reactions}.
Here is a brief summary of the improvements:
\begin{itemize}
\item {\bf Revisit and update hadronic cross sections.}
    We improve the treatment of hadronic cross sections pioneered by \cite{Reno:1987qw, Pospelov:2010cw}.
    In particular, cross sections involving $K^\pm$ and $K_L$ are significantly improved.    
    Within a partial wave analysis, 
    we take a complete set of isospin-related 2-to-2 processes and, if necessary, also include $p$-waves and/or momentum-dependent scattering lengths.
    As elaborated in the next bullet, 
    cross sections at high momenta are important for scattering of $K_L$.
    See Sec.$\,$\ref{sec:x-sec} for a more concrete summary and Appendix~\ref{app:x-sec} for detailed derivations.

\item {\bf Include energetic $K_L$ properly.}
  As pointed out in Refs.\,\cite{Reno:1987qw,Kohri:2001jx}, 
    being electrically neutral, the $K_L$s do not thermalize with the plasma.  
    Therefore, for $K_L$, we must evaluate the momentum distribution to calculate cross sections and time-dilated decay lifetimes. Our fitting of hadronic cross sections finds elastic $K_L N$ scattering to be sizable, especially at high momenta, so we examine how the $K_L$ momentum spectrum inherited from axion decays is modified by the elastic scattering (see Figs.\,\ref{fig:KLmom_reshaped}, \ref{fig:KLmom_lowma_reshaped} and Appendices~\ref{sec:xsec-kaon}, \ref{sec:KLreshape} for more detail). 
    These were not done in previous studies. 
    For instance, Refs.\,\cite{Pospelov:2010cw,Fradette:2017sdd,Boyarsky:2020dzc} used $K_L$ cross sections at the threshold, but as we will show (see Fig.\,\ref{fig:KLNxsec}), the average cross sections are typically about half of the threshold values.

\item {\bf Include secondary hadrons from scattering.}
    In the 5th column in the Table~\ref{Tab:reactions}, we list 
    hadrons produced from scattering, including those from decays of final state particles of scattering. 
    We call them secondary hadrons as opposed to primary hadrons produced from axion decays.
    These secondary hadrons from scattering were never included in the literature. 
    We find that taking into account the secondary hadrons is essential for estimating $\delta X_n$ correctly. For example, $nK^- \to \pi^- \Lambda (\to p \pi^-)$ seems to convert $n$ to $p$, but the two secondary $\pi^-$s do the reverse conversion twice, and therefore the net conversion is actually $p\to n$.
    Not included in our analysis are the secondary hadrons marked by $\ddagger$, where the $\ddagger$ on a hadron means that its contribution is ignored in the Boltzmann equation for that hadron.
    The ``$K_S^\ddagger$'' and ``$K_L^\ddagger$'' are neglected because the initial nucleons and $K^\pm$ do not have enough kinetic energy at $T\sim T_n$ to overcome the negative $Q$ values.
    The ``mesons${}^\ddagger$'' from anti-baryons are ignored because this channel is subdominant, as we can see in the right plot of Fig.$\,$\ref{fig:XnEvo}.

\item {\bf Include temperature dependence of reaction rates.} 
    Since the temperature dependence of cross sections can be significant, using the updated cross sections, 
    we fully incorporate their temperature dependence in the Boltzmann equations.     
\end{itemize}
With these improvements, we will find that the contributions of all the mesons are relevant. This suggests that the community should like to revisit existing studies of hadronically decaying long-lived particles during BBN.

\subsection{Qualitative picture behind our bounds}
\label{sec:QualitativePicture}
We can briefly discuss how hadronic injections with $\tau_a \sim {\cal O}(10^{-2})\,\sec$ change $X_n$ and how robust our bounds are.
In the standard BBN, $X_n$ freezes out when the $n \leftrightarrow p$ conversion processes,
\begin{align}
    n + \nu_e \leftrightarrow p + e^-  
    \quad \text{and} \quad
    n + e^+ \leftrightarrow p + \bar\nu_e \, ,
\end{align}
become slower than the Hubble expansion rate $H$. 
Being mediated by the weak interactions, these reaction rates are small: $\Gamma^{\rm weak}_{n\leftrightarrow p}\sim \langle\sigma v\rangle n_{e,\nu}\sim 10^{-24}\,\GeV\, (T/\MeV)^5$.
This smallness implies that even an exponentially suppressed amount of hadronic injection can still contribute to $n\leftrightarrow p$ conversion.
The conversion rate by an injected hadron $h$ is given by 
\begin{align}
\Gamma^{h}_{n \leftrightarrow p}  
&=\langle\sigma v \rangle_{n \leftrightarrow p} n_h
\sim  \langle\sigma v \rangle_{n \leftrightarrow p}
\frac{{\rm N}_{a\to h} \Gamma_a n_a}{\langle\sigma v \rangle_{\rm dis} n_b + \Gamma_h }
\label{Eq:Gammah_np}
\end{align}
where ${\rm N}_{a\to h}$ is the effective number of $h$'s produced per axon decay, and $\langle \sigma v \rangle_{n \leftrightarrow p}$ is the sum of the average cross sections for the $n \to p$ and $p\to n$ conversions due to scattering with $h$.
The estimate of $n_h$ in the rightmost expression in Eq.\,\eqref{Eq:Gammah_np} comes from balancing the appearance and disappearance rates of $h$.

To estimate $\Gamma^{h}_{n \leftrightarrow p}$ in Eq.\,\eqref{Eq:Gammah_np},  
recall that, as we already discussed below Eq.\,\eqref{eq:nhdot-disappearance}, 
the $\Gamma_h$ term in the denominator of Eq.\,\eqref{Eq:Gammah_np} is smaller than or comparable to the $\langle\sigma v\rangle_{\rm dis}$ term. 
So, we can drop $\Gamma_h$ and then cancel $\langle\sigma v\rangle_{\rm dis}$ with $\langle \sigma v \rangle_{n \leftrightarrow p}$ as they are on the same order of magnitude.
In addition, due to relativistic decoupling, the axion number density is quite large, $n_a\sim [n_\gamma/g_*^{\rm FO}] e^{-t/\tau_a}$ with $g_*^{\rm FO}\sim 100$, 
in comparison with $n_b \sim 6\times 10^{-10} n_\gamma$.
All combined, we obtain
\begin{align}
\Gamma^{h}_{n \leftrightarrow p}  
\sim 10^{-15}\,\GeV \cdot 
e^{-t/\tau_a} \, {\rm N}_{a\to h} \,
\frac{10^{-2}\,\sec}{\tau_a} 
\label{Eq:Gammah_np_2}
\end{align}
which is vastly larger than $\Gamma_{n \leftrightarrow p}^\text{weak}$. 
Therefore, at these early times, hadronic injections from axion decays completely dominate over the SM reactions.

At later times, the $e^{-t/\tau_a}$ suppression of $n_a$ eventually makes $\Gamma^{h}_{n \leftrightarrow p}$ go below $\Gamma_{n \leftrightarrow p}^\text{weak}$, after which the standard weak interactions take over $n \leftrightarrow p$ conversion.  
We can estimate this transition time, $t=t_{\rm wf}$, by solving $\Gamma^{\rm weak}_{n\leftrightarrow p} = \Gamma^h_{n\leftrightarrow p}$, which gives
\bal
t_{\rm wf} \sim \tau_a \cdot 
\left[
18 + 
\log\!\left(
{\rm N}_{a\to h}
\, \frac{10^{-2}\,\sec}{\tau_a}  
\right)\!
\right]
\label{eq:twfsim}.
\eal
Here, ``wf" stands for ``waterfall" since $X_n$ falls rapidly and drastically from a high value due to a large $\Gamma^{h}_{n \leftrightarrow p}$ down to nearly the standard evolution of $X_n$ due to $\Gamma_{n \leftrightarrow p}^\text{weak}$.
Since the $n\leftrightarrow p$ conversion rate is exponentially larger than the standard rate until the time becomes very close to $t_{\rm wf}$, 
we see that any $t_{\rm wf}$ after the standard neutron decoupling time, $t_{\rm wf}>t_n\sim 0.7\,\sec $ ($\tau_a \gtrsim 0.04\,\sec$), is robustly ruled out.
This concludes a qualitative explanation of how our bounds work, 
where we have seen how BBN places a robust bound around $\tau_a \sim 10^{-2}\>\sec$ for heavy QCD axions.

Once the Boltzmann equations as presented in Sec.\,\ref{sec:n-decoupling} are solved to obtain a more accurate bound, we find that the above rough estimate is fairly accurate, only off by a factor of a few.
The accurate bound with all parameters included will be derived in Sec.$\,$\ref{sec:simplified} and presented in Eq.\,\eqref{eq:taua_bound}. 
Here, we present a simplified version with only ${\rm N}_{a\to h}$ dependence shown:  
\bal
\tau_a 
\lesssim 
\dfrac{0.02\>\sec }
{1+ \dfrac{1}{18}\log \left[ {\rm N}_{a\to h} 
\right]}
\,.\label{Eq:bound_qualitative}
\eal
This shows that the lifetime bound is quite insensitive to the modification of the axion decay patterns, indicating that  
our upper bound on the lifetime around $0.02\,\sec$ is nearly model-independent. 
Similarly, the result is also only logarithmically dependent on other parameters such as hadronic cross sections, the initial abundance, and $R_{\Yp}$ in Eq.\,\eqref{Eq:criterion_deltaXn}.


\section{Thermal Production and abundance of Heavy QCD Axions}\label{sec:thermal}

\begin{figure}[t]
    \centering
    \hspace{-1.1em}
    \includegraphics[width=0.48\textwidth]{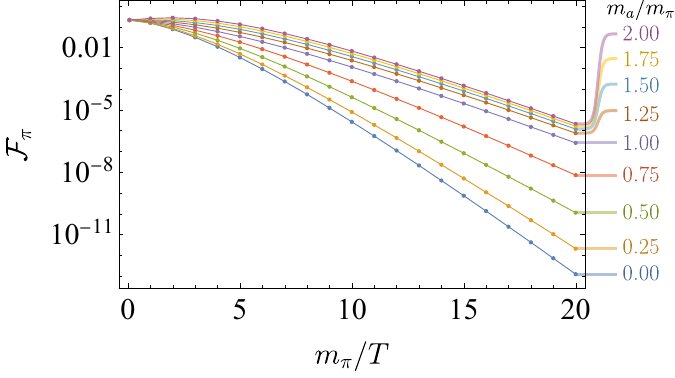}
    \caption{
    ${\cal F}_\pi(m_a, T)$ used in Eq.\,\eqref{Eq:a_pi} for different values of $m_a$. The dots indicate our numerical estimation, and the lines are interpolations of the dots.
    \label{Fig:Fpi}}
\end{figure}

In this work, we estimate the axion abundance by following~\cite{Dunsky:2022uoq}. While their focus is the study of $N_{\rm eff}$ bounds on heavy QCD axions, we share the same axion production mechanism from the $aG\tilde{G}$ coupling. 
We assume that the post-inflation reheating occurs at a sufficiently high temperature that the axions first thermalize with the SM particles and then chemically decouple from the SM sector later, freezing out their number density.
This sets the initial axion abundance for our BBN analysis before the axions begin to decay.
We simply assume an instantaneous freeze-out at $T = T_{\rm FO}$ and adopt a thermal Bose-Einstein distribution (with zero chemical potential) for the axion number density.
As we will see, our bounds are insensitive to the precise value of the axion abundance so this simplification is not a limitation of our analysis. 
Then, below the freeze-out temperature, the Boltzmann equation for the axion number density $n_a$ reads
\bal
\dot n_a + 3H n_a = - \Gamma_{\rm dis} \, (n_a - n_a^{\rm (eq)}) \,,
\eal
where $\Gamma_{\rm dis}$ is the rate of axion disappearance (neglecting processes that reduce the number of axions by more than one), and $n_a^{\rm (eq)}$ is the equilibrium axion number density determined by the temperature and axion mass, where we ignore thermal contributions to $m_a$ and treat it as constant.

Since we are looking at lifetimes of roughly $10^{-2}\>\sec$ or longer, we can completely ignore axion decays in $\Gamma_{\rm dis}$.
Then, before the deconfinement-confinement crossover, $\Gamma_{\rm dis}$ is dominated by $a g \to g g$. 
Its reaction rate is given by
\bal
\Gamma_{ag\to gg}(T) = \frac{16}{\pi}
\left( \frac{g_3^2(T)}{32\pi^2} \right)^{\!\!2} \frac{T^3}{f_a^2}
{\cal F}_g(T) \,,
\label{Eq:Gamma_ag_gg}
\eal
where we take ${\cal F}_g(T)$ from Ref.\,\cite{Salvio:2013iaa, DEramo:2021psx, DEramo:2021lgb}, 
and evaluate the QCD coupling $g_3$ by taking the $\overline{\rm MS}$ scale $\mu=T$.
$g_3(\mu)$ diverges as it approaches to $\mu = \Lambda_{\rm QCD} \simeq 300\,\MeV$, so the perturbative expansion becomes eventually invalid.
Therefore, to ensure the validity of Eq.\,\eqref{Eq:Gamma_ag_gg} we need to restrict $T$ to be sufficiently high, and we choose $T>T_{\rm QCD} \equiv 2\>\GeV$. 

After the deconfinement-confinement crossover, axion interactions are described in terms of hadrons rather than quarks/gluons.
For $\Gamma_{\rm dis}$ we only include the pion-induced process $a\pi \leftrightarrow \pi \pi$, whose rate is given by
\bal
\Gamma_{a\pi \to \pi \pi}(T)
=
\frac{T^5}{f_a^2f_\pi^2} \frac{A^2}{(1-r^2)^2} {\cal F}_\pi (m_a, T) \,,
\label{Eq:a_pi}
\eal
where $r=m_a/m_\pi$ and $A=\frac{1}{3}(m_d-m_u)/(m_d+m_u)$, 
and ${\cal F}_\pi (m_a, T)$ is evaluated by following~\cite{Dunsky:2022uoq}. 
Our evaluation of ${\cal F}_\pi (m_a, T)$ is plotted in Fig.\,\ref{Fig:Fpi}, where different colors correspond to various choices of $m_a$ as indicated to the right of the plot.%
\footnote{Fig.\,16 of~\cite{Dunsky:2022uoq} disagrees with our result
due to an error confirmed by the authors of~\cite{Dunsky:2022uoq}, although the error does not affect their other analysis and results.}

In contrast to pQCD, the approximation $\Gamma_{\rm dis}  = \Gamma_{a\pi \to \pi\pi}$ is good for sufficiently low values of $T$, and we choose $T < T_\pi \equiv  0.1\,\GeV$.
The validity of this approximation requires the following conditions: 
(i) Primakoff processes should be suppressed;  
(ii) the decay and inverse decay rates must be sufficiently small; and 
(iii) the freeze-out temperature must be sufficiently low, e.g., $T_{\rm FO} < T_\pi$, since we have ignored processes involving the other mesons. 
The condition (i) is satisfied with $f_a > 10^5\,\GeV$, 
while the remaining conditions are well satisfied in the parameter space of interest, that is, $m_a \gtrsim 300\,\MeV$ and $\tau_a \lesssim 0.1\,\sec$.

\begin{figure}[t]
    \includegraphics[width=0.48\textwidth]{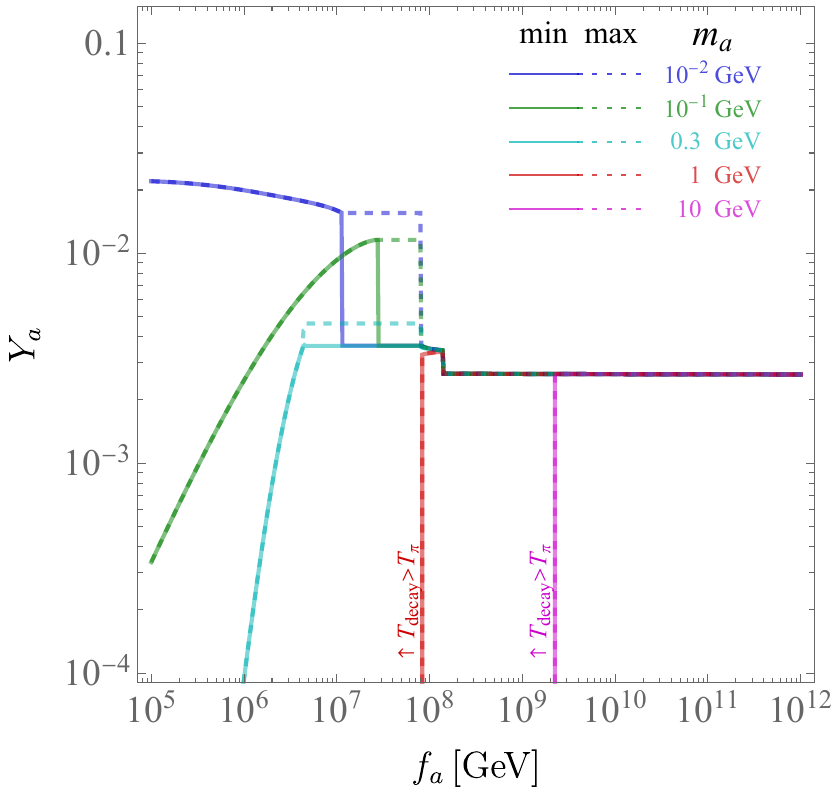}
    \caption{
    $Y_a^{\rm (min)}$ (solid lines) and $Y_a^{\rm (max)}$ (dashed lines) as functions of $f_a$ for various choices of $m_a$ represented by different colors.
    The sharp drops at $f_a \sim 10^8$ and $10^9\,\GeV$ for the $m_a=1$ and $10\,\GeV$ cases, respectively, are our truncations by hand because their respective temperature $T_{\rm decay}$ becomes above $T_{\pi}$ in the region indicated by ``$\leftarrow$''. 
     }
    \label{fig:Ya_individual}
\end{figure}

\begin{figure*}[t]
    \includegraphics[width=0.45\textwidth]{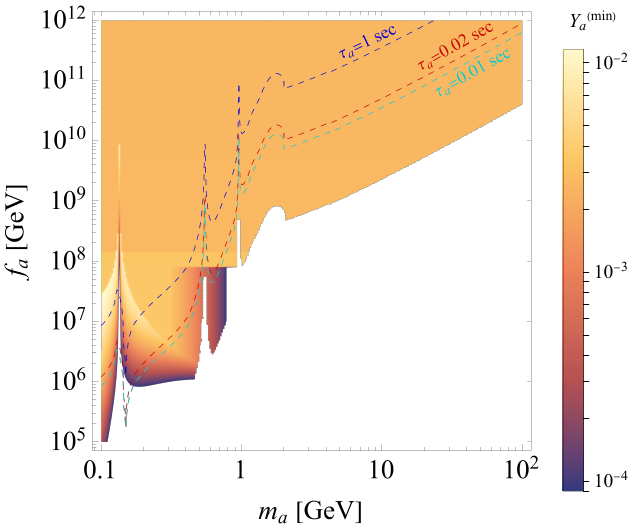}
    \includegraphics[width=0.45\textwidth]{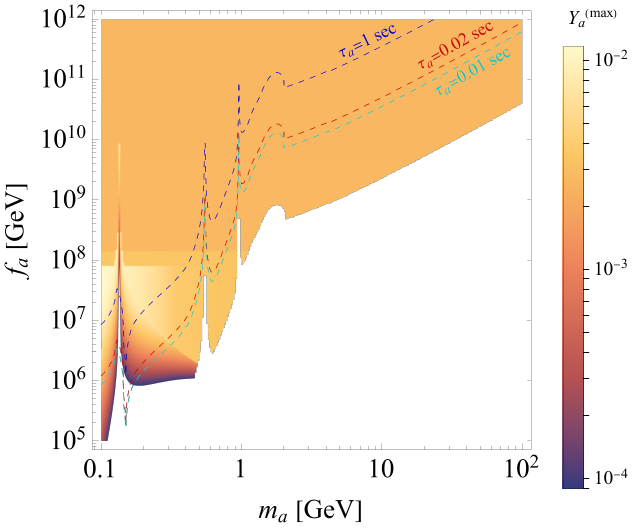}
    \includegraphics[width=0.45\textwidth]{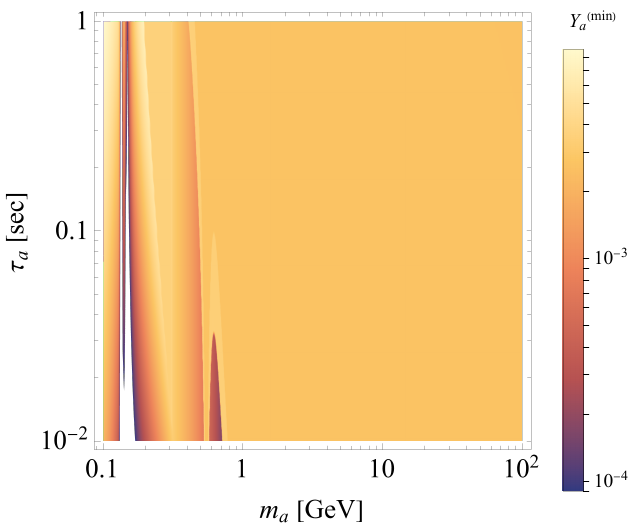}
    \includegraphics[width=0.45\textwidth]{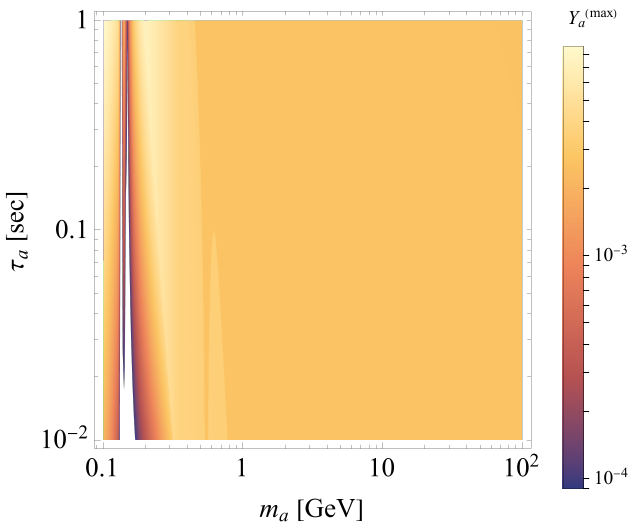}  
    \caption{The left (right) plots show $Y_a^{\rm (min)}$ ($Y_a^{\rm (max)}$) in the $(m_a,\,f_a)$ space (top) and in the  $(m_a,\,\tau_a)$ space (bottom).
    The blue, red, and cyan dashed lines in the upper plots correspond to $\tau_a=1\,\sec$, $0.02\,\sec$, and $0.01\,\sec$, respectively. }
    \label{fig:Ya}
\end{figure*}

Combining the two cases, we take the axion disappearance rate to be given by
\bal
 \Gamma_{\rm dis}(T)  = 
\begin{cases}
    \Gamma_{ag\to gg}(T) & \text{for $T>T_{\rm QCD}$} \,,
    \\
    \Gamma_{a\pi\to \pi\pi}(T) & \text{for $T<T_{\pi}$} \,,
\end{cases}
\eal
but we still do not know $\Gamma_{\rm dis}$  for $T_{\pi}<T<T_{\rm QCD}$.
To deal with this lack of information, we adopt the prescription of~\cite{Dunsky:2022uoq}:%
\begin{enumerate}
\item
First, if $\Gamma_{a\pi\to \pi \pi}(T_\pi) > 3H(T_\pi)$, the freeze-out temperature is determined by $\Gamma_{a \pi \to \pi \pi}(T_{\rm FO})=3H(T_{\rm FO})$.
\item
Otherwise, i.e., if $\Gamma_{a\pi\to \pi \pi}(T_\pi) < 3H(T_\pi)$, we check the freeze-out with $\Gamma_{ag\to gg}$.  
If $\Gamma_{ag\to gg}(T_{\rm QCD}) < 3H(T_{\rm QCD})$, we adopt the freeze-out temperature from $\Gamma_{a g \to g g}(T_{\rm FO})=3H(T_{\rm FO})$. 
\item 
The situation is uncertain if $\Gamma_{a\pi\to \pi \pi}(T_\pi) < 3H(T_\pi)$ and $\Gamma_{ag\to gg}(T_{\rm QCD}) > 3H(T_{\rm QCD})$, which implies that $T_{\rm FO}$ is somewhere between $T_\pi$ and $T_{\rm QCD}$.
To estimate the range of uncertainties due to not knowing $T_{\rm FO}$, we scan $T_{\rm FO}$ between $T_\pi$ and $T_{\rm QCD}$ and take the smallest and largest values of $Y_a = n_a^{\rm (eq)}(T_{\rm FO})/s(T_{\rm FO})$,  denoted as $Y_a^{\rm (min)}$ and $Y_a^{\rm (max)}$, respectively, and will study how results depend on $Y_a^{\rm (max)}$ versus $Y_a^{\rm (min)}$. 
\end{enumerate}

Although our bound is insensitive to the precise value of $Y_a$ as emphasized in Sec.\,\ref{sec:QualitativePicture} and our scheme above is sufficient for our purpose, there are subtleties in a precise estimation of the axion production rate across the QCD deconfinement-confinement crossover. 
This has been studied for conventional axion models in terms of axion-hadron interactions after the crossover\,\cite{Berezhiani:1992rk, Chang:1993gm, Hannestad:2005df, DEramo:2014urw, Kawasaki:2015ofa, Ferreira:2020bpb, DiLuzio:2021vjd} as well as in terms of axion-gluon\,\cite{Masso:2002np, Graf:2010tv, Salvio:2013iaa} and axion-quark\,\cite{Ferreira:2018vjj, Arias-Aragon:2020qtn} interactions before the crossover.
Estimations incorporating all these interactions are performed in~\cite{Giare:2020vzo} and with a smooth interpolation in~\cite{DEramo:2021lgb, DEramo:2021psx}.
In addition, \cite{Notari:2022ffe} points out the importance of strong sphalerons, and \cite{Bianchini:2023ubu} refines it with the DESI result.

Our estimation of $Y_a$ is shown in Fig.~\ref{fig:Ya_individual}.
For each choice of $m_a$, 
$Y_a^{\rm (min)}$ ($Y_a^{\rm (max)}$) is given a function of $f_a$ by a solid (dashed) line. 
Different values of $m_a$ are represented by different colors.
For $m_a\gtrsim 1\,\GeV$, there is no difference between $Y_a^{\rm (min)}$ and $Y_a^{\rm (max)}$.
The sharp drops at $f_a\sim 10^8$ and $10^9\,\GeV$ for the $m_a=1$ and $10\,\GeV$ cases, respectively, are our truncations reflecting the fact that, below those values of $f_a$, the respective decay temperature $T_{\rm decay}$ defined via $\Gamma_a = 3H|_{T=T_{\rm decay}}$ becomes above $T_\pi$, thereby modifying nothing in BBN\@.

In Fig.$\,$\ref{fig:Ya} top-left (top-right), the values of $Y_a^{\rm (min)}$ ($Y_a^{\rm (max)}$) are shown in the $(m_a, f_a)$ space, while a few values of the axion lifetime are indicated by the dashed lines for $\tau_a=1\,\sec$ (blue), $0.02\,\sec$ (red), and $0.01\,\sec$ (cyan). (See Sec.$\,$\ref{sec:axiondecays} for how the lifetime is evaluated.)
In the blank region, BBN is not modified because either $T_{\rm decay}>T_\pi$ at $m_a \gtrsim 1\,\GeV$, or $Y_a$ is exponentially suppressed due to $T_{\rm FO}\ll m_a$ at $m_a \lesssim 1\,\GeV$.

The bottom plots of Fig.$\,$\ref{fig:Ya} again show $Y_a^{\rm (min)}$ and $Y_a^{\rm (max)}$ but this time in the $(m_a,\tau_a)$ space.
Since the axions freeze out while still relativistic for $m_a \gtrsim 1\,\GeV$, 
$Y_a$ is large (about $3\times10^{-3} \gg \eta_b$) in most of the space ($m_a >0.3\,\GeV$ and $0.01\>\sec < \tau_a < 1\>\sec$).
Observe that a noticeable difference between $Y_a^{\rm (min)}$ and $Y_a^{\rm (max)}$ only appears in a narrow region of $0.2\,\GeV\lesssim m_a \lesssim 0.7\,\GeV$.

Due to the large axion yield, its energy density can temporarily dominate the universe as matter until the axions begin to decay into radiation.
The universe is axion matter dominated while the temperature $T$ falls in the range, 
\bal
6\>\MeV \left( \frac{m_a}{3\,\GeV}\right)\! \!\left( \frac{Y_a}{3\times 10^{-3}}\right)
\gtrsim T \gtrsim T_{\rm decay} 
\,,\label{Eq:axion_domination}
\eal
where the beginning temperature comes from comparing $\rho_a = m_a Y_a s(T)$ with $\rho(T) = \frac{4}{3}Ts(T)$. 
This shows that as the axion mass is increased, the axion-dominated period becomes longer, and the standard radiation-dominated background cosmology becomes increasingly inappropriate.
Our numerical codes properly handle the presence of axion domination by solving the Boltzmann equation for the axion number density (after freeze-out) and appropriately modifying the Hubble parameter by the axion energy density.

Additionally, since the baryon asymmetry is diluted during the axion-dominated epoch, we begin with a larger initial baryon asymmetry such that the dilution reduces it to the correct value of $\eta_b$ measured by CMB\@. 
We will give more detail in Sec.$\,$\ref{sec:numerical-steps}.


\section{Hadronic Axion Decays}\label{sec:decay}
\label{sec:axiondecays}
We have two frameworks depending on $m_a$.
For $m_a >2\,\GeV$, we use pQCD and consider two hadronization models: the string fragmentation model used in {\tt PYTHIA} and the cluster model used in {\tt Herwig}.
For $m_a < 2\,\GeV$, both pQCD and these hadronization algorithms fail, so we instead adopt a data-driven method proposed and developed in~\cite{Aloni:2018vki}.
This framework is based on the chiral perturbation theory and vector meson dominance, 
but extends the domain of validity to higher energies than $\sim 4\pi f_\pi$ with the help of experimental data.
While this is not a controlled approximation, it gives sensible results for the limited purpose of estimating the decay width and various branching fractions of the axion, as validated in~\cite{Aloni:2018vki}.

We will present the following quantities relevant for our BBN analysis: (i) the total decay width, (ii) the average number of each hadron per axion decay, and (iii) the energy spectrum of $K_L$.
As we discussed in Sec.$\,$\ref{sec:overall}, the hadrons that can affect the neutron-to-proton ratio are limited to the $\pi^\pm$, $K^\pm$, $K_L$, (anti-)proton, and (anti-)neutron. 
These are sufficiently long-lived to have time to interact with the nucleons in the plasma. 
Therefore, for (ii), we only count the average number of each of these hadrons per axion decay.
Other hadrons, e.g., $K_S$, hyperons, etc., decay before they have chance to scatter with the nucleons.


\begin{figure}[t]
    \includegraphics[width=0.48\textwidth]{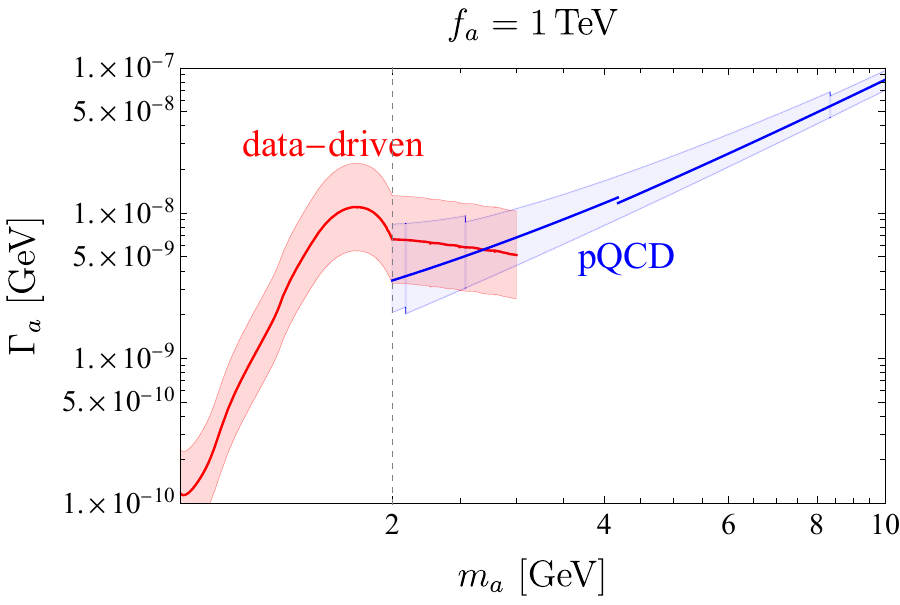}   
    \caption{The total decay width $\Gamma_a$ as a function of $m_a$. Here, $f_a$ is fixed to $1\,\TeV$, but one can rescale the rate by $(\TeV/f_a)^2$ for a different value $f_a$. 
    Estimations based on pQCD and the data-driven method are shown in blue and red, respectively.
    The bands represent the uncertainties of the estimations, where the blue band is determined by varying the renormalization scale from $\mu=m_a/2$ to $\mu=2m_a$, while we assign a factor-of-$2$ uncertainty to the data-driven estimation for the red band.}  
    \label{fig:a-width}
\end{figure}

\begin{figure*}[t!]
    \includegraphics[width=0.48\textwidth]{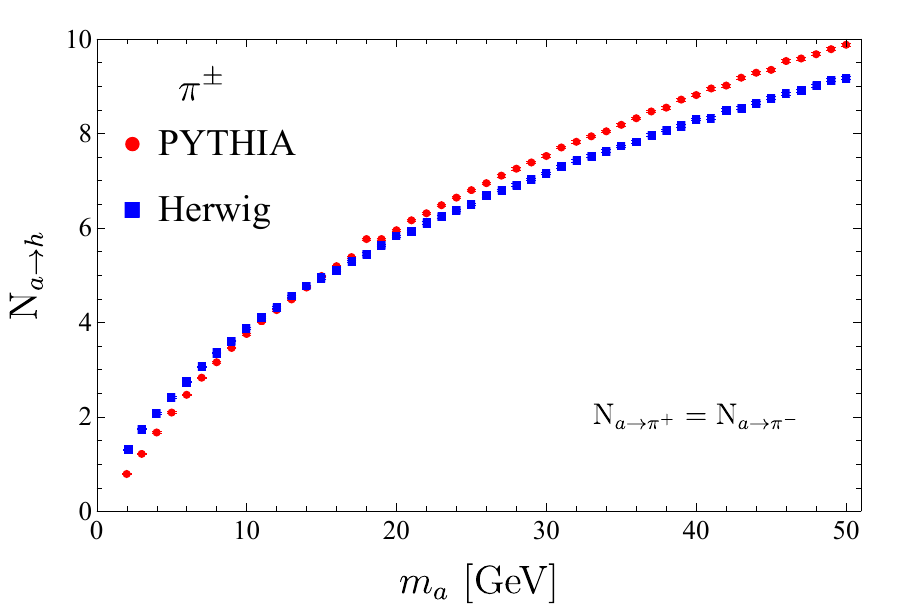}
    \includegraphics[width=0.48\textwidth]{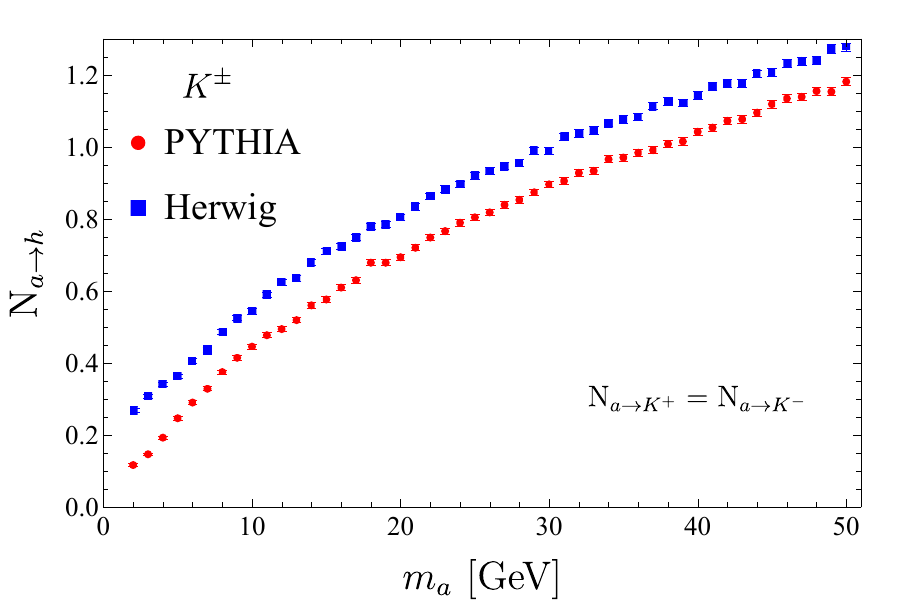}
    \includegraphics[width=0.48\textwidth]{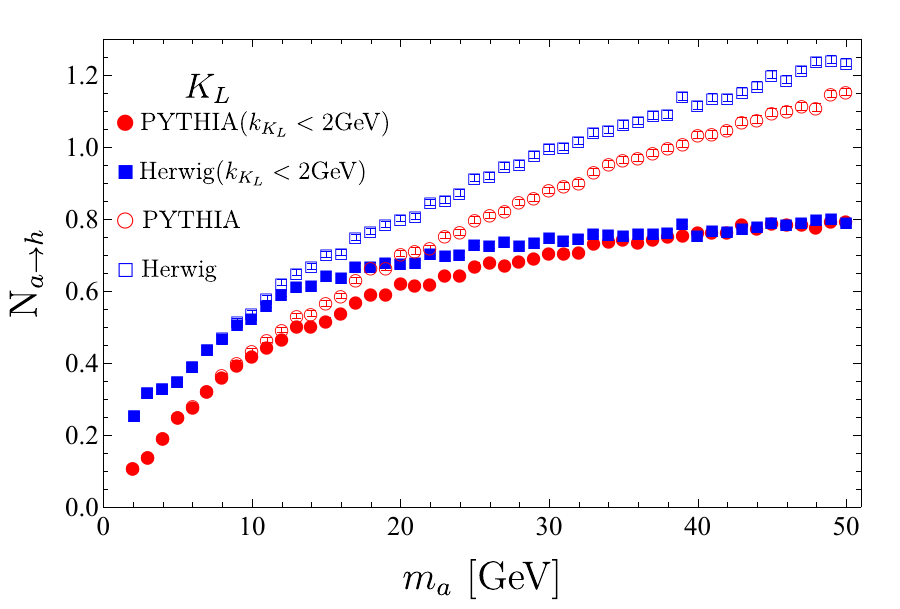}   
    \includegraphics[width=0.48\textwidth]{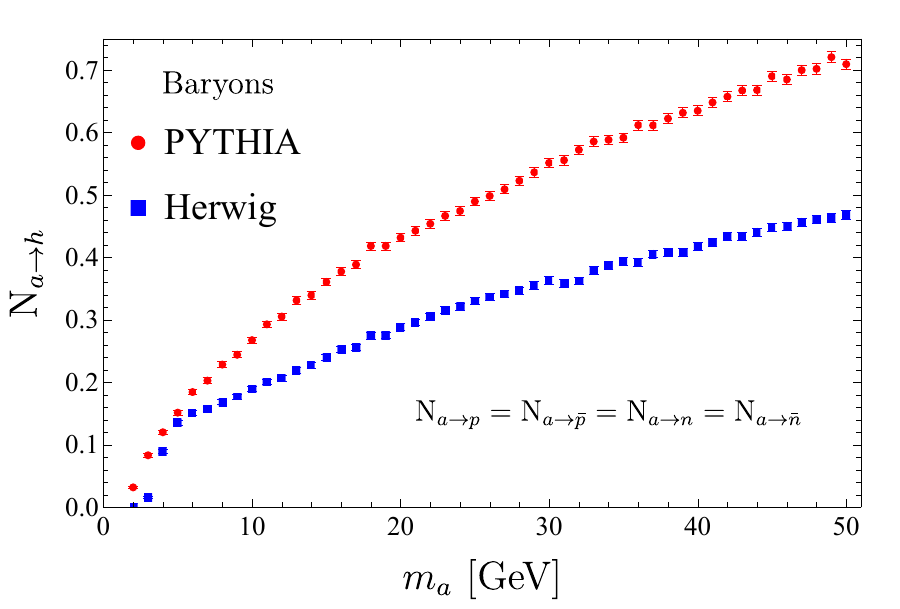}
    \caption{Average numbers of mesons and (anti-)baryons as functions of the axion mass. The error bars are due to statistical uncertainties of finite simulation samples. For $K_L$, the momentum range is restricted as $k_{K_L}<2\,\GeV$ in our analysis, but the original numbers without this cut are shown for comparison.  
    The tabulated samples are given in Ref.\,\cite{github}.
    \label{fig:NhadShower}
    }    
\end{figure*}

\subsection{$m_a>2\,\GeV$}
For $m_a>2\,\GeV$, we obtain the total decay width by using the NNLO calculation of pseudo scalar decaying to gluons in Ref.\,\cite{Chetyrkin:1998mw}.
We obtain
\bal
\Gamma(a\to gg) =& 
\frac{2}{\pi} \left( \frac{\alpha_s(\mu)}{8\pi} \right)^{\! 2}
\frac{m_a^3}{f_a^2} 
\nonumber\\
&\times\left[1+A(\mu)\,\alpha_s(\mu)+B(\mu)\,\alpha_s^2(\mu)\right]
\eal
where $\mu$ is the renormalization scale. 
For the coefficient functions $A(\mu)$ and $B(\mu)$, see Eq.\,\eqref{eq:NNLO} in Appendix~\ref{app:a_decay}. 

We evaluate uncertainties in the total width by varying the renormalization scale from $\mu=m_a/2$ to $\mu=2m_a$, which is shown by the blue band in Fig.\,\ref{fig:a-width}. 
Note that what BBN directly constrains is the axion lifetime, so our bound on the lifetime is robust and independent of this uncertainty.
However, the bound on $f_a$ is affected by the uncertainty when it is translated from the bound on the lifetime through the above relation.

\begin{figure*}[t!]
    \includegraphics[width=0.48\textwidth]{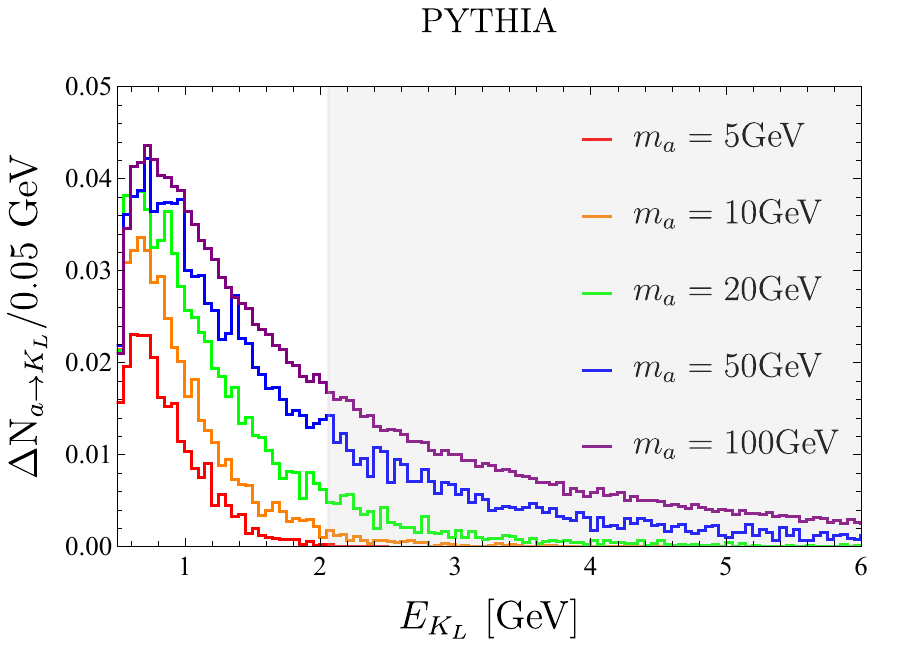}
    \includegraphics[width=0.48\textwidth]{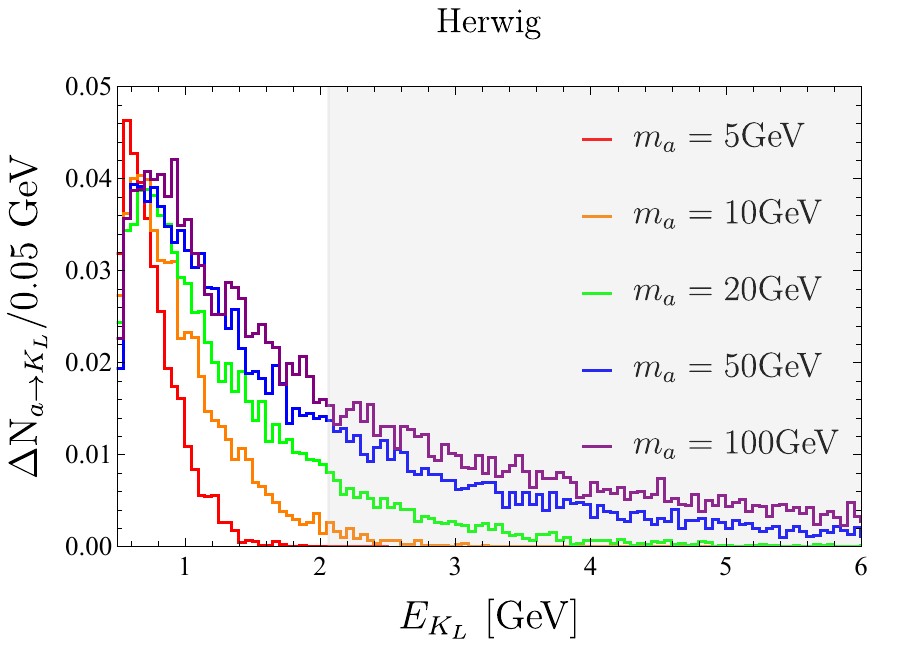}
    \caption{Energy distribution of $K_L$ from the axion decay for $m_a=5,10,20,50$ and 100\,GeV, obtained from {\tt PYTHIA} (left) and {\tt Herwig} (right). 
     \label{fig:KLmom}}    
\end{figure*}
\begin{figure}
    \includegraphics[width=0.48\textwidth]{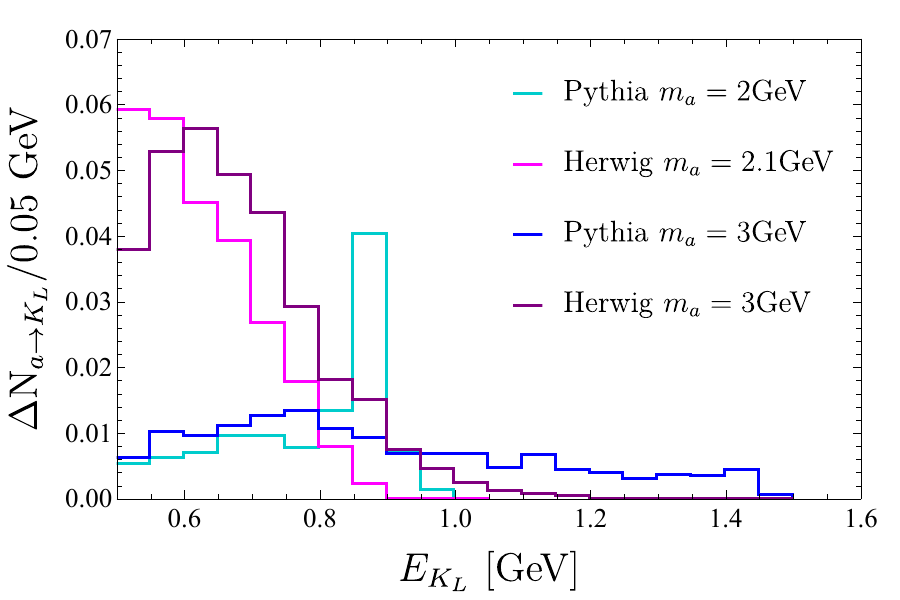}
    \caption{
    Energy distribution of $K_L$ from the axion decay for low axion masses, where
    large discrepancies between {\tt PYTHIA} and {\tt Herwig} predictions are present.
    }   
    \label{fig:KLmom_lowma} 
\end{figure}

For the branching fractions, we use the two most widely used parton-shower and hadronization programs \texttt{PYTHIA v8.306}\,\cite{Bierlich:2022pfr} and \texttt{Herwig v7.3.0}\,\cite{Corcella:2000bw, Bahr:2008pv, Bellm:2015jjp} to obtain the average numbers of $\pi^\pm$, $K^\pm$, $K_L$, $N$ and $\bar N$ from the axion decay.
We generate $10^4$ parton-showering and hadronization events of $a\to gg$ by {\tt PYTHIA} and {\tt Herwig} at each value of the axion mass. Since the predictions from the two programs are sometimes different, we retain the result using both of them.  

In Fig.\,\ref{fig:NhadShower}, the average numbers of hadrons per axion decay are presented, where the estimations from {\tt PYTHIA} and {\tt Herwig} are shown in red and blue, respectively.
The (nearly invisible) error bars represent the statistical uncertainties coming from the finite number of simulated events.
Note that indirect contributions from the decays of unstable hadrons such as $\rho \to \pi\pi$ are also included in the counts.
We find that the average number of pions per axion decay in the plotted axion mass range is $\sim 1$ -- $10$, while the average numbers of charged kaons and (anti-)nucleons are $\sim 0.2$ -- $1.2$ and $\sim 0$ -- $0.7$, respectively. Since our framework of $K_L N$ scattering is within only $2\to 2$ scattering processes, we restrict the $K_L$ momentum to be less than 2\,GeV in the axion rest frame, which will be the rest frame of the background nucleons during BBN. For comparison, we also include the average $K_L$ numbers without this restriction. 
While the two programs agree fairly well for the mesons at large values of $m_a$, ${\cal O}(1)$ discrepancies are present at all $m_a$ for the baryons.

In Figs.\,\ref{fig:KLmom} and \ref{fig:KLmom_lowma}, the predicted energy distributions of $K_L$ by {\tt PYTHIA}  and {\tt Herwig}, respectively, are presented for various values of $m_a$.
As emphasized in Sec.\,\ref{sec:improvements}, this $K_L$'s spectrum is distorted by elastic scattering of $K_L$ with background baryons. The reshaped spectra are shown in Figs.\,\ref{fig:KLmom_reshaped} and \ref{fig:KLmom_lowma_reshaped}.

\begin{figure*}[t]
    \includegraphics[width=0.48\textwidth]{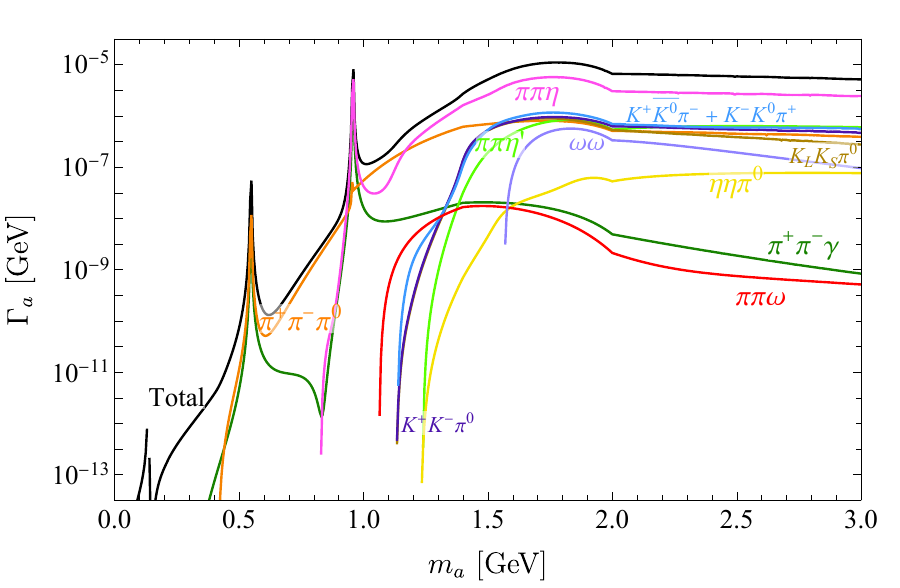}
    \includegraphics[width=0.48\textwidth]{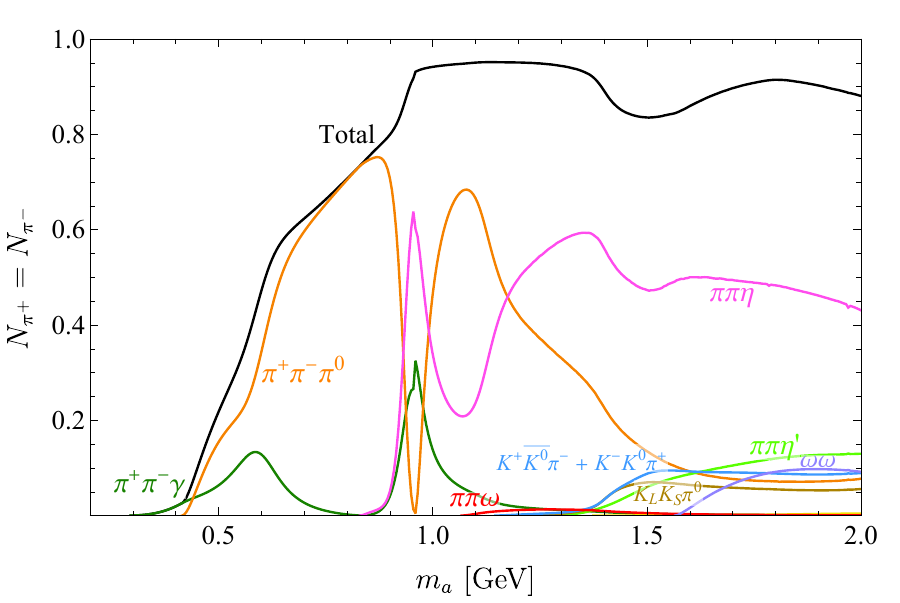}
    
    \caption{{\it Left}: The axion decay widths from the data-driven method for $f_a=1\,\TeV$ as a function of $m_a$, taken from Ref.\,\cite{Bisht:2024hbs}. 
    {\it Right}: Effective number of $\pi^+$ 
    (which is also equal to that of $\pi^-$) from each decay channel.
    The $a\to 3\pi^0$ decay is not shown since the $\pi^0$'s decay rapidly to photons before interacting with the background baryons.   
    \label{fig:decaymode}}
\end{figure*}

\begin{figure}
    \centering
    \includegraphics[width=0.48\textwidth]{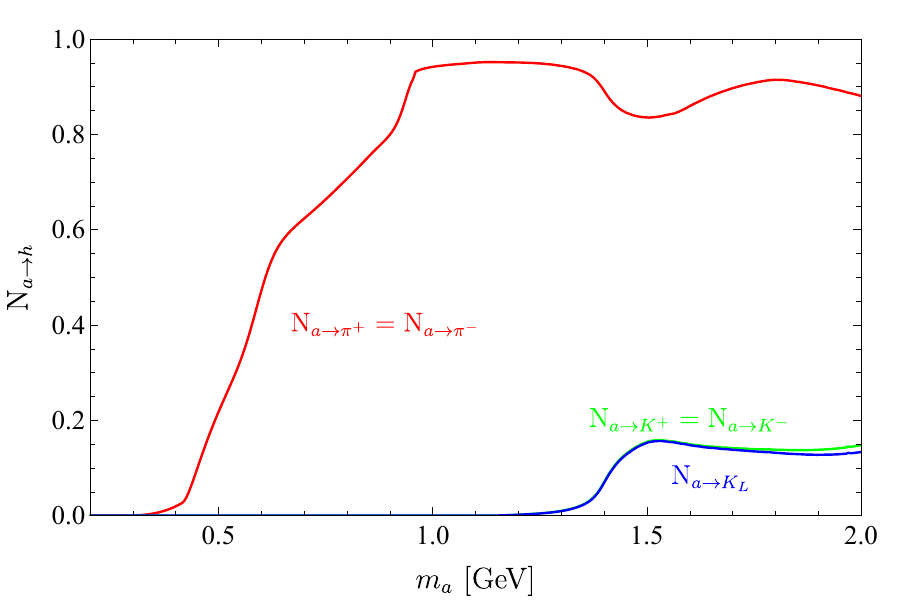}
    \caption{
    Effective number of $\pi^\pm, K^\pm, K_L$ per axion decay as a function of $m_a$. \label{fig:decaynumber}}
\end{figure}

We have also found fundamental problems with hadronization by both programs. For example, neither program respects the fundamental property of the axion that it is a pseudo-scalar. {\tt PYTHIA} allows the axion to decay to two pions. While this particular forbidden decay happens to be absent in {\tt Herwig}, that is not because it respects parity but because of an accidental outcome of kinematical restrictions imposed in its cluster hadronization model.
We add some relevant discussion in Appendix\,\ref{app:hadron}.

Fortunately, none of these subtleties impacts our final result. The $\mathcal{O}(1)$ discrepancy in the baryon counts is inconsequential because our final constraint is dominantly determined by pions and kaons.   
The lifetime upper bound is only logarithmically sensitive to changes in the amounts of hadrons as one has seen in the ${\rm N}_{a\to h}$ dependence in Eq.\,\eqref{Eq:bound_qualitative}.  The $K_L$ spectrum uncertainties affect the averaged cross sections by at most 20\% (see Fig.~\ref{fig:KLNxsec}), but they also only logarithmically impact the lifetime bound as seen in~Eq.\,\eqref{eq:taua_bound}.
The absences of impacts of these subtleties are seen in Fig.$\,$\ref{fig:compare} left; 
also see Sec.$\,$\ref{sec:simplified} for further discussions.

\subsection{$m_a < 2\,\GeV$}

For $m_a < 2\,\GeV$, not only do perturbative QCD calculations become increasingly inaccurate but also the shower/hadronization by \texttt{PYTHIA} and \texttt{Herwig} fail.  
Instead, we adopt a data-driven method based on the chiral perturbation theory and vector meson dominance theory, which was first developed in Ref.\,\cite{Aloni:2018vki} and later refined in Ref.\,\cite{Cheng:2021kjg}.
We take branching fractions necessary for our purpose from Ref.\,\cite{Bisht:2024hbs}.
See also Refs.\,\cite{Bai:2025fvl, Balkin:2025enj} for recent calculations.
Again, recall that our bound on the axion lifetime is not sensitive to the precise values of the branching fractions as can be seen in Eq.\,\eqref{Eq:bound_qualitative}.

Adding up all the decay widths of individual exclusive decay modes from the data-driven method, we obtain the total decay width of the axion, shown in red in Fig.~\ref{fig:a-width}.
Since there is no simple way to estimate uncertainties in the data-driven method, we assign a factor-of-$2$ uncertainty for the total decay width. 
As can be seen in Fig.$\,$\ref{fig:a-width}, the total decay width predictions from the pQCD and data-driven methods overlap quite well when the latter is extended to $m_a = 3\,\GeV$. 

The axion decay modes included in our analysis are $a\to \pi\pi\gamma, 3\pi$, $\eta\pi\pi$, $\eta'\pi\pi$, $KK\pi$, $\rho\rho$, $\omega\omega$, and $\gamma\gamma$, and their partial widths with $f_a=1\,\TeV$ are shown in Fig.~\ref{fig:decaymode} left. 
Unimportant modes like $a\to 3\pi^0$ are not shown there, but they are included in the total width. 
To obtain the average numbers of hadrons per axion decay relevant for BBN, we further decay short-lived particles such as $\rho$ down to $\pi^\pm$, $K^\pm$, and $K_L$.  Fig.~\ref{fig:decaymode} right shows how individual modes contribute to the effective number of $\pi^\pm$. 
The effective numbers of $K^\pm$, $K_L$, and $\pi^\pm$ are shown in Fig.\,\ref{fig:decaynumber}. 

The $K_L$ energy distribution is obtained from  $a\to KK\pi$ decay, where we assume a uniform amplitude in the Dalitz plot for simplicity. For $\pi^\pm$ and $K^\pm$, we only need their average number per axion decay presented in Fig.\,\ref{fig:decaynumber}.

\section{Neutron freeze-out \label{sec:n-decoupling}}

In the standard BBN, neutrinos are the first ones to freeze out, occurring around $T \sim 2\,\MeV$. 
As the universe continues to cool down across $T\sim m_e$, electrons and positrons annihilate into photons, injecting energy and entropy into the plasma. 
At around a similar temperature ($T= T_n\sim  1\,\MeV$), the weak interactions interconverting neutrons and protons become slower than the expansion rate and the neutron fraction $X_n$ freezes out.
After this neutron freeze-out, $X_n$ gradually decreases due to neutron $\beta$-decay until the temperature reaches the so-called deuterium bottleneck temperature $T_D$, above which the deuterium photo-dissociation rate due to the background photons is too high to initiate deuterium synthesis. 

Once $T$ drops below  $T_{\rm D}$ and a sufficient amount of deuterium is synthesized, a chain of nuclear processes happens, but the net result of this is 
that nearly all the neutrons end up in {\He}, with all other elements having exponentially suppressed abundances.
Thus, to an excellent approximation, the number of {\He} is given by half the number of neutrons at $T_{\rm D}$.
The latter can be calculated by solving the Boltzmann equation for $X_n$ without including its fusion processes.
The largest source of uncertainty in our calculation of $\Yp$ is the value of $T_{\rm D}$ because $X_n \propto \exp(-t_{\rm D}/\tau_n) \simeq 1 - t_{\rm D}/\tau_n$,  where $t_{\rm D}\sim 200\,\sec$ is the time corresponding to $T_{\rm D}$ and $\tau_n$ is the neutron lifetime (878.4\,s).
Nevertheless, as we discussed in Sec.$\,$\ref{sec:QualitativePicture}, 
our axion lifetime bound comes from $\delta X_n/X_n^\SM$, which is insensitive to the precise value of $T_{\rm D}$ as $\delta X_n$ and $X_n^\SM$ have approximately the same $T_{\rm D}$ dependence.
To maximally implement this cancellation of $T_{\rm D}$ sensitivities, we define $T_{\rm D}$ via $\Yp(\text{observed}) = 2X_n^\SM(T_{\rm D})$ 
with $X_n^\SM(T_{\rm D})$ calculated by the \emph{same} Boltzmann equations as what we use to calculate $X_n$ with the axion removed.
This insensitivity will be shown explicitly  
in Fig.~\ref{fig:XnEvo} right in Sec.\,\ref{sec:evolutions}.

\subsection{Boltzmann equations and modified background cosmology\label{sec:Boltzmann}}

The Boltzmann equation for $X_n$ depends on the Hubble expansion rate and the $n \leftrightarrow p$ conversion rates of the standard weak interactions and hadronic interactions with the injected hadrons from axion decays. 
We write evolution equations for (i) the photon temperature,  neutrino energy density, and axion number density; (ii) total baryon number density; (iii) number densities of hadrons in Table$\,$\ref{Tab:reactions}, 
and $X_n$. 
As illustrated in Fig.\,\ref{fig:skematic}, 
we solve the systems of equations (i)--(iii) in this order, 
where the temperature ranges are (i) $T_{\rm FO}>T>T_f=64\,\keV$, (ii) $T_i=30\,\MeV>T>T_f$, and (iii) $T_i > T > T_f$. In our Boltzmann equations, $T$ always means the photon temperature.

The evolution of $T$ below $T_i$ is determined by
\bal
\label{eq:rhodot_eg}
&\dot \rho_{\ell\gamma} + 3 H (\rho_{\ell\gamma}+p_{\ell\gamma}) \\
& = \Gamma_a n_a m_a
+  C_{\nu_e}(\rho_{\nu_e}^2-\rho_{\nu_e, {\rm eq}}^2)
+ 2C_{\nu_\mu}(\rho_{\nu_\mu}^2-\rho_{\nu_\mu, {\rm eq}}^2),  \nn 
\eal
where $\rho_{\ell\gamma}$ ($p_{\ell\gamma}$) is the sum of the energy densities (pressures) of $e^\pm$, $\mu^\pm$ and photons, $\rho_{\nu_i}$ ($i=e$, $\mu$) is the energy density of neutrino $\nu_i$,
$C_{\nu_e}= 0.68 \, G_F^2 T$ ($C_{\nu_\mu}= 0.3 \, G_F^2 T$) is the total rate for both charged and neutral current (only neutral current) processes~\cite{Cadamuro:2011fd}, 
and $\rho_{\nu_i, {\rm eq}}= 2 \cdot \frac{7}{8} \frac{\pi^2}{30} T^4$.
Since $\rho_{\ell \gamma}$ and $p_{\ell\gamma}$ are simply functions of $T$, Eq.\,\eqref{eq:rhodot_eg} can be used to convert between $T$ and $t$, as we will see in more detail in the next subsection.   
The factor of $2$ in front of the $C_{\nu_\mu}$ term of Eq.\,\eqref{eq:rhodot_eg} is to take into account $\nu_\tau$.

The Boltzmann equations of the neutrino sector can be written as
\bal
&\dot \rho_{\nu_e} + 4H \rho_{\nu_e} = - C_{\nu_e} (\rho_{\nu_e}^2-\rho_{\nu_e, {\rm eq}}^2), \nonumber\\ 
&\dot \rho_{\nu_\mu} + 4H \rho_{\nu_\mu} = - C_{\nu_\mu} (\rho_{\nu_\mu}^2-\rho_{\nu_\mu, {\rm eq}}^2). \label{eq:Boltzmann_nu}
\eal
We neglect neutrinos that are produced from axion decay chains, e.g., $a \to \pi^+ + \cdots \to \mu^+ \nu_\mu + \cdots \to e^+ \nu_e \nu_\mu + \cdots$ since this contribution is negligible for the lifetimes of our interest, $\tau_a=\mathcal{O}(0.01)\,\sec$; 
a bulk of the axion energy becomes the kinetic energies of primary daughter particles, all of which except $K_L$ quickly lose their energies into the plasma via electromagnetic interactions before they decay. 
Energy injection from axion decays into the neutrino sector could potentially be relevant to Eq.\,\eqref{eq:Boltzmann_nu} only after neutrino decoupling, but the axion energy density then would be very small as most of the axions have decayed by then.  
Given the axion energy injection into $\rho_{\ell\gamma}$ is already as small as $\mathcal{O}(1)$\,\%, energy injection into the neutrino sector from the axion decay chains is even smaller.

The Hubble expansion rate is affected by the axion energy density, 
\begin{align}
{H=\frac{1}{M_{\rm Pl}}
\sqrt{\frac{8\pi}{3}(\rho_{\ell \gamma} + \rho_\nu + m_a n_a)},
\label{eq:Hubble}
}\end{align}
with the Planck mass $M_{\rm Pl} = 1.221 \times 10^{19}\,\GeV$.
We neglect small contributions to the energy density coming from the baryon asymmetry and thermal hadrons.

The evolutions of the baryon and axion number densities are given by their conservation in the comoving box while including the decay rate of the axion;
\bal
&\dot{n}_b + 3H n_b = 0, \quad 
\\
&\dot{n}_a + 3H n_a = -\Gamma_a n_a .
\eal
And the Boltzmann equations for other hadron number densities denoted by $n_{h}$ can be written as
\begin{widetext}
\bal
\dot n_{h}  &= \nn
{\rm N}_{a\to {h}} \Gamma_a  n_a 
- \Big\{3H+
\tilde \Gamma_{h}+
\langle \sigma v (n+ {h} \to \cdots) \rangle n_n  +
\langle \sigma v (p+ {h} \to \cdots) \rangle n_p 
\Big\}  n_{h} 
\nn \\
&\quad 
+ \sum_{h'\neq h} 
\Big\{
{\rm N}_{{h'}\to {h}}  \tilde \Gamma_{h'} 
+
\langle \sigma v (n+ {h'} \to {{h}}\cdots) \rangle n_n  +
\langle \sigma v (p+ {h'} \to {h}\cdots) \rangle n_p 
\Big\} n_{h'} \label{eq:nAdot}
\eal
\end{widetext}
where $h(h')=\pi^+, \, \pi^-, \,  K^+, \,  K^-, \, K_L, \,  \bar p, \, $ and $\bar n$.  $\tilde \Gamma_{h(h')}$ is the effective decay width of $h$($h'$) including the averaged inverse boost factor for $K_L$ as we will discuss in Sec~\ref{sec:x-sec}. We ignore the boost of the other hadrons because they immediately become non-relativistic through thermalization.   
${\rm N}_{a\to {h}}$ is the averaged number of ${h}$ per axion decay (i.e. the sum of branching ratios multiplied by the multiplicity of ${h}$), see Figs.~\ref{fig:NhadShower} and \ref{fig:decaynumber}. The produced hadrons either decay or scatter against the background nucleon, including both \np conversion and non-conversion processes.%

The second line in Eq.\,\eqref{eq:nAdot} accounts for the secondary production of ${h}$ through the decay or scattering of another hadron ${h'}$, where we omit the multiplicity factors to keep the simplicity of the expression; for instance, we multiply a factor of 2 in the process of $nK^- \to Y\pi^-\to p\pi^-\pi^-$ for the case of $h=\pi^-$ and $h'=K^-$. If $K_S$ is produced, its decay product into $\pi^\pm$ is accounted for, assuming the decay is instantaneous.   

Secondary hadrons are necessary to correctly evaluate the net conversion effects, especially for channels initiated by a kaon, as discussed in the third bullet point of Sec.\,\ref{sec:improvements} and shown in Table~\ref{Tab:reactions}. However, we omit the secondary``$K_S^\ddagger$'' and ``$K_L^\ddagger$'' listed in Table~\ref{Tab:reactions} because the processes that produce them are only relevant at temperatures higher than $T_n$.
Secondary mesons from anti-nucleon–nucleon annihilation are also omitted, as this channel is subdominant. While they may become important at higher temperatures, we have checked that the conversion due to these secondary mesons remains subdominant around $T_n$.

Finally, the Boltzmann equation of $X_n$ is given by 
\bal
& \dot X_n = \frac{\dot n_n n_b -\dot n_b n_n}{n_b^2} = -\Gamma_{n\to p}X_n +\Gamma_{p\to n}(1-X_n) .   
\label{eq:Xndot}
\eal
The total conversion rates of $n\to p$ and $p\to n$, including both the SM- and axion-induced processes, are encoded in $ \Gamma_{n\to p}$ and $\Gamma_{p\to n}$, to be described in detail in Eqs.\,\eqref{eq:TotalNtoP} and \eqref{eq:TotalPtoN}.

Note that the above equations remain valid regardless of the axion energy density. 
They hold even during an axion-dominated era, Eq.\,\eqref{Eq:axion_domination}. 
However, our approximation $\rho_a = m_a n_a$ breaks down for very light axions, which are outside the scope of this work.

\subsection{Steps of numerical calculation \label{sec:numerical-steps}}

For interested readers who want to reproduce our results, here we present specific procedures to numerically solve the evolution. 

In order to solve the Boltzmann equations, we redefine various quantities to be dimensionless (we divide or multiply a power of the temperature and denote it with a hat, e.g., $\hat \rho_{\nu_i}= T^{-4}\rho_{\nu_i}$, $\hat n_b = T^{-3} n_b$, etc). The overall routine is outlined in Fig.~\ref{fig:skematic}. 
For the photon temperature $T$, we define a dimensionless variable
\bal
z\equiv m_e/T, 
\eal
which is commonly used in the BBN studies, e.g. see Ref.\,\cite{Pisanti:2007hk, Consiglio:2017pot, Gariazzo:2021iiu}. 
By the chain rule, we have $\dot \rho_{\ell\gamma} = \frac{\partial \rho_{\ell\gamma}}{\partial z} \dot z$, and therefore, from Eq.\,\eqref{eq:rhodot_eg},
\begin{align}
\dot z = \!\!\left( \frac{\partial \rho_{\ell\gamma}}{\partial z} \right)^{\!\!-1} \!
&\bigg\{\!\!-3H(\rho_{\ell\gamma} + p_{\ell\gamma} )+\Gamma_a n_a m_a
\\
&+ C_{\nu_e} (\rho_{\nu_e}^2-\rho_{\nu_i, {\rm eq}}^2)
+ 2C_{\nu_\mu} (\rho_{\nu_\mu}^2-\rho_{\nu_i, {\rm eq}}^2)\bigg\}
.
\nn
\label{Eq:zdot}
\end{align}
Assuming that the visible sector is always in thermal equilibrium, we take
\begin{align}
&\rho_\gamma (z)\!=\! 3 p_\gamma = 2\cdot \frac{\pi^2}{30} \left( \frac{m_e}{z} \right)^4,
\\
&\rho_{\ell} (z) \!= \!
4\left( \frac{m_e}{z} \right)^{\!\! 4}
\!\!
\int_0^\infty \!\!\!
\frac{d \hat p}{2\pi^2}
\frac{\hat p^2 \sqrt{\hat p^2 \!+ \!(\frac{m_\ell}{m_e})^{2}\,z^2}}{e^{\sqrt{\hat p^2 + (\frac{m_\ell}{m_e})^{2}\,z^2}}\!+\!1},   \label{eq:rhoe}
\\
&p_\ell (z) \!= \!4 \left( \frac{m_e}{z} \right)^{\!\! 4}
\!\!
\int_0^\infty \!\!\!
\frac{d \hat p}{2\pi^2}\!
\frac{\hat p^4}{\!3\!\sqrt{\hat p^2 \!+\!(\frac{m_\ell}{m_e})^2\,z^2}}
\frac{1}{e^{\sqrt{\hat p^2 + (\frac{m_\ell}{m_e})^{2}\,z^2}}\!+\!1}. \label{eq:pe}
\end{align}
with $\ell=e$ and $\mu$.

Firstly, we solve the Boltzmann equations for the background cosmology with 
\begin{align}
\label{Eq:Boltzmann_rhonui}
&\hat \rho'_{\nu_i} 
=
\left(
\frac{4}{z}
-\frac{4H}{\dot z}
\right) \hat \rho_{\nu_i}
-\frac{\Gamma_{\nu_i}}{\dot z}(\hat \rho_{\nu_i}^2 - \hat \rho_{\nu_i, {\rm eq}}^2 ) \quad {\rm for}~i=e,\,\mu,
\\
\label{Eq:Boltzmann_na}
&{\hat n}_a' =\left(\frac{3}{z} - \frac{3H}{\dot z} \right)\hat n_a -\frac{\Gamma_a}{\dot z}\hat  n_a , 
\end{align}
where $'$ denotes ${\frac{\partial}{\partial z}}$. From the axion freeze-out down to $T_i$, since the SM particles are in thermal equilibrium, we evaluate only Eq.\,\eqref{Eq:Boltzmann_na}, using $Y_a(T_{\rm FO})$ that is obtained in Sec.\,\ref{sec:thermal} for each ($m_a, f_a$) and $H$ accounting for all relevant degrees of freedom.%
\footnote{
Since it is before the neutrino decoupling temperature, the SM plasma energy density $\rho_{\rm SM}$ simply evolves with $\dot \rho_{\rm SM}+3H(\rho_{\rm SM}+p_{\rm SM})=\Gamma_a n_a m_a$, $\rho_{\rm SM}=\frac{\pi^2}{30}g_*(T)T^4$, $p_{\rm SM} = w \, \rho_{\rm SM}$ and $\Mpl^2H^2=\frac{8\pi}{3}(m_a n_a+\rho_{\rm SM})$.
We take $g_*$ and $w$ from Ref.\,\cite{Saikawa:2018rcs}.}
After $\hat n_a$ at $T_i$ is obtained, we solve the Boltzmann equations from $T_i$ to $T_f$ ($z_f=8$) to extract the necessary information on the modified cosmology, such as $\dot z$ and $H(z)$.

Once $\hat \rho_\nu = \hat \rho_{\nu_e}+2\hat \rho_{\nu_\mu}$ is obtained, we can also estimate $N_{\rm eff}$ by 
\bal
N_{\rm eff} = 
\left( 2\cdot \frac{7}{8}\frac{\pi^2}{30} \Big(\frac{4}{11}\Big)^{\!\! 4/3} \right)^{ \!\!\! -1} \hat \rho_\nu(z_f),
\eal
for a large $z_f \gg 1$.
As a check, we obtain $N_{\rm eff} \simeq 3.040$ for the standard BBN, which agrees well with the PDG estimation $N_{\rm eff} = 3.045$~\cite{ParticleDataGroup:2024cfk}. 

In the presence of axions, $N_{\rm eff}$ is reduced by additional radiation, and we obtain $N_{\rm eff}$ bound following the method of Ref.~\cite{Dunsky:2022uoq}. From the Planck results\,\cite{Planck:2018vyg}, we adopt
$N_{\rm eff}>2.43 \ (\Delta N_{\rm eff}>-0.61)$, while Ref.~\cite{Dunsky:2022uoq} uses $N_{\rm eff}>2.62$ assuming a fixed $\Yp$ value.
More specifically, we obtain the 2$\sigma$ bound, $N_{\rm eff}>2.43$, from Fig.\,41 of Ref.\,\cite{Planck:2018vyg} by projecting the fitting contours without $\Yp$ measurements (blue contours) onto the $N_{\rm eff}$ axis.
We also do not use other stronger bounds in \cite{Planck:2018vyg} that assume underlying physics incompatible with hadronic injections during BBN (see Ref.\,\cite{Ganguly:2025mdi} for related discussion). 

In the next step, we solve for the total baryon number density, 
\begin{align}
\label{Eq:Boltzmann_nb}
&
\hat n_b' = 
\left(\frac{3}{z} - \frac{3H}{\dot z} \right) \hat n_b \ . 
\end{align}
The boundary condition of $\hat n_b$ must be fixed at a low temperature by the CMB data as
\bal
\hat n_b(z_f) = T_f^{-3} \, \eta_b n_\gamma(z_f)
\eal
where $\eta_b(T_{\rm CMB})=6.115\times10^{-11}$ is provided in  Sec.\,24.4 of Ref.~\cite{ParticleDataGroup:2024cfk} based on the Planck measurement. To a good approximation, $\eta_b(T_f)=\eta_b(T_{\rm CMB})$ holds.

Finally, given that $\dot z$, $H$, and $\hat n_b$ are obtained over the relevant temperature,  we can solve the evolution for the neutron fraction, Eq.\,\eqref{eq:Xndot}, which is differentiated with respect to $z$,  
\bal
\label{Eq:Boltzmann_Xn}
&
X_n' = 
\frac{1}{\dot z}
\Big[
-\Gamma_{n\to p} X_n
+\Gamma_{p\to n} (1-X_n)
\Big],
\eal
The  total conversion rates of $n\to p$ and $p\to n$ are given by SM reactions and axion related ones ($\Delta\Gamma$), 
\begin{align}
\Gamma_{n\to p}&=\Gamma_{n\nu_e \to p e^-}\!+\!\Gamma_{ne^+ \to p \bar \nu_e} \!+\! \Gamma_{n\rm{\,decay}}  \!+\!\Delta \Gamma_{n\to p} \, ,\label{eq:TotalNtoP}  \\
\Gamma_{p\to n} &= \Gamma_{p\bar \nu_e \to n e^+} \!+\! \Gamma_{p e^- \to n \nu_e} \!+\! \Delta \Gamma_{p\to n}\ , 
\label{eq:TotalPtoN}
\end{align}
respectively.
Here, we take the beta decay rate $\Gamma_{n\rm{\,decay}} =(878.4\,\sec)^{-1}$\,\cite{ParticleDataGroup:2024cfk}. For the weak interaction rates, we adopt the equations developed in Ref.\,\cite{Weinberg:1972kfs},%
\footnote{Note that we do not assume $\Gamma_{p\to n} = e^{-Q/T} \Gamma_{n \to p}$, which is not satisfied after the neutrino decoupling where $T\neq T_{\nu_e}$.}
\bal
&\Gamma_{n\nu_e\to pe^-}
\simeq \frac{1+3g_A^2}{2\pi^3}G_F^2 Q^5 J_{1}^\infty(T,T_{\nu_e}),
\displaybreak[2]
\\
&\Gamma_{n e^+\to p\bar\nu_e}
\simeq \frac{1+3g_A^2}{2\pi^3}G_F^2 Q^5 J_{-\infty}^{-m_e/Q}(T,T_{\nu_e}),
\displaybreak[2]
\\
&\Gamma_{p\bar\nu_e\to n e^+}
\simeq \frac{1+3g_A^2}{2\pi^3}G_F^2 Q^5 K_{-\infty}^{-m_e/Q}(T,T_{\nu_e}),
\displaybreak[2]
\\
&\Gamma_{pe^-\to n\nu_e}
\simeq \frac{1+3g_A^2}{2\pi^3}G_F^2 Q^5 K_{1}^\infty(T,T_{\nu_e}),
\displaybreak[2]
\eal
where $g_A\simeq 1.27$, and $Q=m_n-m_p-m_e\simeq1.293\,\MeV$.
The $J_a^b$ and $K_a^b$ functions are defined by
\bal
&J_a^b(T,T_{\nu_e})= \!
\int_a^b \!\!\!
\sqrt{1-\frac{(m_e/Q)^2}{q^2}}
\frac{q^2(q-1)^2 dq}
{\! (1+e^{-\frac{Q}{T}q})(1+e^{\frac{Q}{T_{\nu_e}}(q-1)})},
\displaybreak[2]
\\
&K_a^b(T,T_{\nu_e})= \!
\int_a^b \!\!\!
\sqrt{1-\frac{(m_e/Q)^2}{q^2}}
\frac{q^2(q-1)^2 dq}
{\! (1+e^{\frac{Q}{T}q})(1+e^{-\frac{Q}{T_{\nu_e}}(q-1)})}.
\eal
We evaluate the neutrino temperature $T_{\nu_e}$ by solving $\rho_{\nu_e} = {\scriptstyle 2}\frac{7}{8} \frac{\pi^2}{30} T_{\nu_e}^4$ assuming that neutrinos follow the Fermi-Dirac distribution.

\begin{figure*}[t]
    \centering
    \includegraphics[width=0.48\textwidth]{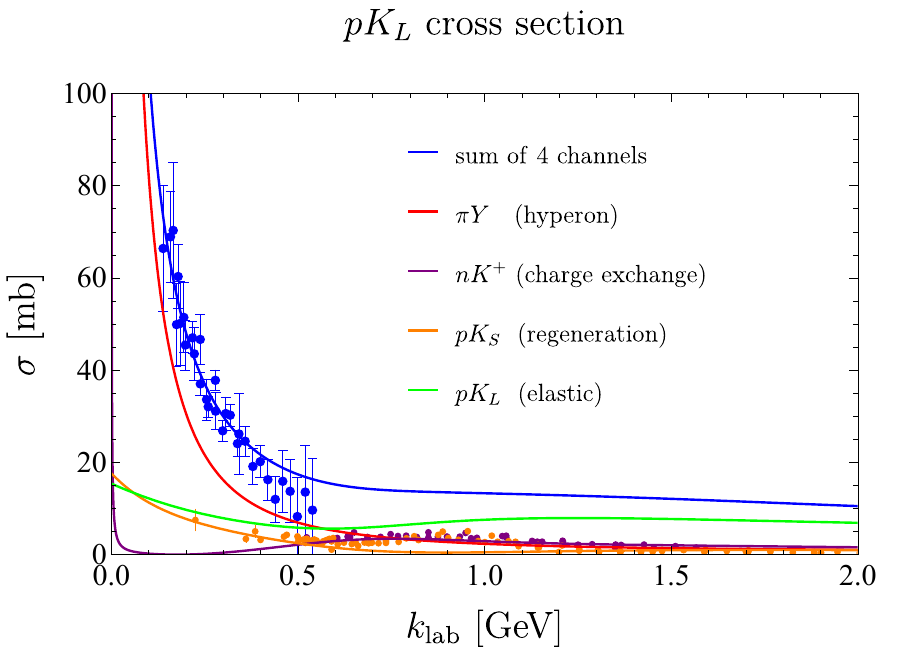}
    \centering
    \includegraphics[width=0.48\textwidth]{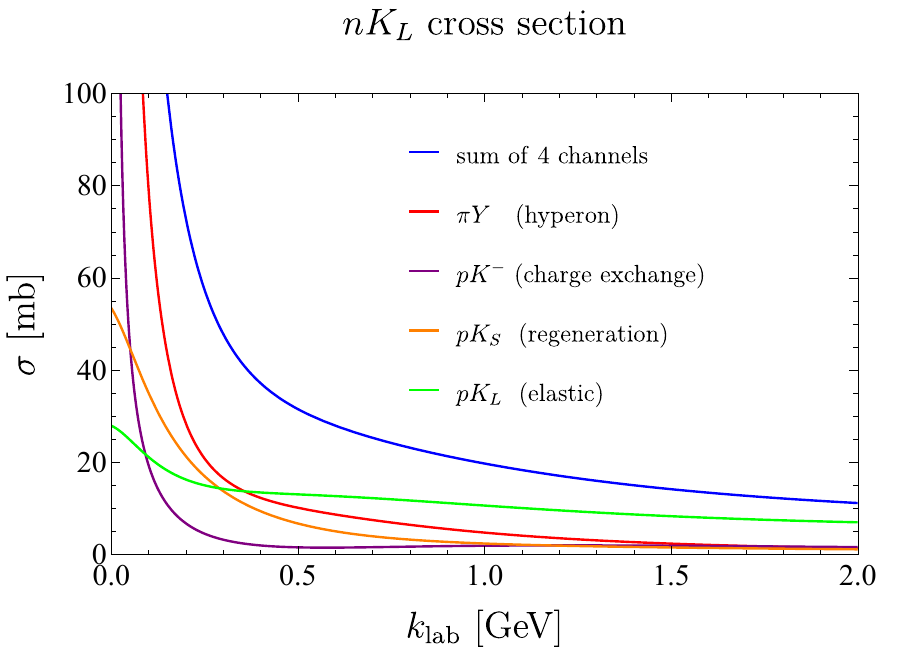}
    \caption{
    {\it Left}:  the cross sections of $p K_L $ into four different final states obtained by the method described in Appendix~\ref{sec:xsec-kaon} and their sum: hyperon channel $\pi Y$  (red), 
    charge exchange $n K^+ $ (purple), 
    regeneration $p K_S $ (orange), 
     elastic scattering $p K_L $ (green), and the sum of four channels (blue). 
    We show relevant datasets as dots with uncertainty bars: the total cross section from Fig.\,4 of Ref.\,\cite{Cleland:1975ex}(which includes data from Ref.\,\cite{Sayer:1968zz}), the charge exchange cross section from Ref.\,\cite{Ferro-Luzzi:1962jcn}, and the regeneration cross section from Fig.\,27 of Ref.\,\cite{GlueX:2017hgs}(Ref.\,\cite{Capiluppi:1982fj}). See Fig.\,\ref{fig:7_fitted_data} for the individual fitted processes.  
    {\it Right}:  $nK_L$ scattering cross sections in the corresponding four different final states and their sum. These are the predictions of our fitting. 
    \label{fig:KLpTotal}}
\end{figure*}

The axion effect through the hadron injection is contained in $\Delta \Gamma_{n\to p}$ and $\Delta \Gamma_{p\to n}$. 
They can be expressed as
\bal
\label{Eq:Delta_Gammanp}
&\Delta \Gamma_{n\to p} = \sum_{h} \langle \sigma v (n+ {h} \to p \cdots) \rangle X_{h} \left( \frac{m_e}{z} \right)^3 \hat n_b
\\
\label{Eq:Delta_Gammapn}
&\Delta \Gamma_{p\to n} = \sum_{h} \langle \sigma v (p+ {h} \to n \cdots) \rangle X_{h} \left( \frac{m_e}{z} \right)^3 \hat n_b,
\eal
where the sum runs over ${h}=\pi^+, \, \pi^-, \,  K^+, \,  K^-, \, K_L, \,  \bar p$ and $\bar n$, and we define $X_{h}\equiv n_{h}/n_b$.  
Note that, unlike in Eq.\,\eqref{eq:nAdot}, only the processes that convert $n \leftrightarrow p$ are relevant here, as indicated by $p$ or $n$ appearing in the \emph{final} state. 
The cross sections are discussed in Sec.\,\ref{sec:x-sec}.

For $X_{h}$ in the presence of axion hadronic decays,  we solve the following Boltzmann equation, 
\begin{widetext}
\bal
\label{Eq:Boltzmann_XA}
X_{h}' =& \frac{1}{\dot z} 
\bigg[
{\rm N}_{a\to {h}} \Gamma_a  X_a
- \Big\{
\tilde \Gamma_{h}+
\Big(
\langle \sigma v (n+ {h} \to p \cdots) \rangle X_n  +
\langle \sigma v (p+ {h} \to n \cdots) \rangle (1-X_n) \Big)
\left( \frac{m_e}{z} \right)^3 \hat n_b
\Big\}  X_{h} 
\nn \\
&\quad 
+ \sum_{h'} 
\Big\{
{\rm N}_{{h'}\to {h}}  \tilde \Gamma_{h'} 
+
\Big(
\langle \sigma v (n+ {h'} \to {h}\cdots) \rangle X_n  +
\langle \sigma v (p+ {h'} \to {h}\cdots) \rangle (1-X_n) \Big)
\left( \frac{m_e}{z} \right)^3 \hat n_b
\Big\} X_{h'}
\bigg]. 
\eal
\end{widetext}
which is modified from Eq.\,\eqref{eq:nAdot} to adopt the new parametrization. 

The boundary conditions of Eqs.\,\eqref{eq:nAdot} and \eqref{Eq:Boltzmann_XA}  specified at $T=T_i$ are set by the quasi-stable solutions where  $X_n'=0$ and $X_{h}'=0$. This choice is merely for numerical stability, and the final results are insensitive to the precise details of the boundary conditions. Both $X_n$ and $X_h$ rapidly settle down to the quasi-stable values after the evolution begins, erasing any dependence on the initial conditions.

In the standard BBN framework, we find $T_{\rm D}=73.7$~keV acts as the effective bottleneck temperature at which we reproduce $\Yp=0.245$~\cite{ParticleDataGroup:2024cfk}, under the assumption of instantaneous neutron capture into deuterium. As discussed in Sec.\,\ref{sec:overview}, we evaluate a more robust observable: the fractional deviation from standard BBN predictions due to new physics, and we impose 
${\delta X_n}/{X_n^{\rm SM}} |_{T=T_{\rm D}} < R_{\Yp}$. 

\section{Updates on hadronic cross sections}
\label{sec:x-sec}
In this section, we briefly discuss the methodology to obtain cross sections  used in Eqs.\,\eqref{Eq:Delta_Gammanp}, \eqref{Eq:Delta_Gammapn} and \eqref{Eq:Boltzmann_XA}  and address the major updates.  The \np conversion processes are summarized in Table\,\ref{Tab:reactions}. More details are explained in Appendix\,\ref{app:x-sec}.

As discussed in Sec.\,\ref{sec:overview}, we need to know the corresponding cross sections often near the threshold. The long-lived hadrons, except for $K_L$, immediately slow down due to the electromagnetic interaction and become highly non-relativistic. We use the partial wave analysis, with $s$ and $p$-waves, and fit the measured data, and then the threshold cross sections are extrapolated.  The higher momentum contribution is important as well to perform thermal averaging. %

However, challenges exist for many processes, often involving $n$ or $K_L$,  because there are no corresponding measurements due to the experimental difficulties. Sometimes, the data of the inverse process exists, for example,  we can use $n\gamma\to p\pi^-$ data to obtain $p\pi^-\to n\gamma$ by time-reversal, while the phase space factor has to be corrected because the mass difference is crucial near the threshold. In addition, we use the isospin symmetry.
For example, in the isospin limit,  we can obtain the $n\pi^+ \to p \pi^0$ cross section by $p\pi^- \to n \pi^0$ measurements, but in reality we need to factor out the Coulomb attraction in addition to correcting the phase space, as it matters more at low momentum. 

For processes initiated by a kaon, we essentially perform the same prescriptions: relating by isospin, manipulating the Coulomb effect, and correcting the phase space factor. Nevertheless, the situation is more complicated because the data is limited to the scattering processes with initial state of $pK^\pm$ or $pK_L$. 
Denoting  the $(\bar K^0, K^-)$ isospin doublet by $\bar K$ and hyperons by $Y$, the amplitudes of processes $N\bar K\to N'\bar K'$ and $N\bar K\to Y \pi$ can be categorized by the total isospin, $I=0$ and $I=1$, following the formalism suggested in Ref.\,\cite{Martin:1970je} (see Appendix~\ref{sec:xsec-kaon} for the details).
Additionally, $K_L$ scattering processes require additional amplitudes of $N  K \to N' K' $, denoting $K=(K^+,  K^0)$,
such as the elastic scattering ($p K^+ \to p K^+$), the charge exchange ($p K_L \to n K^+$) and 
regeneration processes ($pK_L \to p K_S , \ n K_L  \to n K_S $).
We include all available experimental data, to the best of our knowledge, and find the relevant scattering amplitudes by simultaneous fitting. We limit the datasets to $k_{\rm lab}<2\,\GeV$ where $k_{\rm lab}$ is the kaon momentum in the target rest frame because more processes beyond our framework are relevant at high momenta. The restriction of the $K_L$ spectrum in Figs.\,\ref{fig:NhadShower} and \ref{fig:KLmom} is aligned with this limitation.

\begin{figure*}
    \includegraphics[width=0.48\textwidth]{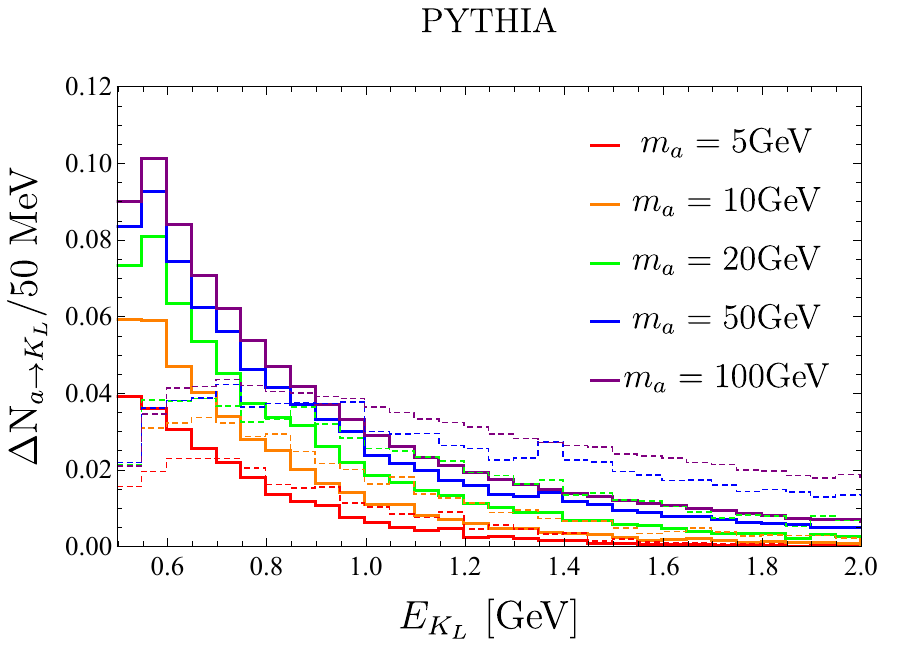}
    \includegraphics[width=0.48\textwidth]{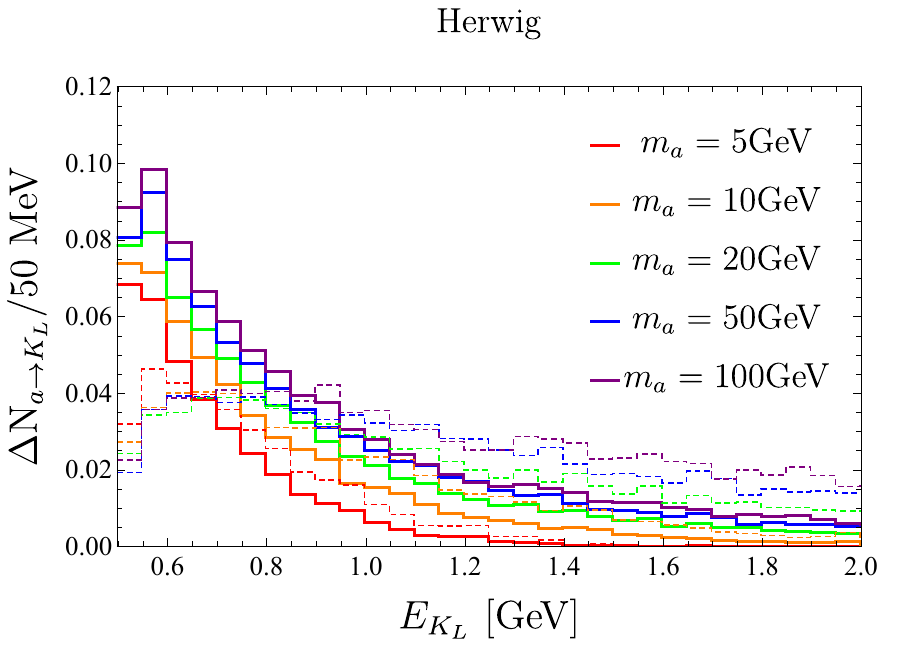}
    \caption{Reshaped  $K_L$ energy spectrum due to elastic scattering. Energy distribution of $K_L$ from the axion decay for $m_a=5,10,20,50$ and 100\,GeV, obtained from {\tt PYTHIA} (left) and {\tt Herwig} (right). These spectra are still different from thermal distributions. 
     \label{fig:KLmom_reshaped}}    
\end{figure*}
\begin{figure}
\includegraphics[width=0.48\textwidth]{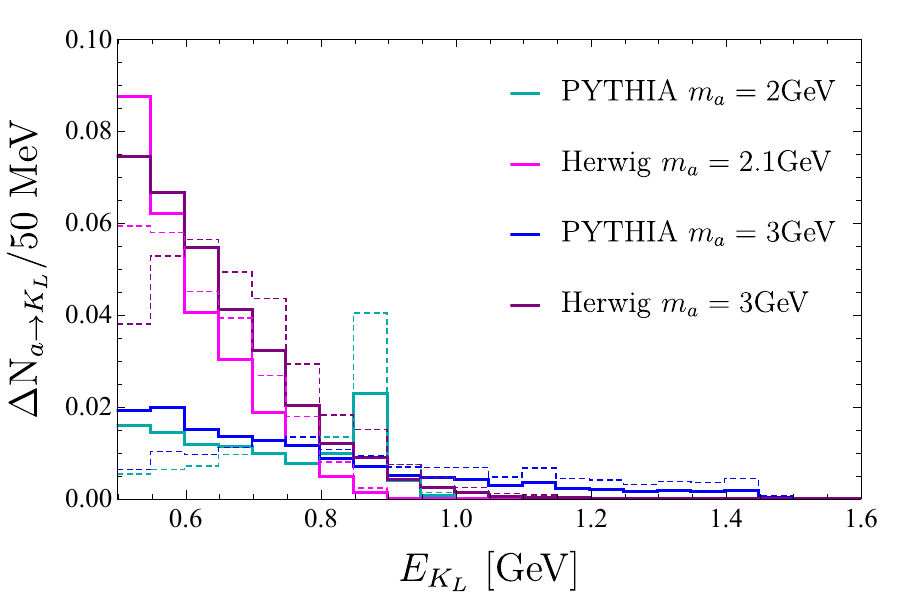}
    \caption{
    Reshaped spectrum due to elastic scattering. Energy distribution of $K_L$ from the axion decay for low axion masses, where
    large discrepancies between {\tt PYTHIA} and {\tt Herwig} predictions are present.
    }   
    \label{fig:KLmom_lowma_reshaped} 
\end{figure}
\begin{figure*}[t]
    \includegraphics[width=0.48
    \textwidth]{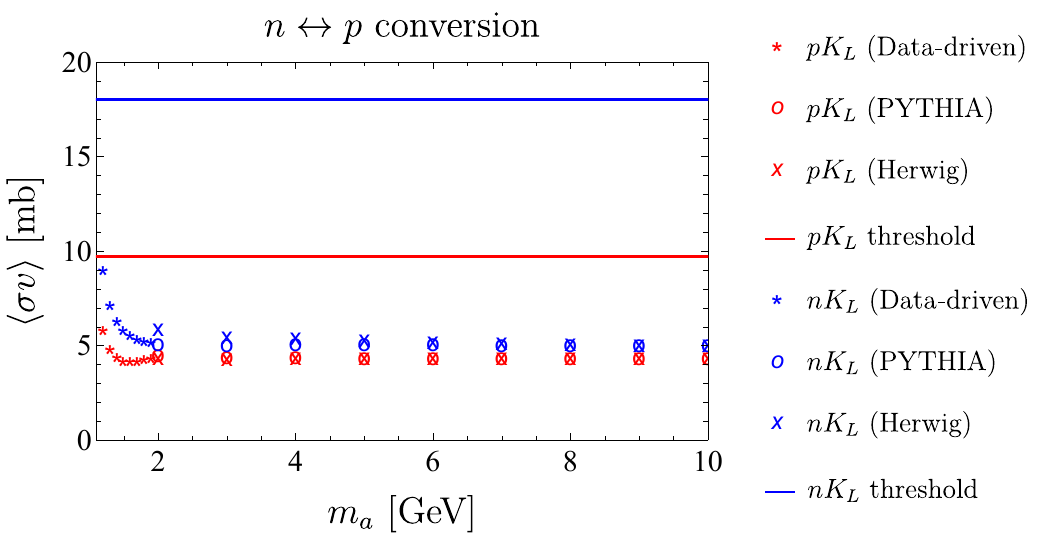}
    \includegraphics[width=0.48
    \textwidth]{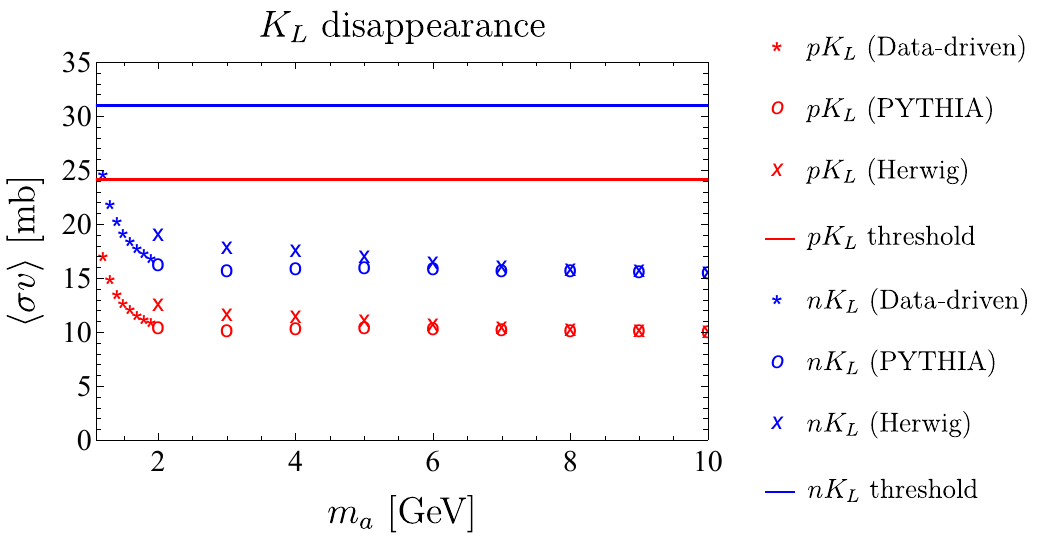}
    
    \caption{
    {\it Left}: Scattering cross sections ($\sigma v$) with incident $p K_L$ (red) or $n K_L$ (blue) weighted by the reshaped $K_L$ distribution from axion decay (Figs.\,\ref{fig:KLmom_reshaped} and \ref{fig:KLmom_lowma_reshaped}). Only $n\leftrightarrow p$ conversion processes are considered. For $m_a\leq 2\,\GeV$, the data-driven method is used (star), and otherwise, {\tt PYTHIA} (circle) or {\tt Herwig} (cross) is used. The horizontal lines show the cross sections at the threshold for comparison.  
    {\it Right}: the averaged cross sections for the $K_L$ disappearance processes, namely both conversion and non-conversion processes. The plot scheme is the same as in the left panel. The breakdown of the processes is found in Figs.\,\ref{Fig:KLn_averaged} and \ref{Fig:KLp_averaged}.  
   \label{fig:KLNxsec}}    
\end{figure*}

Our fitting, as shown in Fig.\,\ref{fig:KLpTotal},  gives various $2\to 2 $ scattering cross sections of $p K_L$ (left) and $n K_L$ (right). The known data for $p K_L$ scattering is overlaid, which shows a good agreement with our fitting. More results are given in Appendix~\ref{sec:xsec-kaon}. We find $p$-wave contributions are significant, which makes the elastic scattering dominant for $k_{\rm lab}\gtrsim 1\,\GeV$.

For our BBN analysis, the elastic scattering cross sections are important for obtaining the correct $K_L$ spectrum.
Although the elastic scattering is not directly involved in $n \leftrightarrow p$ conversion, it modifies the energy distribution of the $K_L$'s from axion decays. 
Accounting for the elastic scattering, we obtain the effective $K_L$ distribution from the one given by the axion decay, as shown in Figs.\,\ref{fig:KLmom_reshaped} and \ref{fig:KLmom_lowma_reshaped}. The reshape scheme is given in Appendix~\ref{sec:KLreshape}. 

Using the reshaped distributions, the averaged cross sections of $NK_L$ as a function of $m_a$ are evaluated as in Fig.\,\ref{fig:KLNxsec}.  
Also, the effective $K_L$ decay width is calculated by averaging this distribution as  $\tilde \Gamma_{K_L}=\langle\gamma^{-1}\rangle\Gamma_{K_L}$ where $\gamma^{-1}=m_{K_L}/E_{K_L}$ which is about 0.6--0.8. 
For example, using {\tt PYTHIA} ({\tt Herwig}), $\langle\gamma^{-1}\rangle =0.70$ $(0.80)$, $0.69$ $(0.69)$, $0.63$ $(0.63)$ at $m_a=3$, $10$, $50$ GeV, respectively.

In the literature, evaluation of the $K_L$ cross sections has been missing, which is partially why the $K_L$ process was omitted or oversimplified. 
The threshold cross section was often used for $N K_L$ scattering, but, as we noted, the high momentum region is important. In fact, cross sections weighted with the $K_L$ momentum profile per axion mass, which are shown in Fig.~\ref{fig:KLNxsec} (stars, crosses, or circles), are quite different from the threshold cross sections (solid lines).  

Finally, let us comment on the annihilation of injected antibaryons with background baryons. 
We take the annihilation cross sections given in Ref.\,\cite{Lee:2015hma} and give details in Appendix~\ref{sec:xsec-baryon}.

\begin{figure*}[t]
    \centering
    \includegraphics[width=0.46\textwidth]{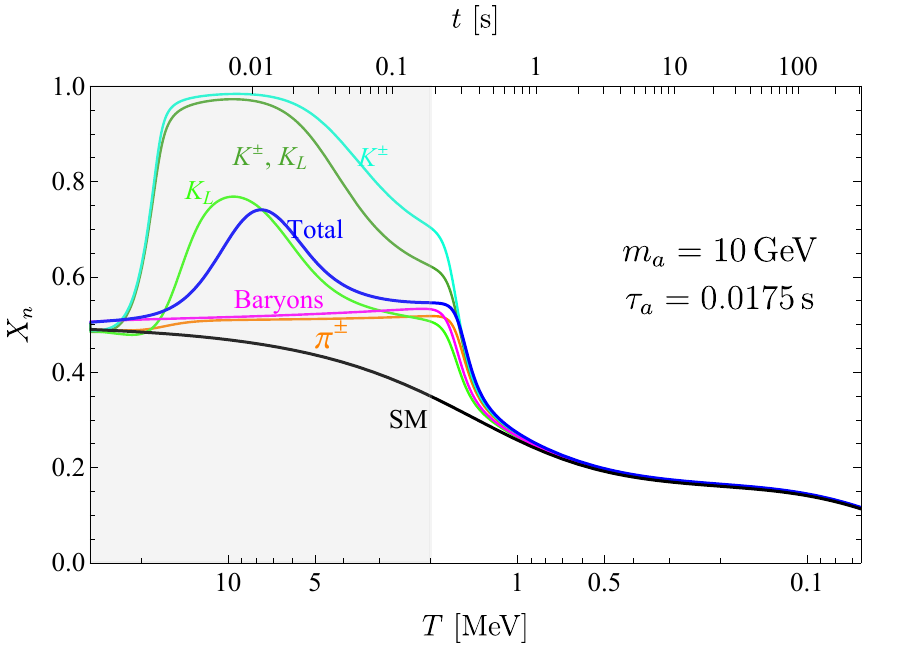}
    \raisebox{-3.5mm}{
    \includegraphics[width=0.52\textwidth]{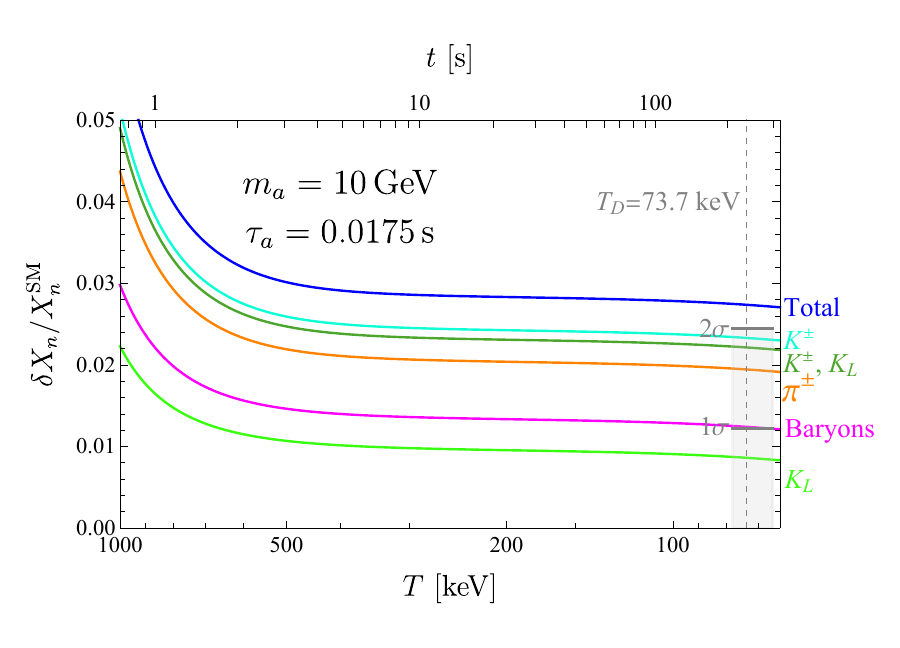}
    }
    \caption{Modification of the $X_n$ evolution in the presence of heavy axion (blue solid line) is compared to the SM scenario (black solid line).
    Here, we take $m_a=10\,\GeV$ and $\tau_a=0.0175\,\sec$ as a benchmark, while $Y_{a}^{\rm (min)}=Y_{a}^{\rm (max)}$.
    Various colored lines other than blue correspond to modified $X_n$ with turning on only a subset of hadronic components from the axion decay; orange, cyan, light green, dark green, and pink lines represent cases of $\pi^\pm$, $K^\pm$, $K_L$, all the kaons ($K^\pm, \,K_L$), and baryons.
    The left and right panels depict the evolutions of $X_n$ and $\delta X_n/X_n^\SM$, respectively. For convenience, the corresponding time is shown on the top axes, assuming SM radiation dominance. }
    \label{fig:XnEvo}
\end{figure*}

\begin{figure}[t]
    \centering    \includegraphics[width=0.45\textwidth]{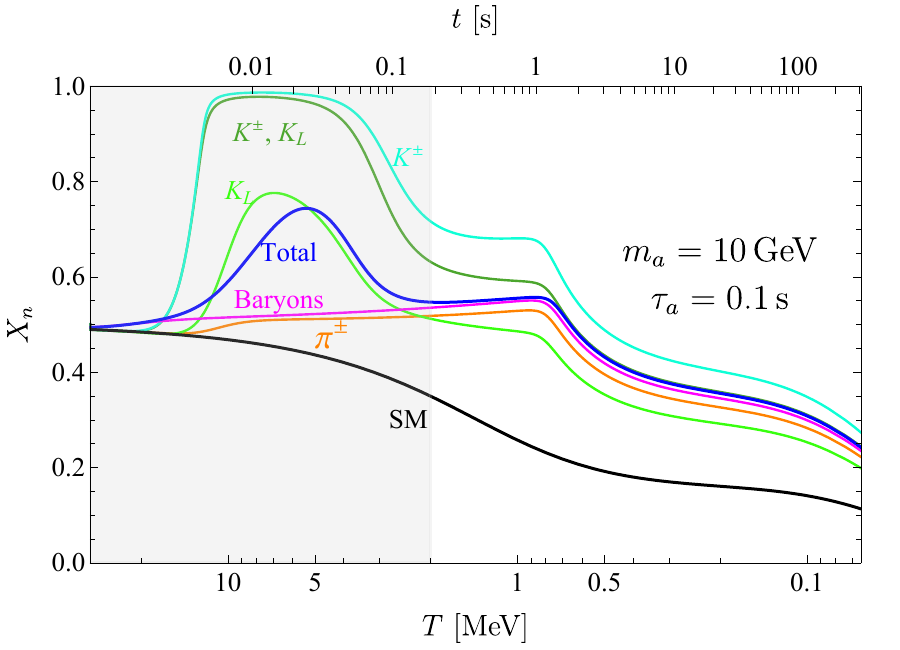}
    \includegraphics[width=0.45\textwidth]{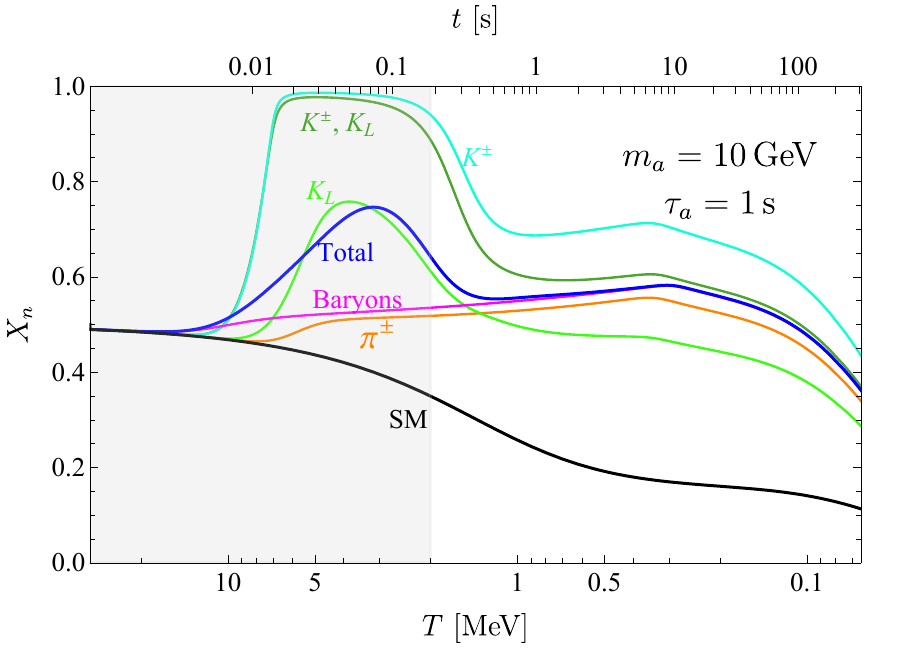} 
    \includegraphics[width=0.45\textwidth]{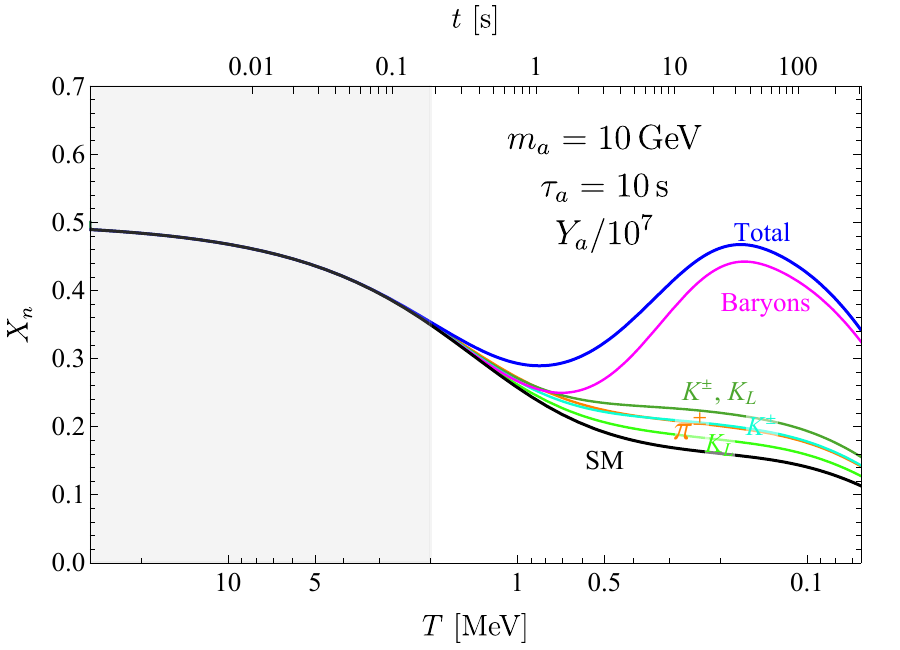}
    \caption{The evolutions of $X_n$ are depicted for different $\tau_a = 0.1\,\sec$ (upper), $1\,\sec$ (middle) and $10\,\sec$ (lower). The color scheme remains the same as that of Fig.\,\ref{fig:XnEvo}.}
    \label{fig:Evos}
\end{figure}

\section{Results \label{sec:result}}

\subsection{Evolutions \label{sec:evolutions}}

Our numerical results on $\delta X_n/X_n^\SM$ at $T_{\rm D}$ are obtained as described in Sec.\,\ref{sec:n-decoupling}. We solve Eq.\,\eqref{Eq:Boltzmann_Xn} with $\Gamma_{n\to p}$ and $\Gamma_{p\to n}$ given in Eqs.\,\eqref{eq:TotalNtoP} and \eqref{eq:TotalPtoN}.
The heavy axion contributions $\Delta \Gamma_{n\to p}$ and $\Delta \Gamma_{p\to n}$ follow Eqs.\,\eqref{Eq:Delta_Gammanp} and \eqref{Eq:Delta_Gammapn} with $X_h$ obtained by solving Eq.\,\eqref{Eq:Boltzmann_XA}.

Fig.\,\ref{fig:XnEvo} shows how the evolution of $X_n$ is modified in the presence of heavy axions (blue solid line) compared to the standard BBN (black solid line, denoted by $X_n^{\SM}$).  We take $m_a=10\,\GeV$ and $\tau_a=0.0175\,\sec$ as a benchmark (note that $Y_{a}^{(\rm min)}=Y_{a}^{(\rm max)}$ in this case).
The left and right panels depict the evolutions of $X_n$ and $\delta X_n/X_n^\SM$, respectively.
Various colored lines other than the blue one correspond to $X_n$ obtained by turning on only a subset of hadronic components from the axion decay while the secondary hadrons are always included. The orange, cyan, light green, dark green, and pink lines represent $\pi^\pm$, $K^\pm$, $K_L$, all the kaons ($K^\pm, \,K_L$), and baryons.

At $T \gtrsim 2\,\MeV$, in the gray shaded region of Fig.\,\ref{fig:XnEvo} left, our curves are not completely correct because we have not included some reactions that can be important at high temperatures.
However, those missing reactions have positive threshold energies greater than several MeV, so they are suppressed at low temperatures.
Thus, all errors from missing those reactions are washed out below $T=2\,\MeV$.

At high temperatures, injected hadrons are so abundant that the reaction rates of $n\to p$ and $p\to n$ are governed by the hadronic cross sections. The solution of $X_n$ in this regime corresponds to the one with $\dot{X}_n\simeq 0$ in \eqref{eq:Xndot}. We call it the {\it quasi-stable} solution where `quasi' indicates that this stable point of $X_n$ changes in time as the hadronic cross sections are temperature dependent.

The quasi-stable behavior is maintained until hadronic injections get exponentially suppressed as most of the axions have decayed.
Afterward, a rapid relaxation occurs, which we call {\it waterfall} (see around $T\simeq 1.5\,\MeV$ in the left panel of Fig.\,\ref{fig:XnEvo}).
Analytical and detailed estimations of the quasi-stable and waterfall behaviors will be provided in the next subsection.
In a nutshell, our final constraint is determined dominantly by how much the waterfall drives the relaxation from the quasi-stable value.
As shown in the right panel, the benchmark results in $\left. \delta X_n/X_n^\SM\right|_{T=T_{\rm D}}$ slightly greater than $2\sigma$ of the $\Yp$ measurement ($1\sigma$ and $2\sigma$ bounds are indicated by the gray bars on the right panel at $T_{\rm D} \simeq 73.7\,\keV$), so this parameter is already ruled out.
Note that $\delta X_n/X_n^\SM$ is quite insensitive to $T_{\rm D}$, so we ignore the uncertainty in the value of $T_{\rm D}$.

From Fig.\,\ref{fig:XnEvo} right, we can also see the importance of individual channels.
As we chose $m_a=10\,\GeV$, all the hadronic channels are open. We observe that the pions and kaons are equally important. 
In terms of ${\rm N}_{a\to h}$, pions have an ${\cal O}(10)$  greater number than kaons, but their cross sections are ${\cal O}(0.1)$ smaller, so they are compensated.
If the axion mass is lighter than about 1.1\,GeV, which forbids  $a\to KK\pi$ decay kinematically, then, only the pion channel remains.

Pions and baryons have their quasi-stable values close to $1/2$ because their approximately isospin symmetric cross sections make the rates of $n\to p$ and $p\to n$ roughly the same.
For instance, we have $\sigma (p \pi^- \to n \pi^0, \, n \gamma ) \simeq \sigma (n \pi^+ \to p \pi^0,\, p\gamma)$, $\sigma_{\rm ann} (p\bar n) = \sigma_{\rm ann} (n\bar p)$, and $\sigma_{\rm ann} (p\bar p) \simeq \sigma_{\rm ann} (n\bar n)$.
On the other hand, kaons severely violate such symmetry between $n\to p$ and $p\to n$, which leads to random quasi-stable values.  
This is because (i) the rapid decay of $K_S$ makes the isospin doublet formation of $K^-$ (or $K^+$) incomplete, and (ii) the $K^-$ and $K^+$ are asymmetric in the $n\leftrightarrow p$ conversion, e.g., $K^-$ produces hyperons while $K^+$ cannot due to strangeness conservation.

If the axion lifetime $\tau_a$ is taken larger, the situation changes as Fig.\,\ref{fig:Evos}.
Here, we take $\tau_a=0.1\,\sec$ (top), $1\,\sec$ (middle) and $10\,\sec$ (bottom) while we reduce $Y_a$ for the last case by multiplying a factor of $10^{-7}$ to make the curves distinguishable. All these parameters are ruled out. 
At $\tau_a=0.1\,\sec$, the waterfall time $t_{\rm wf}$ shifts later proportionally to $\tau_a$, and then neutrons freeze out before a sufficient relaxation to the SM line occurs.  
In the other cases, $\tau_a$ is simply too large to have relaxation.

It's noteworthy to comment on the case of $\tau_a=10\,\sec$ (lower panel) where the baryon channel dominates. There, the mesons produced by the axion decay quickly because the constant decay terms are bigger than the conversion rates which decrease by $T^3$. On the other hand, the baryon channel remains since the produced anti-nucleon $\bar N$ is stable and annihilates with an existing neutron or proton.

\subsection{
Parametric dependences and uncertainty estimation
\label{sec:simplified}}

Before presenting our final results, let us provide an analytic understanding of the evolution.
This allows us to assess the uncertainty of our final constraint in the $\tau_a$ space.
As will be shown in this section, an ${\cal O}(1)$ uncertainty in 
${\rm N}_{a\to h}$ and the hadronic cross sections result in only a few $\%$ level uncertainty in the upper bound of $\tau_a$.
This ensures the reliability and robustness of our final result.
Moreover, one can use our equations presented in this section to quickly estimate constraints on lifetimes of other long-lived particles that have different ${\rm N}_{a\to h}$.

In the following discussion, we assume that the final result is primarily determined by a specific set of hadrons, e.g., $K_L$, $\pi^\pm$, or baryons, so we focus on a single species.
We do not include the effect of the secondary production of other hadrons via scattering/decay for simplicity. 
Furthermore, we introduce notations that shorten many expressions for various cross sections;
\bal
&\llangle n \, h \rrangle ~
\equiv \langle \sigma v (n + h \to \cdots) \rangle,
\\
&\llangle n \, h \rrangle_c
\equiv \langle \sigma v (n + h \to p + \cdots) \rangle.
\eal
where $\llangle n \, h \rrangle$ represents the averaged total cross section of $n$'s consuming $h$ while $\llangle n \, h \rrangle_c$ includes only processes that convert $n$ to $p$.
We also define $\llangle p \, h \rrangle$ and $\llangle p \, h \rrangle_c$ in the same way by interchanging $n$ and $p$.

The hadron number density from the axion decay can be obtained by requiring $\dot n_h\simeq0$ in Eq.\,\eqref{eq:nAdot}, yielding 
\begin{align}
n_h \simeq  \frac{{\rm N}_{a\to {h}} \Gamma_a  n_a }{\tilde \Gamma_{h}+
\llangle n\,h \rrangle 
 n_n  +
\llangle p\,h \rrangle 
 n_p
}
\label{Eq:nh}
\end{align}
with $X_p=n_p/n_b=1-X_n$.
Here, we ignore the Hubble constant since $\tilde \Gamma_h + \llangle n\, h \rrangle n_n + \llangle p\, h \rrangle n_p \gg H$.

The neutron ratio in the quasi-stable regime is given by the ratio of conversion rates, requiring $\dot X_n\simeq0$ in Eq.\,\eqref{eq:Xndot}, 
\begin{align}
X_{n}^{\rm qs} & = \frac{\Gamma_{p\to n}}{\Gamma_{n\to p}+\Gamma_{p\to n}}
\\
&= 
\frac{\Gamma^{\rm weak}_{p\to n}+ \llangle p\,h \rrangle_c n_h}{\Gamma^{\rm weak}_{p\to n}+\Gamma^{\rm weak}_{n\to p}+(\llangle p\,h \rrangle_c  +\llangle n\,h \rrangle_c )n_h}
\end{align}
where $ \Gamma^{\rm weak}_{n\to p} \equiv \Gamma_{n\nu_e \to p e^-}+\Gamma_{ne^+ \to p \bar \nu_e} + \Gamma_{n\rm{\,decay}} $ and 
$\Gamma^{\rm weak}_{p\to n} \equiv \Gamma_{p\bar \nu_e \to n e^+} + \Gamma_{p e^- \to n \nu_e} $. The quasi-stable value, $X_{n}^{\rm qs}$, is determined by the cross section ratio independently of $n_h$;
\begin{align}
X_{n}^{\rm qs} \simeq \frac{ \llangle p\,h \rrangle_c }{\llangle p\,h \rrangle_c  +\llangle n\,h \rrangle_c }.
\label{eq:Xnqs}
\end{align}
when the axion-induced process dominates.%
\footnote{In the presence of multiple hadrons, the quasi-stable value becomes
\begin{align}
X_{n}^{\rm qs} \simeq 
 \frac{ \sum_h\llangle p\,h \rrangle_c  n_h}{\sum_h(\llangle p\,h \rrangle_c  +\llangle n\,h \rrangle_c)n_h }. \nn
\end{align}
As seen from this formula, contributions from multiple hadrons could compensate rather than adding up. For example, $X_{n}^{\rm qs}$ from both $K_L$ and $K^\pm$ is always located between the one from $K_L$ and the one from $K^\pm$.}
This explains why $X_n^{\rm qs}\simeq 0.5$ for $h=\pi^\pm$ or baryons (see Figs.\,\ref{fig:XnEvo} and \ref{fig:Evos}) because their cross sections are nearly symmetric between $n$ and $p$ due to isospin symmetry. 
On the other hand, in the kaon cases, since the isospin symmetry is not respected, $X_{n}^{\rm qs}$ can take any values.%

The solution Eq.\,\eqref{eq:Xnqs} should be interfered at the waterfall time $t_{\rm wf}$ when the weak interaction rate catches up with the axion-induced conversion rate as the hadron number density gets exponentially small. We estimate $t_{\rm wf}$ by solving
$\Gamma^{\rm weak}_{n\to p}+\Gamma^{\rm weak}_{p\to n}
=(\llangle n\, h \rrangle_c+\llangle p\, h \rrangle_c ) n_h $
with Eq.\,\eqref{Eq:nh} and $n_a \simeq Y_a s \, e^{-t/\tau_a}$,
and obtain 
\begin{align}
t_{\rm wf}  \!=\! \tau_a\ln \! \left[ \!\frac{Y_a \Gamma_a s}{(\Gamma^{\rm weak}_{n\to p}\!+\!\Gamma^{\rm weak}_{p\to n})_{t=t_{\rm wf}}} \!\frac
{{\rm N}_{a\to {h}}(\llangle n\,h \rrangle_c +\llangle p\,h \rrangle_c) }{\big(\tilde \Gamma_{h}
\!+\! \llangle n\,h \rrangle n_n  
\!+\! \llangle p\,h \rrangle n_p  \big)} \!\right] \!.
\label{eq:twf}
\end{align}
A rough evaluation with typical parameters such as $Y_a \sim 10^{-3}$, $n_b/s \simeq 8.7 \times 10^{-11}$ and $N_{a\to h} \sim 1$ gives the logarithm from $15$ to $18$.%
\footnote{Unlike $X_{n}^{\rm qs}$, the multiple hadron contributions constructively add up for $t_{\rm wf}$ as
\begin{align}
t_{\rm wf}  \!=\! \tau_a\ln \! \left[ \!\frac{Y_a \Gamma_a s}{(\Gamma^{\rm weak}_{n\to p}\!+\!\Gamma^{\rm weak}_{p\to n})_{t=t_{\rm wf}}} \!\sum_h\frac
{{\rm N}_{a\to {h}}(\llangle n\,h \rrangle_c +\llangle p\,h \rrangle_c) }{\big(\tilde \Gamma_{h}
\!+\! \llangle n\,h \rrangle n_n  
\!+\! \llangle p\,h \rrangle n_p  \big)} \!\right] \!.
\nn
\end{align}
}

After the exit of the quasi-stable regime, the weak interaction dominates, 
and the axion-induced effect can be treated as 
a perturbation from the SM scenario.  
The analytic solution of $\delta X_n \equiv X_n-X_n^{\rm SM}$, can be written as 
\begin{align}
&\delta X_n(t) 
= \left(X_{n}(t_{\rm wf})-X_n^{\rm SM}(t_{\rm wf})\right) G(t)
\label{eq:delta_Xn_sol}
\\
&+\! G(t) \!\!\! \int_{t_{\rm wf}}^{t} \!\!\! \Big(\! \llangle p\,h \rrangle_c X_p^{\rm SM}\!(t') \! -\! \llangle n\,h \rrangle_c X_n^{\rm SM}\!(t') \!\Big) n_h(t')G(t')dt'  ,
\nn
\end{align}
where 
\begin{align}
G(t)=\exp\left[-\int_{t_{\rm wf}}^{t} \Big(\Gamma_{p\to n}+\Gamma_{n\to p}
\Big)dt'
\right].
\end{align}
As we are interested in the solution at $t>t_{\rm wf}$, we neglect the second line in Eq.\,\eqref{eq:delta_Xn_sol} since $n_h \propto e^{-t/\tau_a}$ is rapidly decreasing with $t/\tau_a \gtrsim 15$. 
By taking a junction from the quasi-stable solution Eq.\,\eqref{eq:Xnqs} to this at $t_{\rm wf}$, i.e., $X_n(t_{\rm wf}) \simeq X_{n}^{\rm qs}(t_{\rm wf})$, we obtain
\begin{align}
\delta X_n&(t) \! \simeq \!\Big(X_{n}^{\rm qs}(t_{\rm wf})\!-\! X_n^{\rm SM}(t_{\rm wf})\!\Big) G(t).
\label{eq:delta_Xn_app}
\end{align}
Thus, the waterfall solution is nothing but the relaxation from a deviation set by the axion-induced effect.

$\delta X_n$ gets suppressed until the relaxation rate becomes smaller than the Hubble rate, and then it becomes frozen.
Since the relaxation rate is approximately given by the sum of the weak interaction rates, we evaluate $\delta X_n$ at $t=t_n\simeq 0.73\,\sec$ ($T=T_n=1.0\,\MeV$), at which $\Gamma^{\rm weak}_{p\to n}+\Gamma^{\rm weak}_{n\to p} = 3H$, as its freeze-out value.
More precisely speaking, we evaluate the ratio $\delta X_n/X_n^\SM$ since $\delta X_n$ does not get completely frozen because of the neutron decay.

\begin{figure*}[t]
    \centering
    \includegraphics[width=0.48\textwidth]{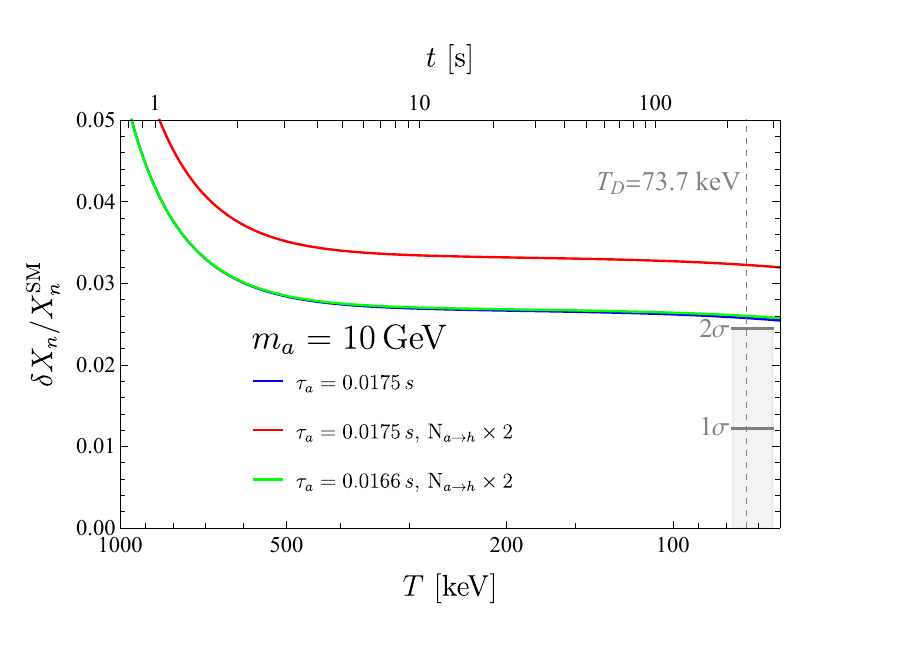}
    \hspace{-30pt}
    \includegraphics[width=0.48\textwidth]{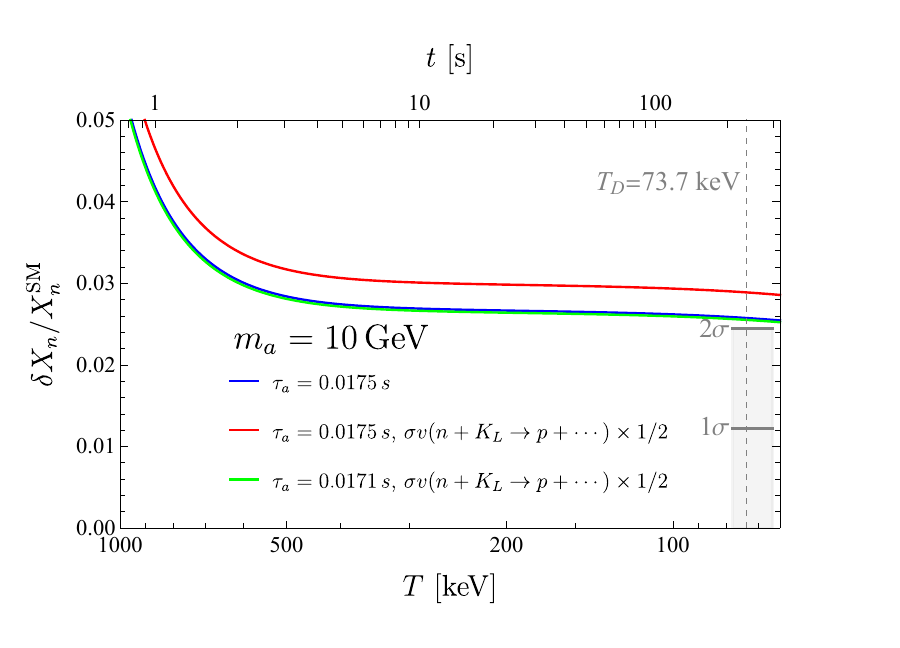}
\vspace{-15pt}
    \caption{ $\delta X_n/X_n^\SM$ including all the hadrons with $m_a=10\,\GeV$ and $\tau_a=0.0175\,\sec$ is depicted by the blue line, while the red lines correspond to that with artificial changes of ${\rm N}_{a\to h}$ (left) and $\sigma v (n+K_L \to p+ \cdots)$ (right) by a factor of 2 to see parametric dependence.
    As the green lines show, a few percent modification of $\tau_a$ makes the evolution back to the blue line.
    }
    \label{fig:dXn_Uncertainties}
\end{figure*}

With $\Gamma^{\rm weak}_{p\to n}+\Gamma^{\rm weak}_{n\to p} \propto T^5$, the exponent of $G(t_n)$ can be approximated as
\begin{align}
&-\int_{t_{\rm wf}}^{t_n} (\Gamma^{\rm weak}_{p\to n}+\Gamma^{\rm weak}_{n\to p})\,dt' 
\simeq-\left(\left(\frac{t_n}{t_{\rm wf}}\right)^{\!\! \frac{3}{2}}-1\right)
\end{align}
where we take $\Gamma^{\rm weak}(T) \simeq  \Gamma^{\rm weak}(T_n) (T/T_n)^5$, $H(T) \simeq H(T_n) (T/T_n)^2$ and $t \simeq 1/2 H$. 
The bound we adopt, $\delta X_n/X_n^{\rm SM}|_{T=T_D}\simeq \delta X_n/X_n^{\rm SM}|_{T=T_n} <2.45\%\equiv R_{\rm Y_p}$,
can now be rearranged as
\begin{align}
t_{\rm wf} \lesssim t_n\bigg(1+\ln\left[\frac{1}{R_{\rm Y_p}} \frac{\delta X_{n}(t_{\rm wf})}{X_n^{\rm SM}(t_n)}\right]\bigg)^{\!\! -\frac{2}{3}},
\end{align}
where the logarithm is roughly $3.7$ as $\delta X_n(t_{\rm wf})/X_n^{\rm SM} \sim {\cal O}(1)$.

Finally, replacing $t_{\rm wf}$ on the left hand side by  Eq.\,\eqref{eq:twf} leads to the bound on the axion lifetime, 
\begin{widetext}
\begin{align}
\tau_a &\lesssim \frac{t_n}{\left(1 \!+\!\ln\!\left[\!\dfrac{1}{R_{\rm Y_p}} \dfrac{\delta X_{n}(t_{\rm wf})}{X_n^{\rm SM}(t_n)}\!\right]\right)^{\!\!\frac{2}{3}}\,\ln \!\left[ \!\dfrac{{Y_a} \Gamma_a s}{\Gamma^{\rm weak}_{n\to p}+\Gamma^{\rm weak}_{p\to n}} \dfrac
{{\rm N}_{a\to {h}}(\llangle n\,h \rrangle_c+\llangle p\,h \rrangle_c) }{\big(\tilde \Gamma_{h}+\llangle n\,h \rrangle n_n +\llangle p\,h \rrangle n_p \big)} \!\right]} \,.
\label{eq:taua_bound}
\end{align}
\end{widetext}
This formula works quite well. 
Putting the typical numerical values mentioned below Eq.\,\eqref{eq:twf}, we obtain $\tau_a \lesssim 0.014$\,--\,$0.018\,\sec$ from Eq.\,\eqref{eq:taua_bound}, which agrees well with our full numerical result. 
Note that $\delta X_n(t_{\rm wf})$ in the first logarithmic factor depends on cross sections as they appear in Eq.\,\eqref{eq:Xnqs} and $\delta X_n(t_{\rm wf}) \simeq X_n^{\rm qs}(t_{\rm wf})-X_n^\SM(t_{\rm wf})$.

Now, we can estimate the uncertainties of our final constraint.
In our calculation, $Y_a$, $N_{a\to h}$, and hadronic cross sections have the largest uncertainties, so let us estimate their impact on our final constraint in terms of $\tau_a$ by using Eq.\,\eqref{eq:taua_bound}.

Suppose the hadron yield from the axion decay is modified by the change of $Y_a\, {\rm N}_{a\to {h}}\to Y_a' \, {\rm N}_{a\to {h}}'$.
The shift of $t_{\rm wf}$ from this change can be obtained from Eq.\,\eqref{eq:twf} as
\begin{align}
&t_{\rm wf}'-t_{\rm wf} =\tau_a \ln \frac{Y_a' {\rm N}_{a\to {h}}'}{Y_a {\rm N}_{a\to {h}}},
\end{align}
and consequently, the neutron fraction is modified as
\begin{align}
\left.\frac{\delta X_n'}{\delta X_n}\right|_{t=t_D} \!\!\!\! &\simeq  
\exp\left[-\int_{t_{\rm wf}'}^{t_{\rm wf}} (\Gamma^{\rm weak}_{p\to n}+\Gamma^{\rm weak}_{n\to p})dt\right]
\\
&\sim 
\left(\frac{Y_a' \,{\rm N}_{a\to {h}}'}{Y_a \, {\rm N}_{a\to {h}}} 
\right)
^{(\Gamma^{\rm weak}_{p\to n}(t_{\rm wf})+\Gamma^{\rm weak}_{n\to p}(t_{\rm wf}))\tau_a}
\label{eq:dXn_uncertainty_from_Nah}
\end{align}
where the exponent is roughly
\begin{align}
\label{eq:dXn_exponent_in_Nah}
{(\Gamma^{\rm weak}_{p\to n}(t_{\rm wf})+\Gamma^{\rm weak}_{n\to p}(t_{\rm wf}))\tau_a}
\sim 0.3  \left(\frac{\tau_a}{0.0175\,\sec}\right)^{-\frac{3}{2}}
\end{align}
Clearly, it is not linearly scaling, unlike the naive expectation. 
We checked and confirmed this scaling behavior with our full numerical code.

Conservatively, we estimate the uncertainty of ${\rm N}_{a\to h}$ as $\sim 100\,\%$ since \texttt{PYTHIA} and \texttt{Herwig} can have a difference of at most $100\%$ depending on $m_a$ and channels (except for the baryonic channel in the region very close to $2\,\GeV$).
If we multiply a factor of 2 to every ${\rm N}_{a\to h}$, this only changes $\delta X_n|_{t=t_D}$ by $23\,\%$ as indicated in Eqs.\,\eqref{eq:dXn_uncertainty_from_Nah} and \eqref{eq:dXn_exponent_in_Nah}.
Then, the upper bound of $\tau_a$ is only affected by $4\,\%$ as Eq.\,\eqref{eq:taua_bound} shows.

Our numerical check of this feature is depicted in the left panel of Fig.\,\ref{fig:dXn_Uncertainties}, where the blue line is the evolution of $\delta X_n/X_n^\SM$ including all the hadrons, while the red line is obtained by multiplying a factor of 2 to ${\rm N}_{a\to h}$.
As expected, the red line is increased by about $25\,\%$ from the blue line.
Then, $\delta X_n$ with roughly $5\,\%$ smaller $\tau_a$ is depicted by the green line, which comes close to the blue line.

On the other hand, the difference between $Y_a^{\rm (min)}$ and $Y_a^{\rm (max)}$ is negligible in most of the parameter space, but it can be large in a particular region,  $ m_a  \lesssim 0.7\,\GeV$.
The difference can be as large as a factor of about $10$ around $m_a \sim 0.7\,\GeV$.
However, as the $Y_a$ dependence is still in the logarithm in Eq.\,\eqref{eq:taua_bound}, the upper bound of $\tau_a$ changes only about $10\,\%$.
This change will be shown in the next subsection.

Analyzing the impact of uncertainties in hadronic cross sections is slightly more complicated as they contribute to both logarithms in Eq.\,\eqref{eq:taua_bound}.
As implied by Eqs.\,\eqref{eq:Xnqs} and \eqref{eq:twf}, a multiplication of an overall factor to all hadronic cross sections has almost zero impact, ignoring the decay width $\tilde \Gamma_h$, which we confirmed numerically. 
What matters in the end is the ratio of $p\to n$ and $n\to p$ cross sections.
Once the ratio is changed, it modifies both the quasi-stable value and the waterfall timing.

For the kaon cross sections, uncertainties are dominated by those from $K_L$, and we consider, conservatively, a factor of 2 uncertainty on the $K_L$ cross sections. For example, in the right panel of Fig.\,\ref{fig:dXn_Uncertainties}, 
we show the evolution of $\delta X_n/X_n^\SM$ with (red) and without (blue) multiplying the cross sections of $K_L$-induced $n\to p$ conversion by a factor of 1/2. This changes $\delta X_n/X_n^\SM$ by $12\,\%$. We can bring the modified curve back if $\tau_a$ is increased by $2\,\%$.

Therefore, we conclude that uncertainty in our upper bound on $\tau_a$ is on the order of a few percent.

\begin{figure*}[t]
    \centering
    \includegraphics[width=0.47\textwidth]{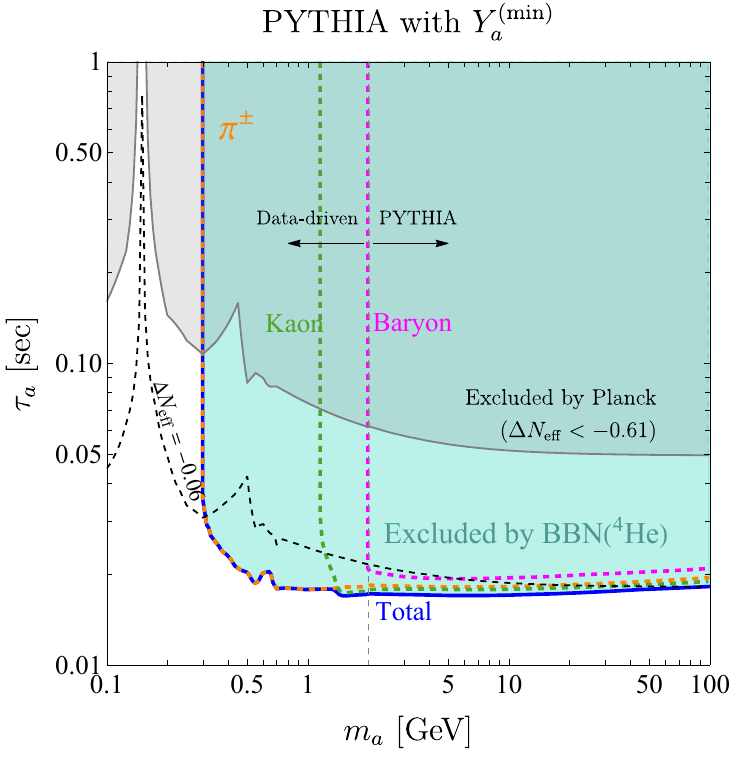}
    \includegraphics[width=0.47\textwidth]{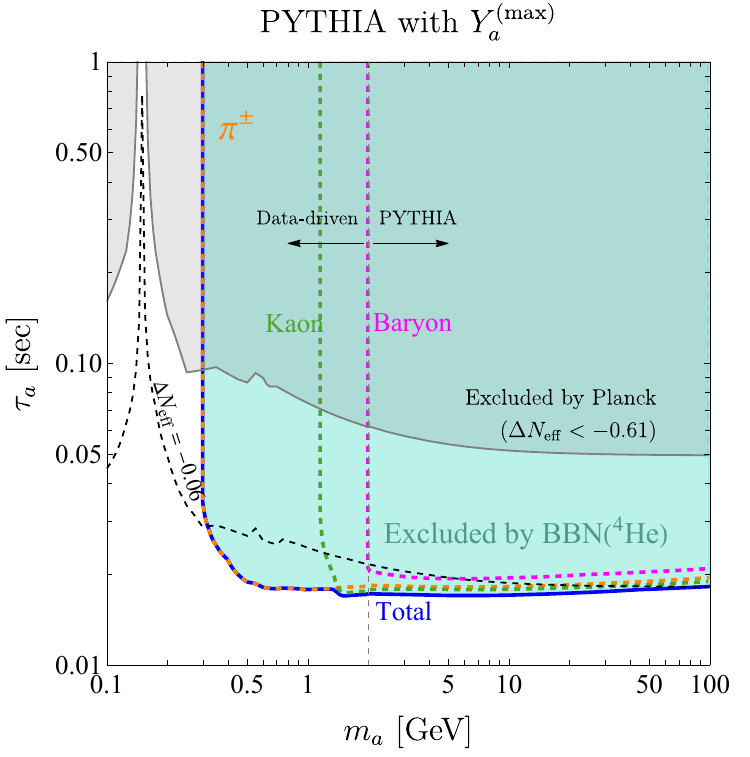} 
    \vspace{0.3cm}
    \includegraphics[width=0.47\textwidth]{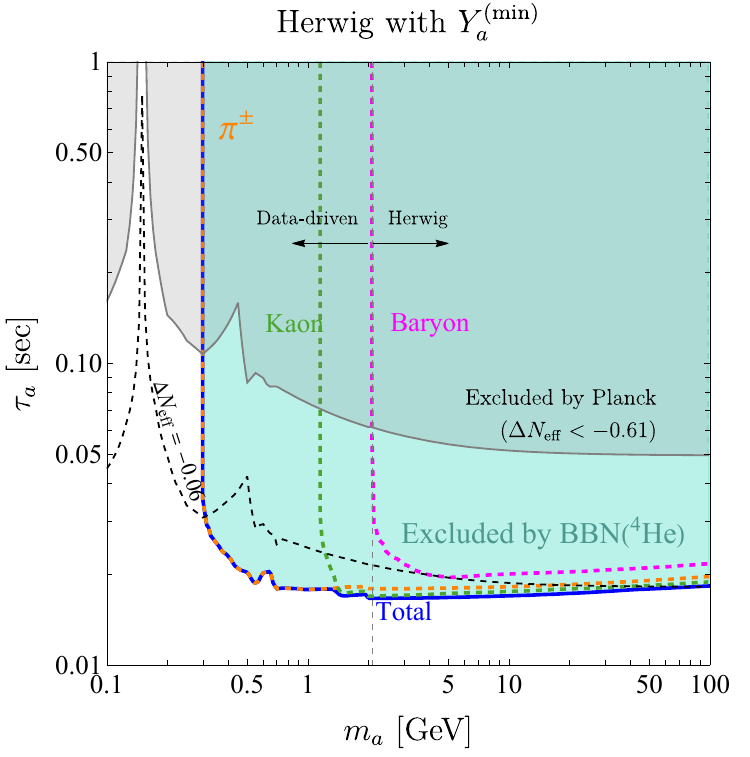}
    \includegraphics[width=0.47\textwidth]{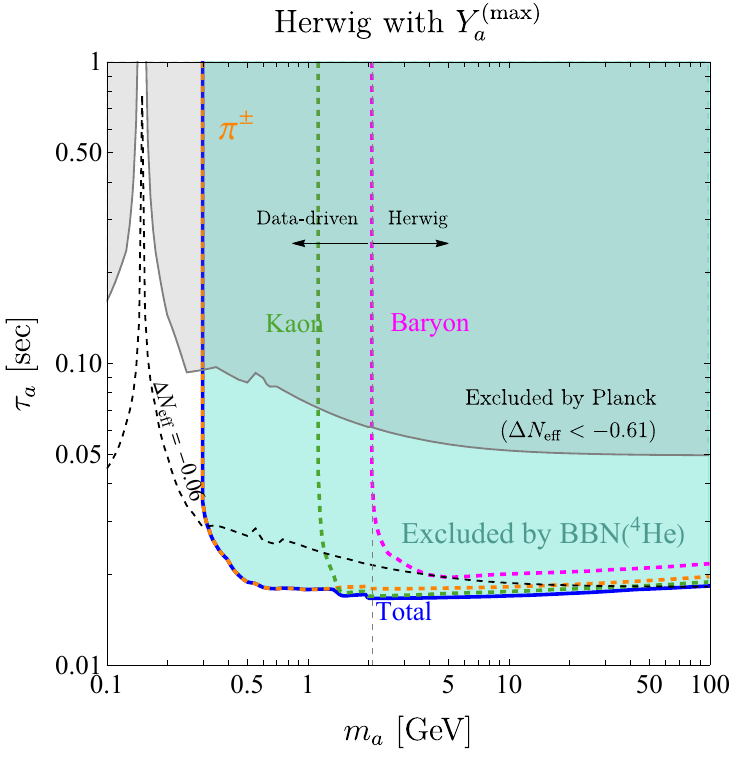}
    \caption{The exclusion by our BBN calculation is given by the cyan-shaded region surrounded by the blue contour in the $(m_a,\,\tau_a)$ space.
    Left and right panels take $Y_a^{\rm (min)}$ and $Y_a^{\rm (max)}$ and upper and lower panels use \texttt{PYTHIA} and \texttt{Herwig}, respectively.
    The orange, dark green, and pink dashed contours represent constraints estimated from individual channels of $\pi^\pm$, kaons, and baryons.
    The gray vertical dashed line at $m_a\simeq2\,\GeV$ indicates the boundary of changing our scheme of estimating the axion decay rate and branching ratios; we use the data-driven method for $m_a<2\,\GeV$ and \texttt{PYTHIA} or \texttt{Herwig} for $m_a>2\,\GeV$. }
    \label{fig:resultintaua}
\end{figure*}

\begin{figure*}[t]
    \centering
    \includegraphics[width=0.47\textwidth]{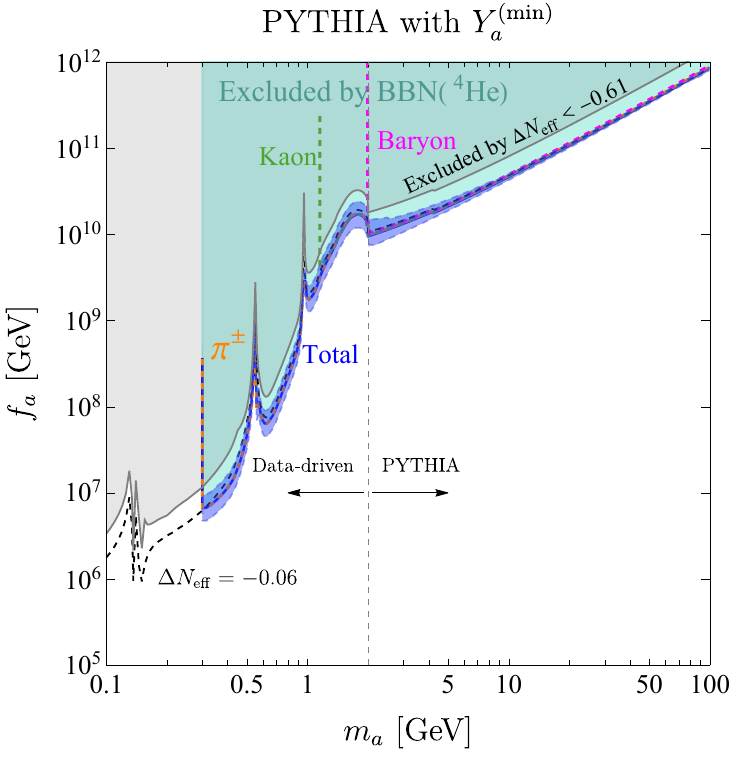}
    \includegraphics[width=0.47\textwidth]{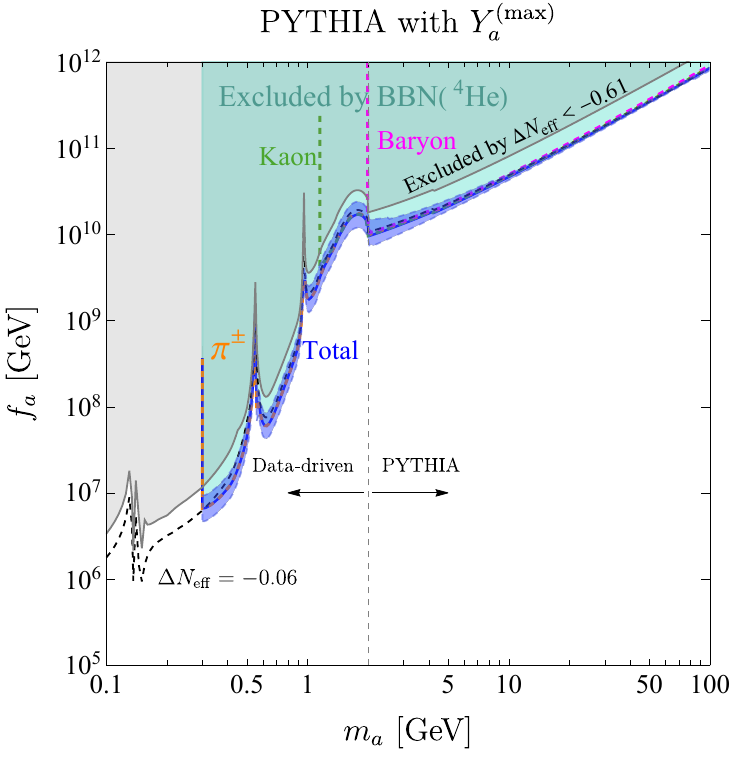}
    
    \vspace{0.3cm}    
    \includegraphics[width=0.47\textwidth]{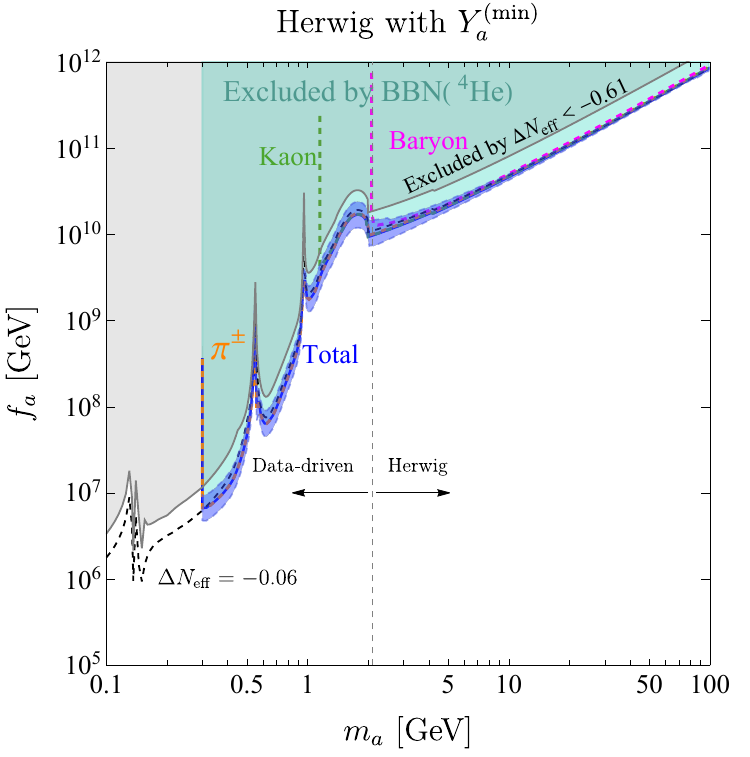}
    \includegraphics[width=0.47\textwidth]{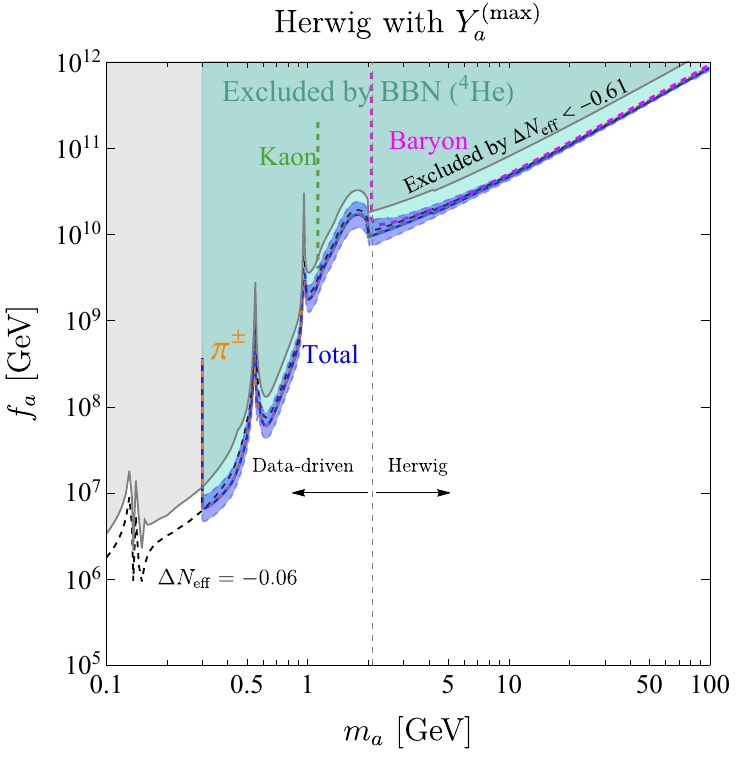}
    
    \caption{
    The exclusion by our BBN calculation is now given in the $(m_a,\,f_a)$ space, where the color scheme is unchanged from Fig.\,\ref{fig:resultintaua}.
    Here, we also demonstrate the large uncertainty coming from translating $\tau_a$ into $f_a$ by the blue band (we omit the uncertainty bands for the CMB constraint contours to avoid overcrowding).
    }
    \label{fig:resultinfa}
\end{figure*}

\subsection{Bounds on the axion lifetime and decay constant}

In our calculation, we have two possibilities of the axion abundance, $Y_a^{\rm (min)}$ and $Y_a^{\rm (max)}$, and two programs for the parton-shower and hadronization (\texttt{PYTHIA} and \texttt{Herwig}).
Moreover, we have two ways of presenting our constraint: $\tau_a$ vs $f_a$.
Therefore, there are eight combinations in presenting our results (see Figs.\,\ref{fig:resultintaua} and \ref{fig:resultinfa}). 

Fig.\,\ref{fig:resultintaua} shows our results in $(m_a,\,\tau_a)$ space, where left and right panels take $Y_a^{\rm (min)}$ and $Y_a^{\rm (max)}$ and upper and lower panels use \texttt{PYTHIA} and \texttt{Herwig}, respectively.
The cyan-shaded region surrounded by the blue contour is the region excluded by our BBN constraint ({\He}), while the orange, dark green, and pink dashed contours represent constraints estimated from individual channels of $\pi^\pm$, kaons, and baryons (with secondary hadrons turned on).
The gray vertical dashed line at $m_a\simeq2\,\GeV$ indicates the boundary of changing our scheme to estimate the axion decay rate and branching ratios; we use the data-driven method for $m_a<2\,\GeV$ and \texttt{PYTHIA} or \texttt{Herwig} for $m_a>2\,\GeV$.

Our upper bound on $\tau_a$ is almost flat although $N_{a\to h}$ increases as $m_a$ increases (e.g. see Fig.\,\ref{fig:NhadShower}).
The bound becomes even weaker at higher mass, although this dependence is tiny.
This counter-intuitive feature is because a heavier axion actually induces a longer axion-dominated period, decreasing $N_{\rm eff}$ and the Hubble rate.
It consequently delays the neutron freeze-out, leading to longer relaxation time, and decreases $X_n$. 
This effect competes with the enhancement of $X_n$ due to larger ${\rm N}_{a\to h}$.

The gray-shaded region is excluded by the $N_{\rm eff}$ bound from the CMB fitting at Planck collaboration\,\cite{Planck:2018vyg}.
Our $N_{\rm eff}$ bound is slightly different from that of Ref.\,\cite{Dunsky:2022uoq} because (i) we use a weaker criterion $N_{\rm eff}>2.43$,  as discussed in Sec.\,\ref{sec:numerical-steps},
and (ii) we take the different total decay width 
as described in  Sec.\,\ref{sec:decay}%
\footnote{We take the result of Ref.\,\cite{Bisht:2024hbs} for $m_a <2\,\GeV$ and the ${\rm NNLO}$  expression \cite{Chetyrkin:1998mw} for $m_a >2\,\GeV$ while Ref.\,\cite{Dunsky:2022uoq} used the result of Ref.\,\cite{Aloni:2018vki} for $m_a <2\,\GeV$ and the ${\rm NLO}$ expression for $m_a >2\,\GeV$.}
that causes modification of $Y_a$ in the $\tau_a$ space (the axion disappearance rate relevant for $Y_a$ is evaluated in terms of $f_a$, not $\tau_a$). 
The black dashed line ($\Delta N_{\rm eff}=-0.06$) represents the potential sensitivity of $N_{\rm eff}$ bound expected at the CMB-S4 experiment with about 1\% precision \cite{CMB-S4:2022ght}.
As shown in the plot, our BBN constraint, if present, is stronger than the projected CMB bound.

The $N_{\rm eff}$ bound is essential for constraining the lifetimes in the low mass region where the hadronic decay is forbidden. Potentially, there would be other BBN constraints, such as the one from the primordial deuterium measurement, which is sensitive to the late-time photodissociation.  However, we do not consider those bounds because the $N_{\rm eff}$ bound is substantially stronger.

These constraints are then transferred to the $(m_a,\,f_a)$ space in Fig.\,\ref{fig:resultinfa}; again, left and right panels show the results with $Y_a^{\rm (min)}$ and $Y_a^{\rm (max)}$ and upper and lower panels are obtained by using \texttt{PYTHIA} and \texttt{Herwig}, respectively.
As we discussed in Sec.\,\ref{sec:decay} with Fig.\,\ref{fig:a-width}, there is a large uncertainty in the axion total decay width estimation in terms of $m_a$ and $f_a$.
We depict this uncertainty by the blue band for our BBN constraint, while we do not present similar uncertainty bands for the $N_{\rm eff}$ constraint and sensitivity contour, as they would make the plot unreadable.

\begin{figure*}[t]
    \centering
    \includegraphics[width=0.47\textwidth]{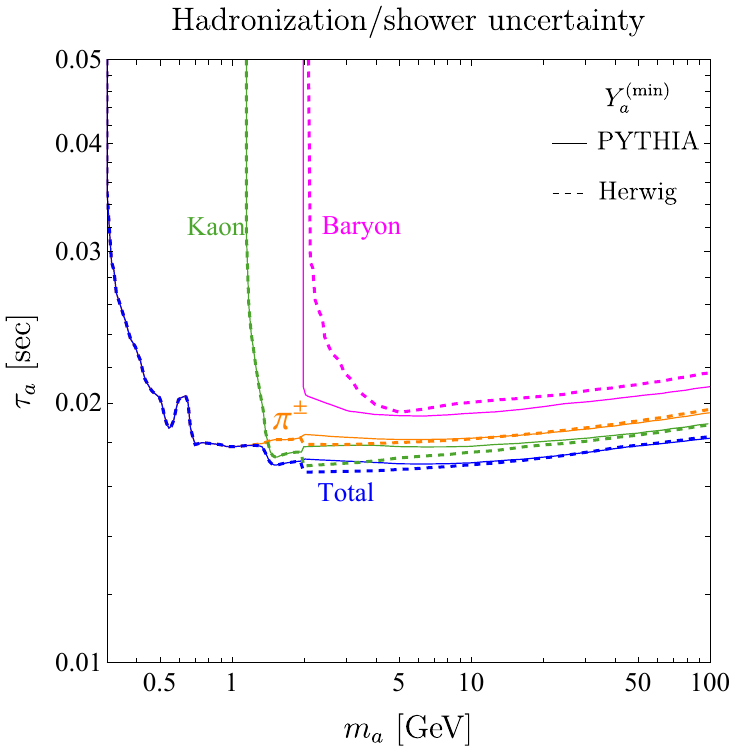}
    \hspace{10pt}
    \includegraphics[width=0.47\textwidth]{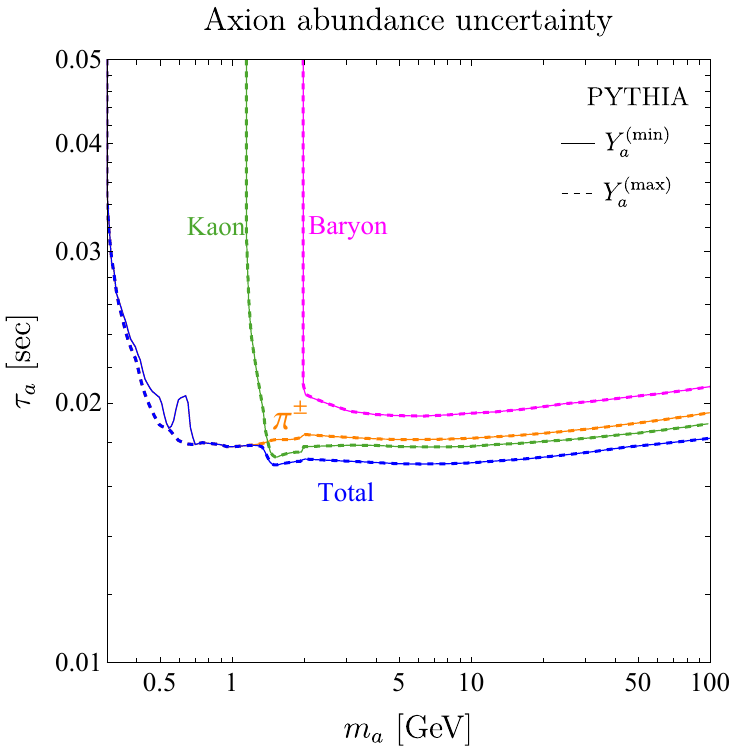}
    \caption{
    Comparison of our final constraints among different sets of $Y_a$ and parton-showering/hadronization programs.
    In the left panel, we fix the axion yield by $Y_a^{\rm (min)}$, and compare \texttt{PYTHIA} (solid lines) vs \texttt{Herwig} (dashed lines).
    In the right panel, we use \texttt{PYTHIA}, and compare $Y_a^{\rm (min)}$ (solid) vs $Y_a^{\rm (max)}$ (dashed).
    The color scheme is unchanged from Fig.\,\ref{fig:resultintaua}.
    \label{fig:compare}}
\end{figure*}

In the left panel of Fig.\,\ref{fig:compare}, we depict the results with \texttt{PYTHIA} (solid lines) and \texttt{Herwig} (dashed line) with $Y_a^{\rm (min)}$ while the color scheme remains unchanged.
As we discussed in the earlier subsection, their impact on $\tau_a$ is small except for the baryon channel around $m_a\sim 2\,\GeV$, where a large discrepancy appears as the thresholds for the baryonic channel do not match due to their different hadronization algorithms as shown in Fig.\,\ref{fig:NhadShower}.
Although the discrepancy in the baryon channel is somewhat large, our final result is dominated by the charged pions and kaons, which are much more stable.

In the right panel of Fig.\,\ref{fig:compare}, we compare the results using $Y_a^{\rm (min)}$ (solid) or $Y_a^{\rm (max)}$ (dashed), where  \texttt{PYTHIA} is used.
For $\tau_a \sim 0.02\,\sec$, the difference in $Y_a$ only appears at $m_a \lesssim 0.7\,\GeV$ (see the bottom panels in Fig.\,\ref{fig:Ya}).
The largest discrepancy appears around $0.5\,\GeV \lesssim m_a \lesssim 0.7\,\GeV$, where $Y_a^{\rm (max)}/Y_a^{\rm (min)}$ can be as large as $10$.
As we argued in the previous subsection, however, this leads to only $20\,\%$ discrepancy in $\tau_a$.

\section{Conclusion \label{sec:concl}}

In this work, we have estimated the BBN constraint on the heavy QCD axion that decays hadronically. 
The constraint is derived by computing the axion-induced modification to the neutron-to-proton ratio, which directly determines the primordial {\He} abundance.
The axion yield is evaluated by its freeze-out value, assuming a large reheating temperature, and its hadronic branching ratios are obtained by using the data-driven method at $m_a<2\,\GeV$ and \texttt{PYTHIA} or \texttt{Herwig} at $m_a > 2\,\GeV$.
With these input quantities, we solve the Boltzmann equation for $X_n$, and obtain the constraint in the $(m_a,\,\tau_a)$ space.
Our constraint is also depicted in the $(m_a,\,f_a)$ space, although there is an $\mathcal{O}(1)$ uncertainty in the conversion of $\tau_a$ and $f_a$.

Our analysis incorporates several key updates on hadronic injection scenarios during BBN that can be applied to other models.
We especially include $K_L$ contributions using isospin relations to obtain their cross sections, and account for their momentum distributions as $K_L$ does not get thermalized kinetically.
We also trace secondary hadrons produced from decays and scatterings, ensuring a more consistent treatment. 
Our methodology can also be applied to a broader class of long-lived particles that decay into hadrons. 

Based on these improvements, we have derived a robust upper bound on the axion lifetime, $\tau_a \lesssim 0.02\,\sec$, across a wide range of axion masses above $0.3\,\GeV$. 
We find that our BBN constraint is more stringent than existing and even projected CMB constraints via $N_{\rm eff}$.
This highlights the importance of the BBN analysis for hadronically decaying long-lived particles.

Our bound is shown to be quite insensitive to the branching fractions, hadronic cross sections, and the initial axion yield, as the dependence of the  $\tau_a$ bound on these parameters appears in a logarithm with a small coefficient. As discussed thoroughly in Sec\,\ref{sec:simplified}, even 100\% modification of the initial axion yield or the branching fractions due to different models would lead to only 4\% shift in the lifetime bound. 
Hence, our constraint is robust and nearly model independent.

\section*{Acknowledgments}
We thank Sabyasachi Chakraborty, Sean Dobbs, David Dunsky, Keisuke Harigaya, Kazunori Kohri, Takeo Moroi, Yotam Soreq, Sheng-Quan Wang, and Seokhoon Yun for useful discussions. 
We thank Sabyasachi Chakraborty for correspondence regarding \cite{Bisht:2024hbs}, David Dunsky for correspondence regarding \cite{Dunsky:2022uoq}, and Jonas Frerick for pointing out errors in an earlier version of Appendix~\ref{sec:xsec-pion}. 
This work was supported in part by IBS under the project code, IBS-R018-D1,  the US Department of Energy grant DE-SC0010102, and the FSU Bridge Funding 047302. 

\section*{Data availaility  \label{sec:data}}
The data that support the findings of this article are openly available~\cite{github}, embargo periods may apply.

\begin{appendix}

\section{Axion decay rate at ${\rm NN LO}$ \label{app:a_decay}}
The QCD corrections to the pseudo-scalar decay to gluons are calculated in Ref.\,\cite{Chetyrkin:1998mw}. Adopting the relevant corrections ($\tilde C_1^2$ part)
\bal
\label{eq:NNLO}
&\Gamma(a\to gg) = 
\frac{2}{\pi} \left( \frac{\alpha_s(\mu)}{8\pi} \right)^{\! 2}
\frac{m_a^3}{f_a^2} 
\\
& \times \!
\Bigg[ 1 \!+\! 
\frac{\alpha_s(\mu)}{\pi} 
\Big( 
\frac{97}{4} \!-\! \frac{7n_f}{6}
\!-\! \big(\frac{11}{2} \!-\! \frac{n_f}{3}\big)
\log \frac{m_a^2}{\mu^2}
\Big) 
\nn \\
\displaybreak[2]
& \quad + 
\Big( \frac{\alpha_s(\mu)}{\pi} \Big)^{\!\! 2}
\Bigg(
\frac{51959}{96}-\frac{363}{8}\zeta(2)
-\frac{495}{8}\zeta(3) 
\nn \\
\displaybreak[2]
& \qquad \qquad \qquad \quad
+ \Big( \!\! -\! \frac{469}{8}+\frac{11}{2} \zeta(2)+\frac{5}{4}\zeta(3)
\Big) n_f
\nn \\
\displaybreak[2]
& \qquad \qquad \qquad \quad
+ \Big( \, \frac{251}{216}-\frac{1}{6}\zeta(2) \Big) n_f^2
\nn\\
\displaybreak[2]
& \qquad \qquad \qquad \quad
+ \Big(\frac{3405}{16}-\frac{73}{3}n_f + \frac{7}{12}n_f^2 \Big) \log \frac{\mu^2}{m_a^2}
\nn\\
\displaybreak[2]
&\qquad \qquad \qquad \quad
+ \Big( \frac{363}{16} - \frac{11}{4} n_f +\frac{1}{12} n_f^2 \Big) \log^2 \frac{\mu^2}{m_a^2}
\Bigg)
\Bigg],
\nn
\eal
where $\alpha_s(\mu)$ is obtained by solving the renormalization group equation and matching condition, Eqs.\,(9.3, 9.4) of \cite{ParticleDataGroup:2024cfk}, and $\alpha_s(m_Z) = 0.1177$.  $n_f$ is the number of quark flavors lighter than $\mu$. We vary the renormalization scale by a factor of 2 to estimate the uncertainty.

\section{Hadron yields from {\tt PYTHIA} and {\tt Herwig}
\label{app:hadron}}
We use the simulation programs {\tt PYTHIA} and {\tt Herwig} to obtain the effective hadron yields from axion decays. These programs automatically simulate the decays of unstable particles, such as vector mesons and hyperons. For each axion mass, we generate 10,000 events and compute the average hadron yields.
We observe that both programs give consistent results for charge-conjugate pairs, such as $\pi^+$/$\pi^-$ and $K^+$/$K^-$, as expected, so we take the average of them for each hadron type. The tabulated data files are available in \cite{github}.

We specifically record the information of $K_L$ energy spectrum, as shown in Figs.~\ref{fig:KLmom} and \ref{fig:KLmom_lowma}. Although $K_L$ does not slow down by the electromagnetic interaction,  the $K_L$ spectrum is modified by the $NK_L$ elastic scattering (Appendix~\ref{sec:xsec-kaon}), and the reshaped spectrum is presented in Figs.~\ref{fig:KLmom_reshaped} and \ref{fig:KLmom_lowma_reshaped}, following the procedure described in Appendix~\ref{sec:KLreshape}. 
We do not retain similar information for other hadrons because they are kinetically thermalized rapidly.

Let us comment on the instabilities of the two generators, which lead to discrepancies in the intermediate steps, such as the effective number of hadrons and the cross section weighted with the $K_L$ spectrum. However, the final result, the bound on the axion lifetime, is almost insensitive to these issues, see Fig.~\ref{fig:compare} left.

\begin{itemize}
    \item {\bf Peak from $a \to \bar{K}^0 K^*,K^0 \bar K^*$ for $m_a\lesssim$ 2.5\,GeV}
    \\
    The $K_L$ distributions obtained by the two generators are unstable for the low mass  $m_a\lesssim$ 3\,GeV. There is an abnormal peak of $K_L$ spectrum in {\tt PYTHIA} for  $m_a\lesssim$ 2.5\,GeV (e.g., see Fig.\,\ref{fig:KLmom_lowma}). The peak comes from direct two-body decay events of $a \to K^0 K^*$. We observe {\tt PYTHIA} does not generate three-body decay of $a\to KK\pi $ at $m_a=2\,\GeV$ even though it is kinematically allowed. However, this feature fades away as the axion mass increases to 3\,GeV. This kind of behavior leads to some discrepancies in the average $NK_L$ cross sections in Fig.\,\ref{fig:KLNxsec}

    \item {\bf Parity Violation}
    \\
    The simulation programs do not handle polarization information properly. Therefore, sometimes they generate unphysical processes that violate parity. For example, we find up to 30\% of events are unphysical for $m_a=2\,\GeV$. Although both {\tt PYTHIA} and {\tt Herwig} have such flaws, they are still the representative shower programs. We use both to solve the system, and the difference between their results can be taken as unknown systematics. 

    \item {\bf Baryon yield discrepancy}
    \\
    The two programs predict the number of nucleons, but their predictions differ by about a factor of 2, and the difference becomes larger as $m_a$ increases. However, this baryon decay channel is subdominant in affecting the neutron freeze-out dynamics, so this discrepancy is negligible for the final result. 
    
\end{itemize}

\section{Updates on hadronic cross sections \label{app:x-sec}}

In this appendix, we present the schemes to obtain the hadronic cross sections which are necessary for the Boltzmann equations addressed in Sec.\,\ref{sec:n-decoupling}. We present general treatments in Sec.\,\ref{app:xsec-general}: how we treat the average of cross sections, the Coulomb correction, and conversion of the phase space factors involved in the time reversal or isospin transformations.

Then, we apply these techniques to the cross sections involving injected pions in Sec.\,\ref{sec:xsec-pion} and kaons in Sec.\,\ref{sec:xsec-kaon}. We use Ref.\,\cite{Lee:2015hma} for the baryon annihilation cross sections, which is given in Sec.\,\ref{sec:xsec-baryon} for completeness.

\subsection{General treatments \label{app:xsec-general}}

\noindent
{\bf \underline{Kinematically averaged cross sections}:}
As shown in Ref.\,\cite{Kohri:2001jx}, hadrons injected from the axion decay, except for $K_L$,  get quickly thermalized kinetically via electromagnetic interactions with the background photons.
The time scale of the kinetic thermalization is much shorter than that of number changing processes, and therefore, we take their kinetic distributions as the thermal distribution determined by the photon temperature, while their number densities are solved via the Boltzmann equation. 

Because the hadron masses are much greater than the BBN temperature range, we can take the non-relativistic limit where the momentum distribution is given by the Maxwell-Boltzmann distribution $f_{\rm MB}(p,m) \propto \exp[-\frac{1}{T}(m+\frac{p^2}{2m})]$. 
In this approximation, we can rewrite the phase space integration as
\bal
&d^3 p_1 d^3 p_2 f_{\rm MB}(p_1; m_1) f_{\rm MB}(p_2; m_2)
\nn
\\
&= 
d^3 P d^3 p_{\rm cm}
f_{\rm MB}(P; m_1+m_2)
f_{\rm MB}(p_{\rm cm}; \bar\mu(m_1,m_2))
\eal
where $\vec P=\vec p_1 + \vec p_2$, $\vec p_{\rm cm}=\bar\mu\left({\vec p_1}/{m_1}-{\vec p_2}/{m_2} \right)$ with $\bar\mu(m_1,m_2) = (m_1^{-1} + m_2^{-1})^{-1}$.
Note that $\vec p_{\rm cm}$ ($-\vec p_{\rm cm}$) is the momentum of the particle $1$ ($2$) in the center-of-momentum frame.
Therefore, the averaged cross-sections can be obtained by
\bal
&\langle \sigma v (N X \to N' X')\rangle
\nn
\\
&= \frac{1}{\cal C} \int \frac{d^3 p_{\rm cm}}{(2\pi)^3} 
f_{\rm MB}(p_{\rm cm} ; \bar\mu(m_N, m_X) )
\,
\sigma v (N X \to N' X') ,
\eal
with $N (N') =p, n$, $X(X')=p,\bar p, n, \bar n, \pi^\pm, K^\pm$ and
${\cal C}= \int \frac{d^3 p}{(2\pi)^3} f_{\rm MB}(p; \mu)$.

The relative velocity $v$ should be the M{\o}ller velocity\,\cite{Gondolo:1990dk}.
It is defined as 
\bal
v &\equiv \frac{|\vec p_{\rm cm}| \Ecm}{E_1 E_2}
= \frac{1}{E_1 E_2}\sqrt{(p_1 \cdot p_2)^2 - m_1^2 m_2^2}
\nn
\\
&
= \frac{1}{E_1 E_2} \sqrt{(E_1 E_2 - |\vec p_1| |\vec p_2| \cos \theta)^2 - m_1^2 m_2^2},
\label{eq:Moller_velocity}
\eal
with the center-of-mass energy $\Ecm$.
This velocity becomes a familiar form, $v \simeq |\vec p_{\rm cm}|/\bar\mu \simeq |\vec v_1 - \vec v_2|$, in the non-relativistic limit where $E_1\simeq m_1 +\frac{|\vec{p_1}^2|}{2m_1}$ and $E_2\simeq m_2 +\frac{|\vec{p_2}^2|}{2m_2}$ although we do not employ this approximation in the numerical integral.

As the thermalization of $K_L$  is highly suppressed\,\cite{Kohri:2001jx}, we take the energy distribution determined by the axion decay and subsequent elastic scattering (see Sec.\,\ref{sec:x-sec} for its distribution for different $m_a$ and  Appendix \,\ref{sec:KLreshape} for relevant discussion) and denote it $f_{K_L}(p_{K_L} ; m_a)$.
Therefore, averaged cross-sections for $K_L$-involved processes are estimated by
\bal
\langle \sigma v (N K_L \to N' X) \rangle
=& \frac{1}{{\cal C}'} \int \frac{d^3p_N}{(2\pi)^3} 
\frac{d^3p_{K_L}}{(2\pi)^3}
f_{\rm MB}(p_N ; m_N)
\nn
\\
&f_{K_L}(p_{K_L} ; m_a)
\,
\sigma v (N K_L \to N' X) 
\eal
where ${\cal C}'=\int \frac{d^3p_N}{(2\pi)^3} f_{\rm MB}(p_N ; m_N) \cdot \int \frac{d^3p_{K_L}}{(2\pi)^3}f_{K_L}(p_{K_L} ; m_a)$.

\vspace{0.2cm}

\noindent
{\bf \underline{Coulomb correction}: }
We take into account the Coulomb correction by multiplying the Sommerfeld enhancement/suppression factor 
\bal
F(Z,v) = \frac{|\psi(0)|^2}{|\psi_0(0)|^2}
\eal
where $\psi(x)$ and $\psi_0(x)$ are the wave functions with and without the Coulomb potential.
The general form of the correction can be written as
\bal
F(Z,v)
=
9\cdot 2
(1+S)
(2p R)^{-2(1-S)}
\frac{|\Gamma(S+i\eta)|^2}
{\Gamma(2S+2)}
e^{\pi \eta}
\label{Eq:Fermi_function}
\eal
which is known as the Fermi function\,\cite{Fermi:1934hr},
where $S=(1-\alpha^2 Z^2)^{1/2}$, $p$ is the momentum of relative motion, and $R\simeq 1\,{\rm fm}$ is the proton radius that provides the UV cutoff. $\eta=Z\alpha/v$ is the Sommerfeld parameter with the relative velocity $v=p/E$ (which approximates the M{\o}ller velocity \eqref{eq:Moller_velocity}).
As $S\simeq 1+{\cal O}(\alpha^2 Z^2) $,
we can approximate $|\Gamma(S+i\eta)|^2 \simeq |\Gamma(1+i\eta)|^2 = \pi \eta/\sinh \pi \eta$, and obtain
\bal
F(Z,v)
\simeq
\frac{2\pi \eta}
{1-e^{-2\pi\eta}}
+ {\cal O}(\alpha^2 Z^2).
\label{Eq:Coulomb}
\eal
Note that the $R$ contribution appears in the $Z^2 \alpha^2$ order with logarithmic dependence as $\sim Z^2 \alpha^2 \log p R$, so we ignore it.
We use  Eq.\,\eqref{Eq:Coulomb} since it is more stable numerically than using the full expression of Eq.\,\eqref{Eq:Fermi_function}.

\begin{figure*}[t!]
    \centering
    \includegraphics[width=0.45\textwidth]{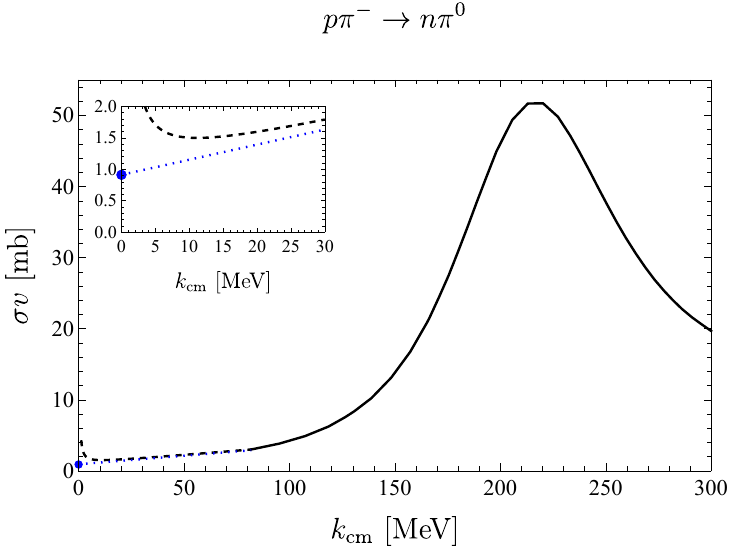}
    \includegraphics[width=0.45\textwidth]{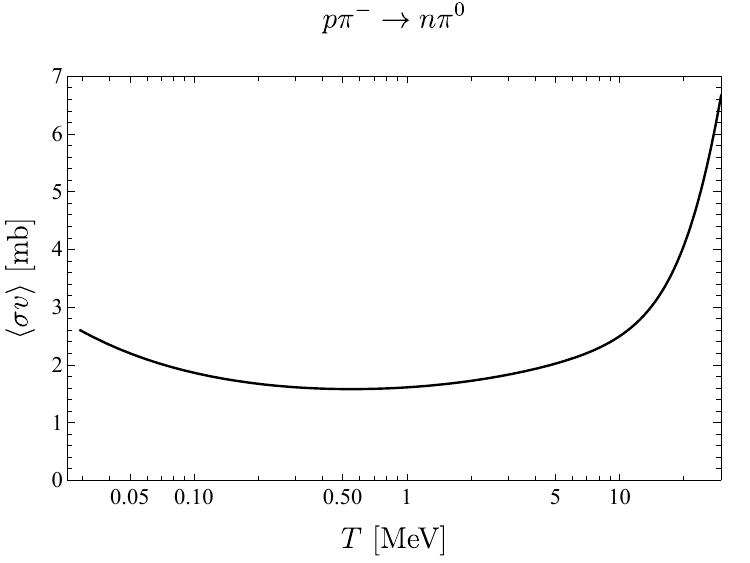}
    \caption{ Cross sections for $p \pi^- \to n \pi^0$. In the left panel, $k_{cm}$ is momentum in the initial system ($p \pi^-$). The small panel in the left is to magnify the threshold region. The threshold value (blue point) is inferred from analysis of $1S$ bound state of $\pi^- p$. The dashed line is a linear interpolation from the SAID program data to the threshold value. Right panel shows thermally averaged cross section in the relevant temperature range.
    \label{Fig:ppim_pi0}}
    \centering
    \includegraphics[width=0.45\textwidth]{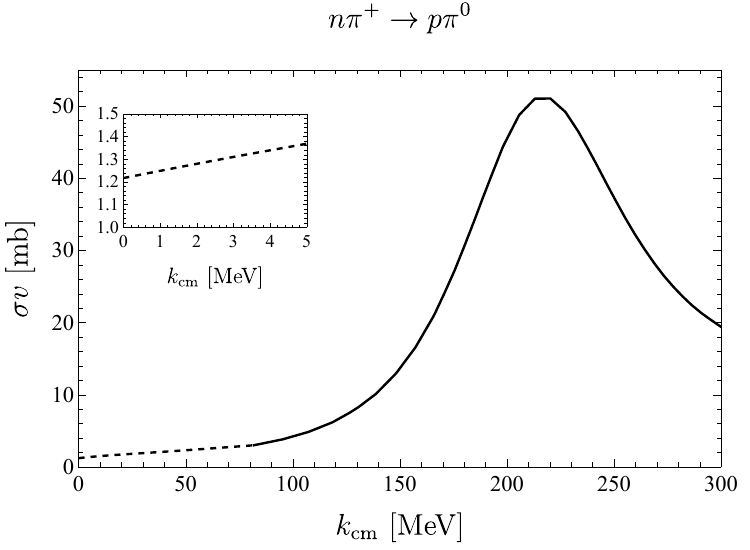}
    \includegraphics[width=0.45\textwidth]{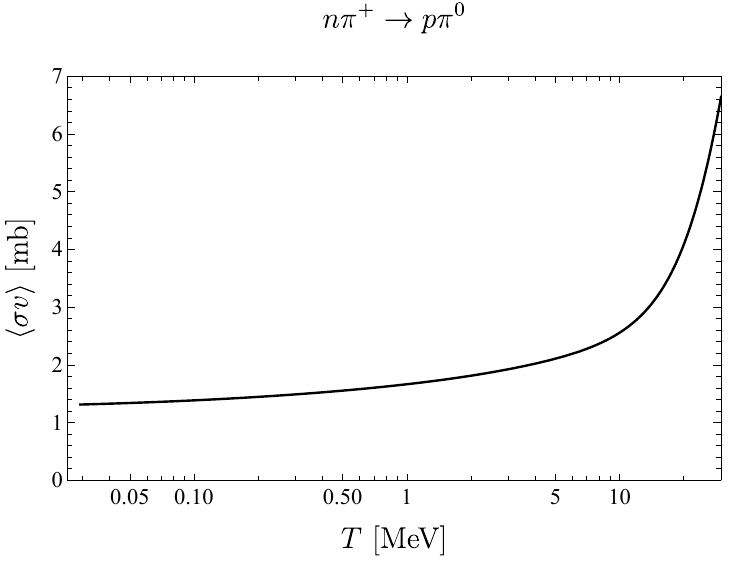}
    \caption{ Similar plots as in Fig.\,\ref{Fig:ppim_pi0}, but for  $n \pi^+ \to p \pi^0$. The dashed line in the left panel is an interpolation of the SAID program data to the threshold value after phase space correction.
    \label{Fig:npip_pi0}}
\end{figure*}

\vspace{0.2cm}

\noindent
{\bf \underline{Phase space factors}: }
There are many processes whose experimental measurements or fitted functions do not exist.
Those cross sections can be inferred either by taking data from the reverse process or from a combination of isospin transformations.
We relate scattering amplitudes by time-reversal or isospin ignoring the mass differences while we still need to correct the phase space difference, which is crucial near the threshold.

More explicitly, we approximate the cross section to contain the simple phase space factor as $\sigma(1\,2\to 3\,4) \propto \frac{1}{\Ecm^2}\frac{\kcm(3,4)}{\kcm(1,2)}$ and obtain the relation, 
\bal
\sigma (1 \, 2 \to 3 \, 4) 
\simeq
\sigma (1' 2' \to 3' 4') 
\left[ g \cdot  
\frac{\PS(1 2 \to 3 4)}{\PS(1' 2' \to 3' 4')} 
\right]_{X}
\eal
where $g$ accounts for the Coulomb correction as well as the change of internal degrees of freedom.
The subindex $X$ emphasizes that we need to fix a kinematic variable $X$ depending on cases; it could be $\kcm(1,2)=\kcm(1',2')$ or $\Ecm(1,2)=\Ecm(1',2')$.
We take $X=\Ecm$ when we utilize the reverse process since we are using the invariance of the amplitude under the time reversal.
On the other hand, when we use the relations under isospin transformations for non-relativistic scattering processes,
we take $X=k_{\rm cm}$ since the non-relativistic scattering amplitude should not care about the total mass.

The ratio of the phase space factors is given by
\bal
&\left[ 
\frac{\PS(1\,2\to 3\,4)}{\PS(1' 2' \to 3' 4')} 
\right]_X
\nn
\\
=&
\left(
\frac{\kcm(3,4)}{\kcm(1,2)}
\right)_{\!\!X}
\left(
\frac{\kcm(1', 2')}{\kcm(3', 4')}
\right)_{\!\!X}
\left( \frac{\Ecm(1',2')^2}{\Ecm(1,2)^2} \right)_{\!\!X} \,.
\eal
For two cases of $X$, we use
\bal
&X=\Ecm: 
\nn
\\
&\kcm(1,2) \nonumber\\
\displaybreak[2]
&=
\frac{\Ecm}{2}
\sqrt{
\left(1-\frac{(m_1+m_2)^2}{\Ecm^2} \right)
\left(1-\frac{(m_1-m_2)^2}{\Ecm^2} \right)
}\quad ,
\\\nn
&X=\kcm: 
\\
& \Ecm(1,2)= \sqrt{m_1^2 + \kcm(1,2)^2}+\sqrt{m_2^2+\kcm(1,2)^2} \quad . 
\eal

Experimental data are often given in $\Tlab$, the kinetic energy of an injected particle in the lab frame where the target is fixed.
Therefore, it is useful to write down explicit formulas for $\Ecm$ and $\kcm$ in terms of $\Tlab$.
Denoting the particle $2$ as the beam particle in the lab frame for the $1\,2\to 3\,4$ process, we obtain
\bal
&\Ecm = \sqrt{(m_1+m_2)^2 +2m_1 \Tlab}\quad ,
\\
&\kcm(1,2) = \frac{m_1 \sqrt{\Tlab(2m_2+\Tlab)}}{\sqrt{(m_1+m_2)^2+2m_1 \Tlab}}\quad .
\eal

\subsection{Injected pions \label{sec:xsec-pion}}

\begin{figure*}
    \centering
    \includegraphics[width=0.4\textwidth]{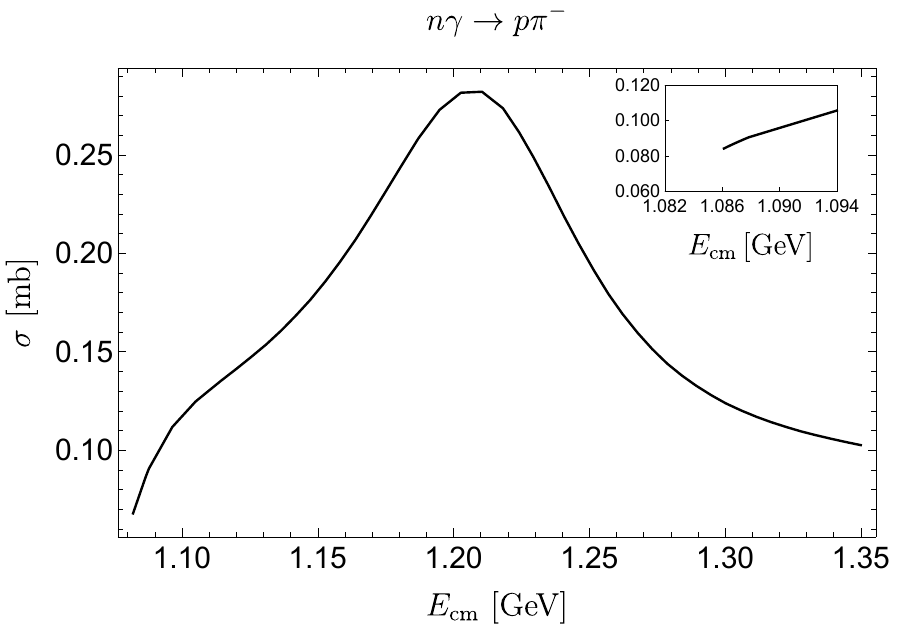}\\\vspace{-10pt}
    \includegraphics[width=0.4\textwidth]{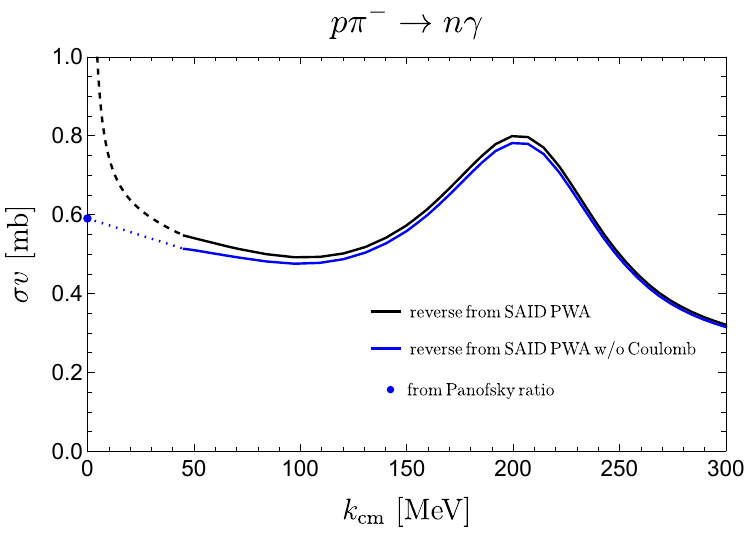}
    \includegraphics[width=0.4\textwidth]{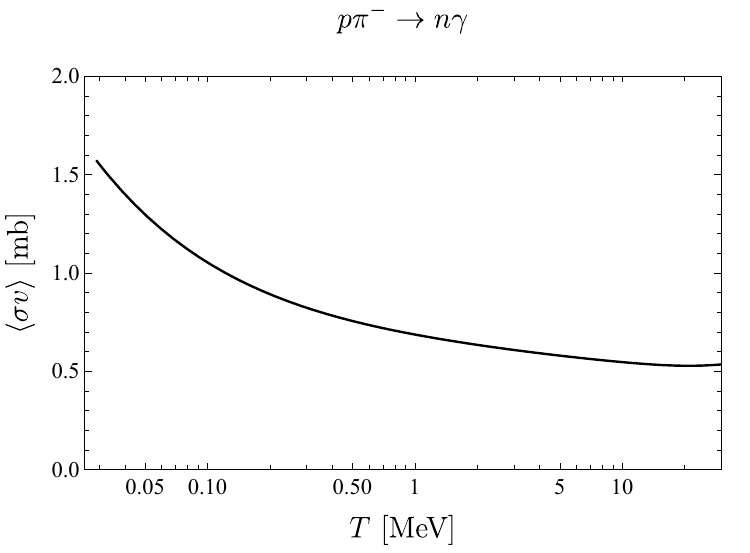}
    \caption{ Cross sections for $n \gamma \to p \pi^-$ (top) and $p \pi^- \to n \gamma$ (bottom). In the top panel, the data is from SAID program with a magnified view near threshold in the inset. 
    The bottom-left panel shows the cross section obtained by time reversal, Eq.~\eqref{Eq:ppim_TO_ngamma_conversion}; the blue point is the threshold value in Eq.~\eqref{eq:threshold_ppi-},  the blue dotted line is the linear interpolation to it, and the (dashed) black line includes the Coulomb attraction. The bottom-right panel shows the thermally averaged cross section in the relevant temperature range.
    \label{Fig:ppim_photon}
    }
    \centering
    \includegraphics[width=0.4\textwidth]{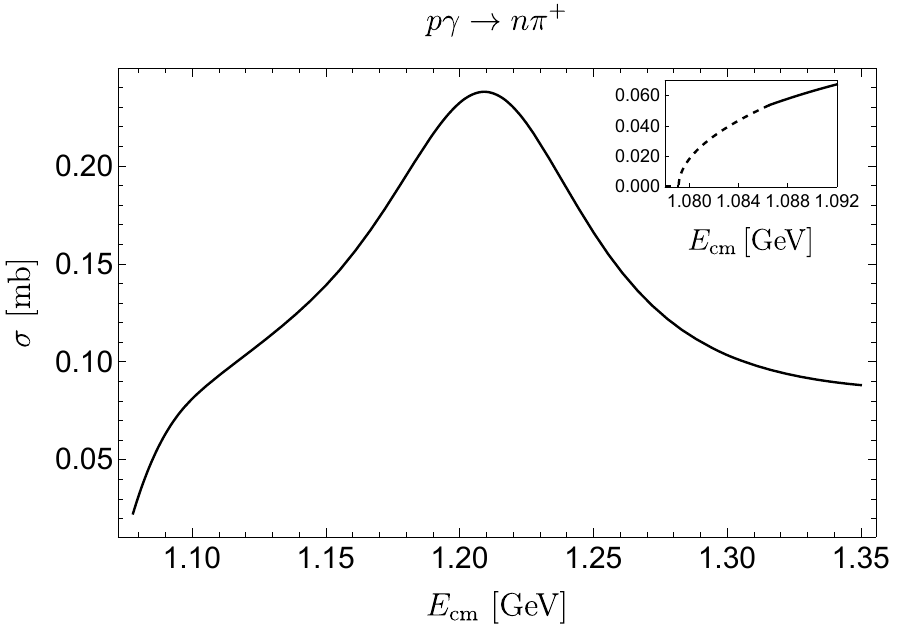}\\\vspace{-10pt}
    \includegraphics[width=0.4\textwidth]{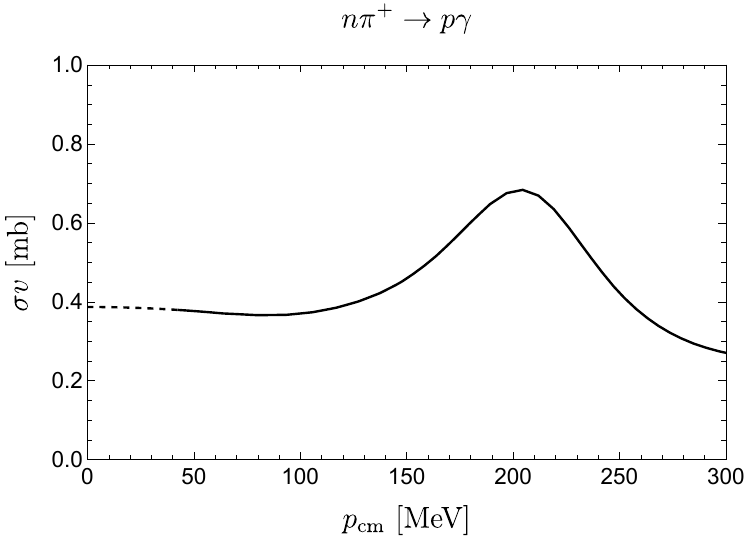}
    \includegraphics[width=0.4\textwidth]{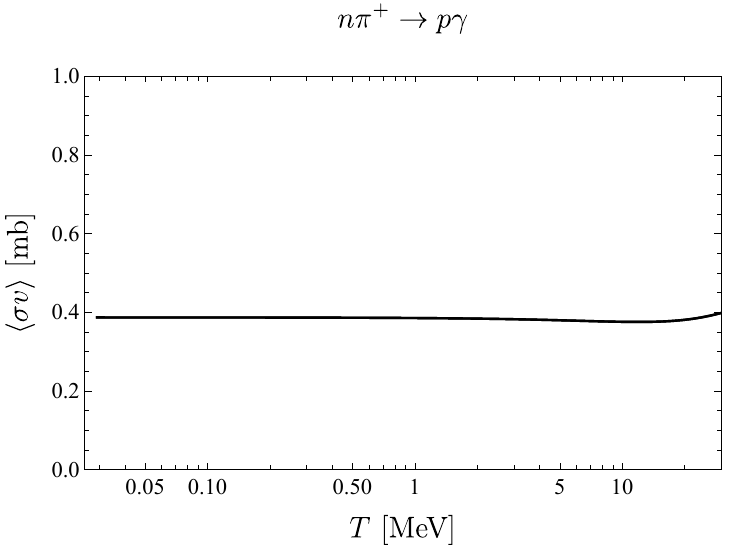}
    \caption{ Similar plots as in Fig.\,\ref{Fig:ppim_photon}, but for $p \gamma \to n \pi^+$ (top) and $n \pi^+ \to p \gamma$ (bottom). In the top panel, the curve is from the SAID program. The inset in the top right shows a zoomed view near the threshold. The dashed line in the top panel is the fit with $\sigma \propto v_{\pi,f}-v_{n,f}$.
    \label{Fig:npip_photon}
    }
\end{figure*}

\underline{
$\mathbf{p \pi^- \to n \pi^0}$
} (Fig.\,\ref{Fig:ppim_pi0}):
We take the partial wave analysis (PWA)\,\cite{Workman:2012hx} provided by the George Washington University SAID program\,\cite{Arndt:2003fj, GWU:SAID}.
Direct measurements are available down to a lab-frame kinetic energy of $\Tlab=27.4~\MeV$, and we use the SAID fit down to $\Tlab = 30~\MeV$. Since this corresponds to an incident CM momentum of $81.9~\MeV$, which is well above the momenta relevant for BBN, we separately infer the threshold cross section from the width of pionic hydrogen, following Ref.\,\cite{Gasser:2007zt}. 

The cross section of $\pi^- p$ scattering via the strong interaction can be obtained from the width of the pionic hydrogen 1S state,
\begin{align}
    &\Gamma_{p\pi^-} = \frac{4\pi}{\mu_{p\pi^-}} |\Psi_{10}(0)|^2 {\rm Im} \ A(p\pi^- \to  p\pi^-), 
\end{align}
where $A$ is the scattering length via the strong interaction, $|\Psi_{10}(0)|^2 =\alpha^3 \mu_{p\pi^-}^3/\pi$ is the squared wave function at the origin, and $\mu_{p\pi^-} =0.1214~ \GeV$ is the reduced mass.

The imaginary part of the scattering length gives the total inelastic cross section,
\begin{align}
  \sigma v(p\pi^-\to n\pi^0 \ {\rm or\ } n \gamma)
  = \frac{4\pi}{\mu_{p\pi^-}}\,{\rm Im}\,A(p\pi^-\to p\pi^-),
\end{align}
where $v=k_{\rm cm}/\mu_{p \pi^-}$. 

For the pionic hydrogen width, we use the most recent measurement, $\Gamma_{p\pi^-} = 0.856\times 10^{-9}\,\GeV$~\cite{Hirtl:2021zqf}, which is slightly larger than the 2006 value, $0.823\times 10^{-9}\,\GeV$~\cite{Curceanu:2006sq}, adopted in Ref.~\cite{Pospelov:2010cw}. This gives $\sigma v(p\pi^- \to  n\pi^0 \ {\rm or\ } n \gamma) \approx 1.501\, \mb$.

To unfold the $n \pi^0$ and $n \gamma$ final states, we use the Panofsky ratio~\cite{Panofsky:1950he}measured in Ref.~\cite{Spuller:1977ve}
(see Ref.~\cite{Flugel:1999gr} for a review),
\begin{align}
    P_{\rm Panofsky}=\frac{\sigma(p\pi^- \to n \pi^0)_{\rm th}}{\sigma(p\pi^- \to n\gamma)_{\rm th}} = 1.546\ .
\end{align}
This gives the individual threshold cross sections due to the strong interaction,
\begin{align}
    \sigma v(p\pi^- \to  n\pi^0 )= 0.912 \,{\rm mb} ,\  \sigma v(p\pi^- \to  n \gamma)= 0.590 \,{\rm mb}. 
    \label{eq:threshold_ppi-}
\end{align}
These values are slightly larger than those in Ref.~\cite{Pospelov:2010cw},
$\sigma v(p\pi^-\to n\pi^0) = 0.88\,\mb$ and $\sigma v(p\pi^-\to n\gamma) = 0.57\,\mb$,
because of the updated width.

We then linearly interpolate between this threshold value and the SAID fit,
both with the Coulomb attraction removed (blue dotted line in
Fig.~\ref{Fig:ppim_pi0}). Finally, we restore the Coulomb attraction by
multiplying the result by $F(1,v)$ (black line). The resulting $\sigma v$ in the center-of-mass frame as a function of $\kicm$
and its thermal average are shown in the left and right panels of
Fig.~\ref{Fig:ppim_pi0}, respectively.


\noindent 
\underline{
$\mathbf{n \pi^+ \to p \pi^0}$
} (Fig.\,\ref{Fig:npip_pi0}):
We do not have experimental data for this process.
Therefore, we take an isospin rotation of $p\pi^- \to n \pi^0$.
In this case, as all the external particles are non-relativistic, we fix the initial center-of-mass momentum $\kicm$;
\bal
&\sigma (n \pi^+ \to p \pi^0) (\kicm)
\nn
\\
&
\!\!=\sigma (p\pi^- \to n\pi^0) (\kicm)\times
\left[ 
\frac{1}{F(1,v)}\cdot
\frac{\PS(n \pi^+ \to p \pi^0)}{\PS(p\pi^- \to n\pi^0)} 
\right]_{\kicm}
\!\!. \label{eq:npi+TOppi0}
\eal
The $1/F(1,v)$ factor accounts for the absence of the Coulomb enhancement in the $n\pi^+$ initial state.
The phase-space correction, on the other hand, is small. 

The left panel of Fig.~\ref{Fig:npip_pi0} shows the resulting $\sigma v$ in
the center-of-mass frame as a function of $\kicm$. The dashed part is a linear interpolation between the low-momentum end of the SAID curve and the threshold value obtained from Eq.~\eqref{eq:npi+TOppi0}. The right panel shows
the thermally averaged cross section.



\noindent 
\underline{
$\mathbf{p \pi^- \to n \gamma}$
and
$\mathbf{n \pi^+ \to p \gamma}$
}(Figs.\,\ref{Fig:ppim_photon} and \ref{Fig:npip_photon}):
Since these processes have not been measured directly, we use the reverse
processes, $n\gamma\to p\pi^-$ and $p\gamma\to n\pi^+$, whose PWA fits are
available~\cite{Workman:2012hx, Arndt:2003fj, GWU:SAID}. We infer the cross sections by the time-reversal transformation, 
\bal
&\sigma (p \pi^- \to n \gamma) (\Ecm)
\nn
\\
&=
\sigma (n \gamma \to p \pi^-) (\Ecm)
\times
\left[ 
2\cdot \frac{\PS(p\pi^-\to n\gamma )}{\PS(n\gamma \to p \pi^-)} 
\right]_{\Ecm}
,
\label{Eq:ppim_TO_ngamma_conversion}
\\
&
\sigma (n \pi^+ \to p \gamma)(\Ecm)
\nn
\\
&=
\sigma (p \gamma \to n \pi^+)(\Ecm)
\times
\left[ 
2 \cdot \frac{\PS(n\pi^+\to p\gamma )}{\PS(p\gamma \to n \pi^+)} 
\right]_{\Ecm},
\label{Eq:npip_TO_pgamma_conversion}
\eal
where both sides are evaluated at the same $\Ecm$, and the factor of 2
is due to the two photon polarization states.

Near threshold, the two processes are treated differently. For
$p\pi^-\to n\gamma$, we linearly interpolate between the threshold value in Eq.~\eqref{eq:threshold_ppi-} and the time-reversed SAID cross section (blue dotted line in the bottom-left panel of Fig.~\ref{Fig:ppim_photon}).
For $n\pi^+\to p\gamma$, the threshold cross section is unknown, we fit the near-threshold $p\gamma\to n\pi^+$ cross section with
$\sigma\propto v_{\pi^+,f}-v_{n,f}$ (inset of the top panel of Fig.~\ref{Fig:npip_photon}), and then apply the time-reversal, Eq.\,\eqref{Eq:npip_TO_pgamma_conversion}. The resulting cross sections and their thermal
averages are shown in Figs.~\ref{Fig:ppim_photon} and \ref{Fig:npip_photon}.

\subsection{Injected kaons}\label{sec:xsec-kaon}

The  $2\to 2$ scattering processes are considered in our analysis, but their cross sections are not fully known for the momentum from the threshold to ${\cal O}(1)\,\GeV$ because relevant $p$-wave contributions were not consistently investigated and the measurements involving $n$ or $K_L$ are limited. Here, we utilize the known datasets with $k_{\rm lab}<2\,\GeV$ cut, relate the amplitude by isospin, and perform a simultaneous fit of the scattering lengths up to $p$-wave for the first time.

Our method involves 14 parameters. Four of them are adopted from Ref.\,\cite{Martin:1970je}, and the remaining ten are fitted. 
We consider scattering amplitudes initiated by ${\bar K}N$ and $KN$ with the isospin channels $I=0$ and $1$, including $s$- and $p$-waves. In the ${\bar K}N$-initiated processes, the scattering lengths are complex, resulting in eight parameters, and the $s$-wave parameters are given in Ref.\,\cite{Martin:1970je}, leaving four $p$-wave parameters. 
Similarly, in the $KN$-initiated processes, we have four scattering lengths ($s$- or $p$-waves and $I=0$ or $1$), but they are taken to be real as hyperons are not produced. To improve the overall fit, we introduce a linear momentum dependence to the $s$-wave scattering length, such as $a_s=a_s^{(0)}(1+\kicm a_s^{(1)})$. Thus, the $KN$ processes are fixed by six fitting parameters.

The adopted $s$-wave scattering lengths of $\bar K N$ amplitudes from Ref.\,\cite{Martin:1970je} are  
\bal
A_{0,s}^{\bar K N} &=a_{0,s}^{\overline{K} N} +i b_{0,s}^{\overline{K} N}
\simeq (-1.74 +  0.70\,i){\rm fm},
\label{Eq:A0}
\\
A_{1,s}^{\bar K N} & = a_{1,s}^{\overline{K} N} + i b_{1,s}^{\overline{K} N}
\simeq
(-0.05 +  0.63\,i){\rm fm},
\label{Eq:A1}
\eal
where various hyperon productions and also $pK^- \to n\overline{K}^0$ with $k_{\rm lab}<0.3\,\GeV$ were fitted.

In the following, we present the formulae we use for the scattering cross sections, datasets, and fitting scheme. 

\noindent 
\underline{
\bf $\bar K N$ cross sections}:
For the  $s$-wave, we rely on the $K$-matrix analysis with parameters obtained in Ref.\,\cite{Martin:1970je}, which includes the effects coming from charge and mass differences between $n$ and $p$ as well as $K^-$ and $\bar K^0$\,\cite{Dalitz:1960du}. We add the $p$-wave separately, and we check that this contribution is very small in the datasets used in \cite{Martin:1970je}. 

Let us first consider $I_3=0$ of hyperon production channels, where $(\bar K N)_0 = p \, K^- $ and $n \, \bar K^0$, 
\bal
p \, K^-  ~{\rm or}~ n\, \bar K^0 \to 
\begin{cases}
    \pi^{\pm} \Sigma^{\mp} \\
    \pi^0 \Sigma^0 \\
    \pi^0 \Lambda
\end{cases}
%
\eal
When we replace $\bar K^0 \to K_L$, we evaluate the cross sections with an additional factor of $1/2$.  
From the representation in the isospin space with proper Clebsch-Gordan coefficients,
we can parameterize cross sections as
\bal
&\sigma((\bar K N)_{0} \to \pi^\pm \Sigma^\mp)
=\frac{1}{6}\sigma_{0,s}^{(\overline K N)_0}
+\frac{1}{4}(1-\epsilon)\sigma_{1,s}^{(\overline K N)_0} 
\nn
\\
&\quad \pm \kappa \left[\frac{1}{6}(1-\epsilon)\sigma_{0,s}^{(\overline K N)_0}\sigma_{1,s}^{(\overline K N)_0}
\right]^{1/2}
\cos\phi 
+\frac{1}{2}\sigma_{0,p}^{(\overline K N)_0} \ ,
\\
&\sigma((\bar K N)_0 \to \pi^0 \Sigma^0)
=\frac{1}{6}\sigma_{0,s}^{(\overline K N)_0}
+\frac{1}{2}\sigma_{0,p}^{(\overline K N)_0}\ , 
\\
&\sigma((\bar K N)_0 \to \pi^0 \Lambda)
=\frac{1}{2}\epsilon \sigma_{1,s}^{(\overline K N)_0}
+\frac{3}{2}\sigma_{1,p}^{(\overline K N)_0}\ , 
\eal 
where $\kappa = +1$ ($-1$) for $(\bar KN)_0 = pK^-$ ($n\bar K^0$), and $\epsilon\simeq 0.34$ denotes the ratio of $\pi\Lambda$ production over the total hyperon production ($\pi\Lambda $ and $\pi\Sigma^0 $) within $I=1$ $s$-wave cross sections. The similar quantity with $I=1$ and $p$-wave is assumed to be 1 since it is dominated by the $\pi\Lambda$ channel~\cite{Martin:1970je}. 
For the $s$-wave results from \cite{Martin:1970je}, 
$\sigma_{0(1),s}^{(\overline{K} N)_0}$ are given by
\bal
\sigma_{0(1),s}^{(\overline{K} N)_0}
\!= 
\!\!\begin{cases}
\frac{4\pi b_{0(1),s}^{\overline{K} N}}{k}
F(1,v)
\left|
\frac{1-i k_0 A_{1(0),s}^{\overline{K} N}}{D}
\right|^2 
& \!\!\!{\rm for}~  p \, K^-
\\
\frac{4\pi b_{0(1),s}^{\overline{K} N}}{k_0}
\left|
\frac{1}{1-i k_0 A_{0(1)}^{\overline{K} N}} \right|^2
& \!\!\!{\rm for}~  n \, \bar K^0
\end{cases}
\eal
with 
$k$ and $k_0$ being the $p\, K^-$ and $n\, \bar K^0$ momenta in the center-of-mass frame for a given center-of-mass energy ($k_0$ is taken as $i|k_0|$ below the $n\, \bar K^0$ threshold)\,\cite{Dalitz:1960du}. $b_{0,1}^{\bar K N}$ is the imaginary part of scattering length $A_{0,1}^{\bar K N}$.
Here, $D$ is given by
\bal
D =&
1-\frac{i}{2}(A_0^{\overline{K} N} + A_1^{\overline{K} N})
[k_0+k F(1,v) \, (1-i \lambda)]
\nn
\\
&-k_0 k F(1,v) \, (1-i\lambda) A_0^{\overline{K} N} A_1^{\overline{K} N},
\\
\lambda =&
-\frac{2}{k B F(1,v)}
\left[\log(2kR)
+ {\rm Re}\Big[ \frac{\Gamma'(i/k B)}{\Gamma(i/k B)} \Big]+ 2 \gamma_E 
\right],
\eal
where $\gamma_E$ is the Euler constant, $B=(\alpha_{\rm EM} \, \mu)^{-1}$ is the Bohr radius of the $p \, K^-$ system, $R\simeq 0.4\,\fm$ is the interaction radius,
and
\bal
\phi = \phi_{\rm th} + {\rm Arg}\left[\frac{1-ik_0 A_1^{\bar K N}}{1-i k_0 A_0^{\bar K N}} \right].
\eal
From the fitting to experimental data of $p \, K^-$ cross sections in Ref.\,\cite{Martin:1970je}, $\phi_{\rm th} \simeq -52.9^\circ$.


The $p$-wave cross sections, $\sigma_{0(1),p}^{(\bar K N)_0}$, are given by
\bal
\sigma_{0(1),p}^{(\bar K N)_0}
\!= 
\!\!\begin{cases}
4\pi
F(1,v)(1+\eta^2)\frac{k {\rm Im}[(A_{0(1),p}^{\overline K N})^3]}{|1-i k^3 (A_{0(1),p}^{\overline K N})^3|^2}
& \!\!\!{\rm for}~  p \, K^-
\\
4\pi\frac{k_0 {\rm Im}[(A_{0(1),p}^{\overline K N})^3]}{|1-i k_0^3 (A_{0(1),p}^{\overline K N})^3|^2}
& \!\!\!{\rm for}~  n \, \bar K^0
\end{cases}
\eal
We verify that this $p$-wave is sub-leading for $k_{\rm lab}<280\,\rm{MeV}$, which agrees with statement in \cite{Martin:1970je}.
The factor of $(1+\eta^2)$ for $pK^-$ corresponds to the correction to the Coulomb factor for $p$-wave contributions\,\cite{Cassel:2009wt}.

\begin{figure*}
    \centering
    \includegraphics[width=0.36\textwidth]{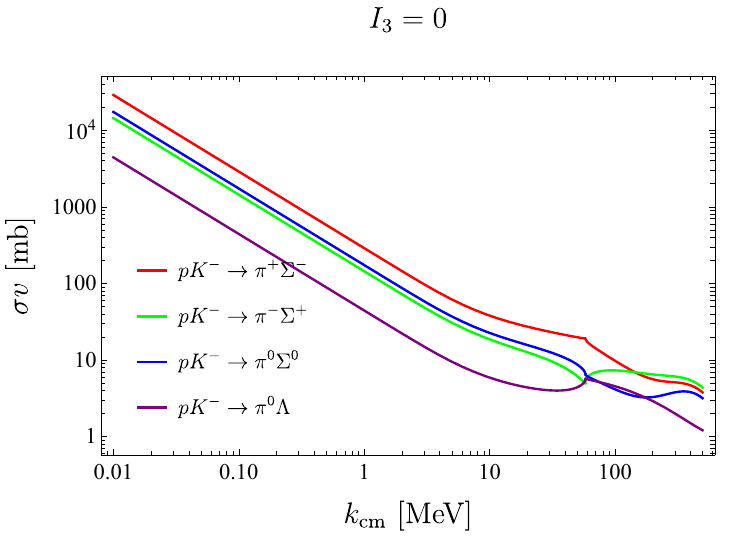}
    \hspace{1cm}
    \includegraphics[width=0.35\textwidth]{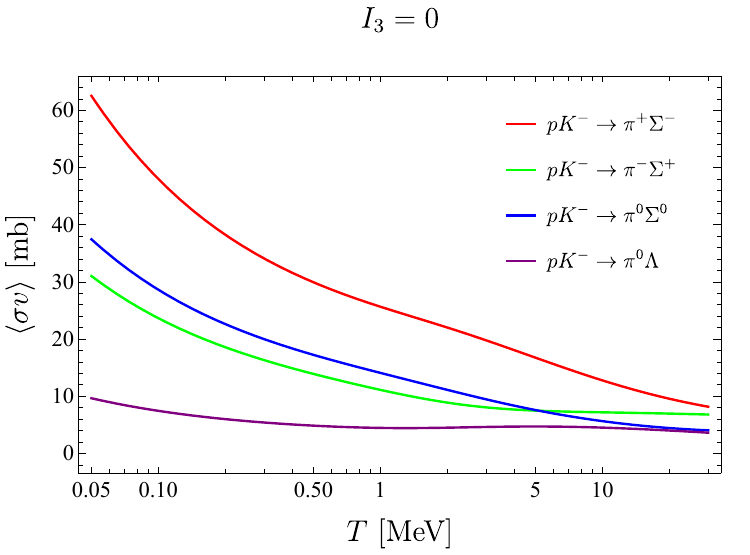}
    \caption{Cross sections for $p\,K^- \to \pi Y$. The kink at $k_{cm}=58.23\,\rm{MeV} $ is the division of $k_0$ being real or imaginary.
    \label{Fig:pKm_Y}
    }
%
    \centering
    \includegraphics[width=0.36\textwidth]{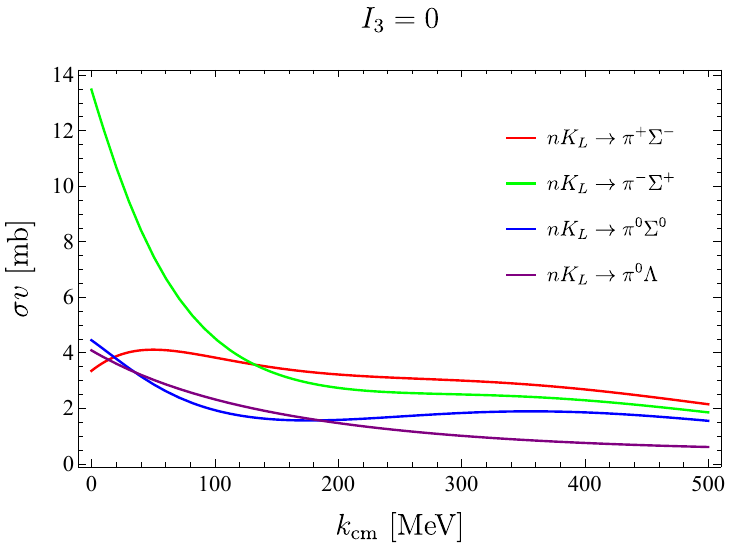}
    \hspace{1cm}
    \includegraphics[width=0.36\textwidth]{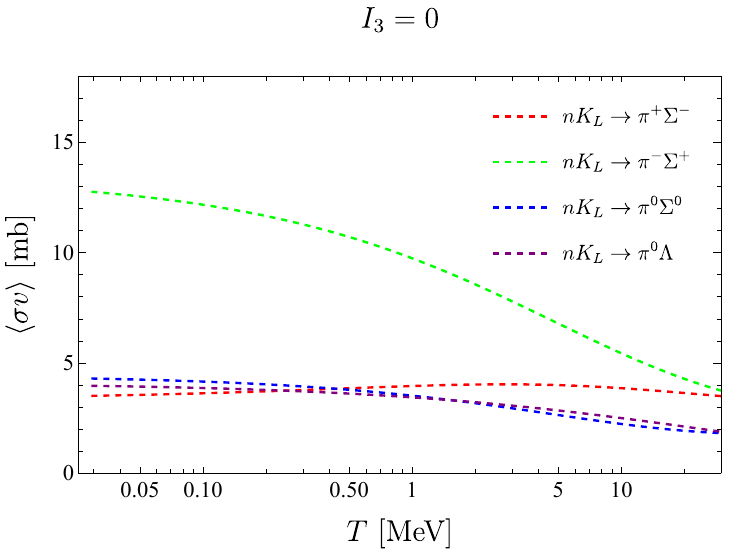}
    \caption{Cross sections for $n\,K_L \to \pi Y$. Note that the right panel assumes the Maxwell–Boltzmann distribution; therefore, it is not used in our calculation because the $K_L$s are not thermalized. One should use the momentum spectrum determined from the axion decay. The cross sections we use are shown in Fig.\,\ref{Fig:KLn_averaged}.
    \label{Fig:nKL_Y}
    }
    \includegraphics[width=0.36\textwidth]{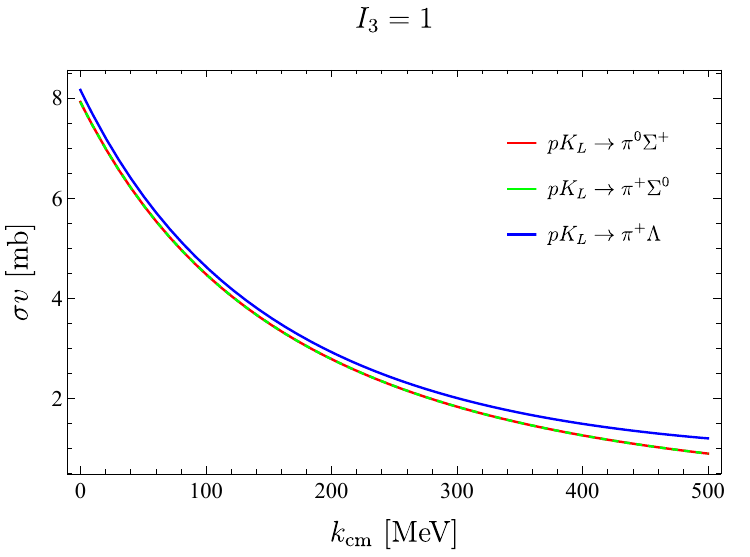}
    \hspace{1cm}
    \includegraphics[width=0.36\textwidth]{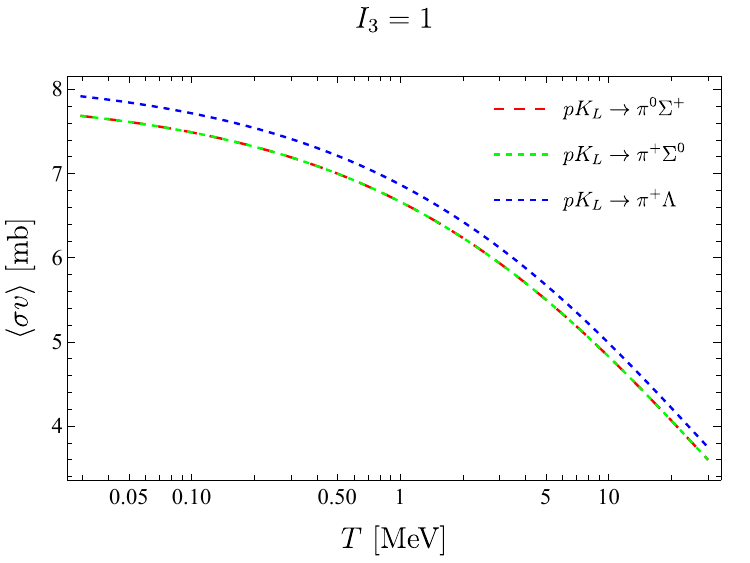}
    \caption{Cross sections for $p\,K_L \to \pi Y$. The cross sections for $p\,K_L \to \pi^+ \Sigma^0$ and $p\,K_L \to \pi^0 \Sigma^+$ are the same up to the mass difference. The right panel is not used, as explained in the caption of Fig.\,\ref{Fig:nKL_Y}. The cross sections we use are shown in Fig.\,\ref{Fig:KLp_averaged}.}
    \label{Fig:pKL_Y}
    \centering
    \includegraphics[width=0.36\textwidth]{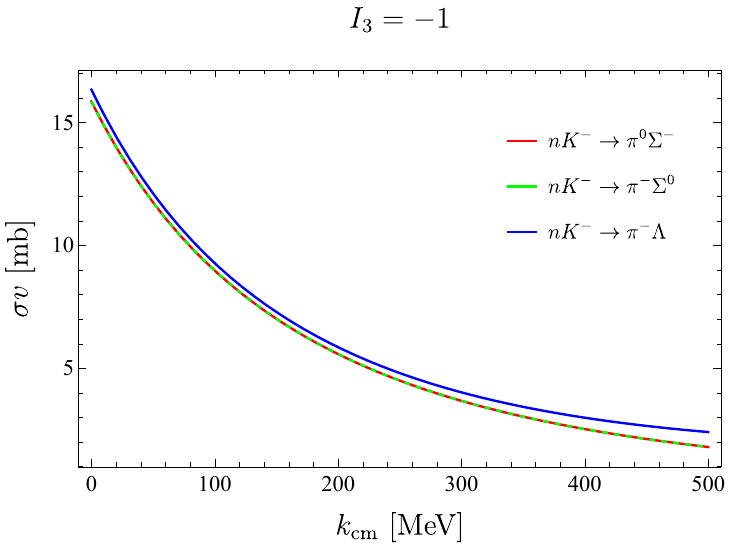}
    \hspace{1cm}
    \includegraphics[width=0.36\textwidth]{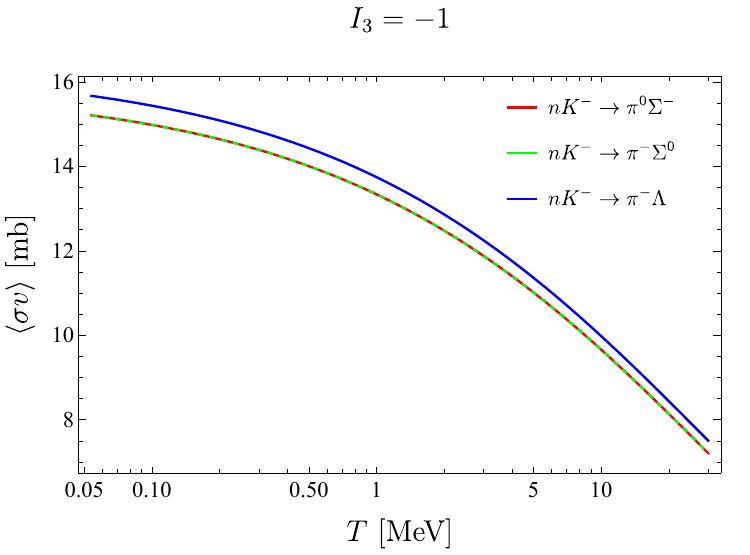}
    \caption{Cross sections for $n\,K^- \to \pi Y$. The cross sections for $n\,K^- \to \pi^- \Sigma^0$ and $n\,K^- \to \pi^0 \Sigma^-$ are the same up to the small mass difference.}
    \label{Fig:nKm_Y}
\end{figure*}

The cross sections of $p K^- \to \pi Y$ and $n K_L \to \pi Y$ scattering in the center-of-mass frame as well as the thermally averaged cross sections are shown in Figs.\,\ref{Fig:pKm_Y} and \ref{Fig:nKL_Y}.

Hyperons can be also produced by the $I_3=\pm 1$ processes where $( \bar N K)_{+} = p \bar K^0$ and $( \bar K N)_{-} = n K^-$,   
\bal
\sigma(p \, \bar K^0 \to \pi^0 \Sigma^+)
&=\sigma(p \, \bar K^0 \to \pi^+ \Sigma^0)
\nn
\\
&=(1-\epsilon) \frac{2\pi}{\kicm} \frac{b_{1,s}^{\overline{K} N}}{|1 - i \kicm A_{1,s}^{\overline{K}  N}|^2},
\\
\sigma(p \, \bar K^0 \to \pi^+ \Lambda)
&= \epsilon \frac{4\pi}{\kicm} \frac{b_{1,s}^{\overline{K} N}}{|1 - i \kicm A_{1,s}^{\overline{K}  N}|^2}
\nn \\
&+12\pi \frac{\kicm {\rm Im}[(A_{1,p}^{\overline K N})^3]}{|1-i (\kicm)^3 (A_{1,p}^{\overline K N})^3|^2},
\eal
\bal
\sigma(n K^- \to \pi^0 \Sigma^-)
&=\sigma(n K^- \to \pi^- \Sigma^0)
\nn
\\
&= (1-\epsilon) \frac{2\pi}{\kicm} \frac{b_{1,s}^{\overline{K} N}}{|1 - i \kicm A_{1,s}^{\overline{K}  N}|^2},
\\
\sigma(n K^- \to \pi^- \Lambda)
&= \epsilon \frac{4\pi}{\kicm} \frac{b_{1,s}^{\overline{K} N}}{|1 - i \kicm A_{1,s}^{\overline{K}  N}|^2}
\nn \\
&
\!\!+12\pi \frac{\kicm {\rm Im}[(A_{1,p}^{\overline K N})^3]}{|1-i (\kicm)^3 (A_{1,p}^{\overline K N})^3|^2},
\eal
The $K_L$ cross section is just given by $\sigma(N K_L\to \pi Y)=\frac{1}{2}\sigma(N \bar K^0\to \pi Y)$. 

The cross sections of $p \bar{K}^0 \to \pi Y$ and $n K^- \to \pi Y$ scattering in the center-of-mass frame as well as the thermally averaged cross sections are shown in Figs.\,\ref{Fig:pKL_Y} and \ref{Fig:nKm_Y}.

Now we consider $N \bar K\to N'\bar K' $ processes. 
To proceed, it is useful to define scattering amplitude expressions as
\bal
& T_s (A_s)=\frac{A_s}{1-i \kicm A_s}, \\
& T_p (A_p)= T_s \!\left( (\kicm)^2 A_p^3  \right).
\eal
Note that, in our convention, $A_p$ still remains in the dimension of length, while $A_p^3$ has the dimension of volume.

\begin{figure*}
    \centering
    \includegraphics[width=0.45\textwidth]{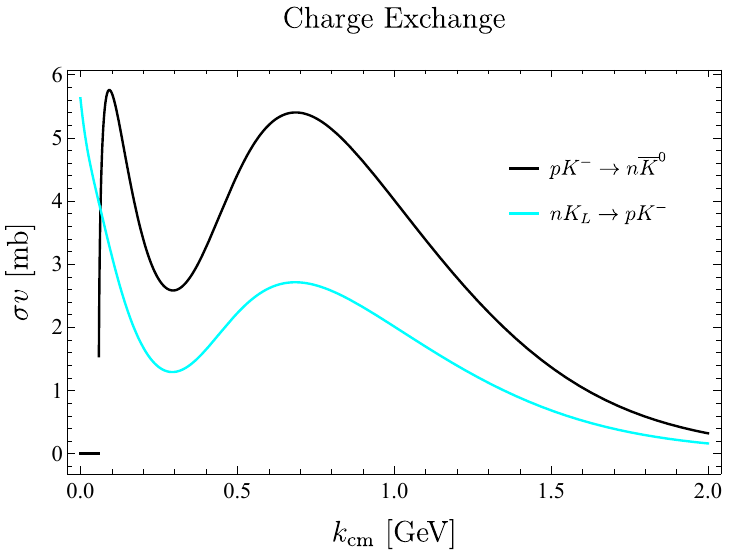}
    \includegraphics[width=0.45\textwidth]{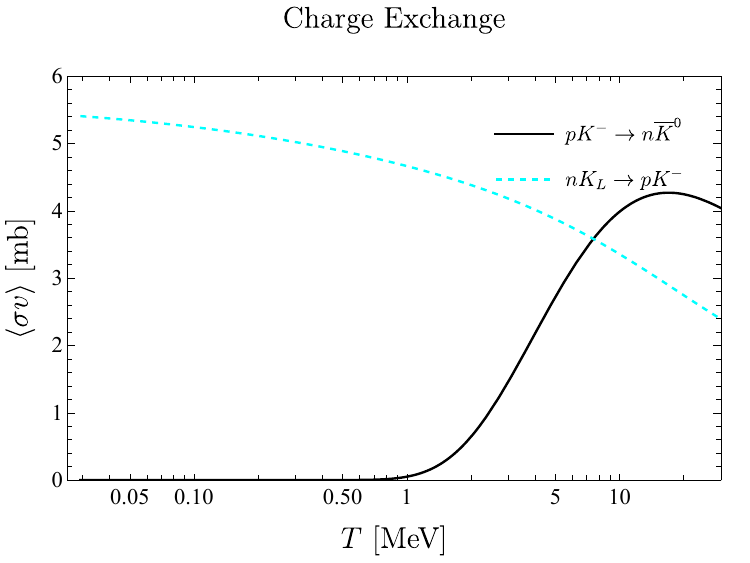}
    \caption{Cross sections for $p\,K^- \leftrightarrow n\,K_L$ as a function of $k_{\rm cm} $ (left) and the thermally averaged cross sections (right). Thermally averaged cross section for $n K_L \to p K^-$ is not used, as explained in the caption of Fig.\,\ref{Fig:nKL_Y}. The correct cross sections are shown in Fig.\,\ref{Fig:KLn_averaged}.}
    \label{Fig:pKm_K0}
    \centering
    \includegraphics[width=0.45\textwidth]{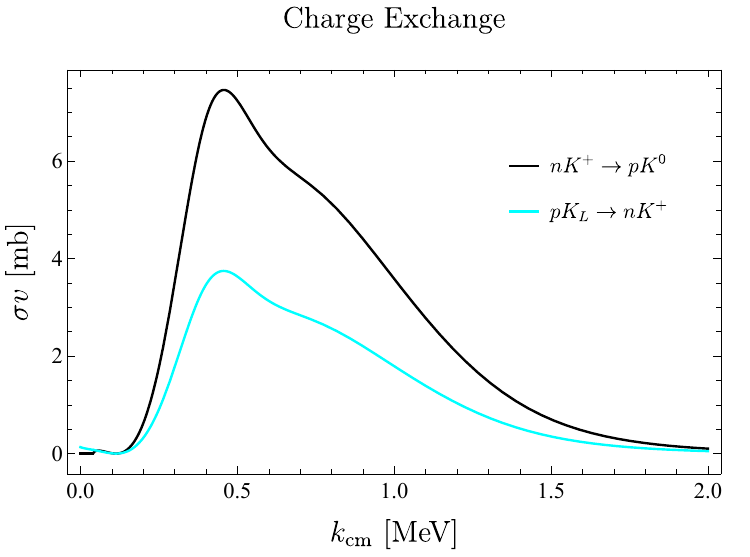}
    \includegraphics[width=0.45\textwidth]{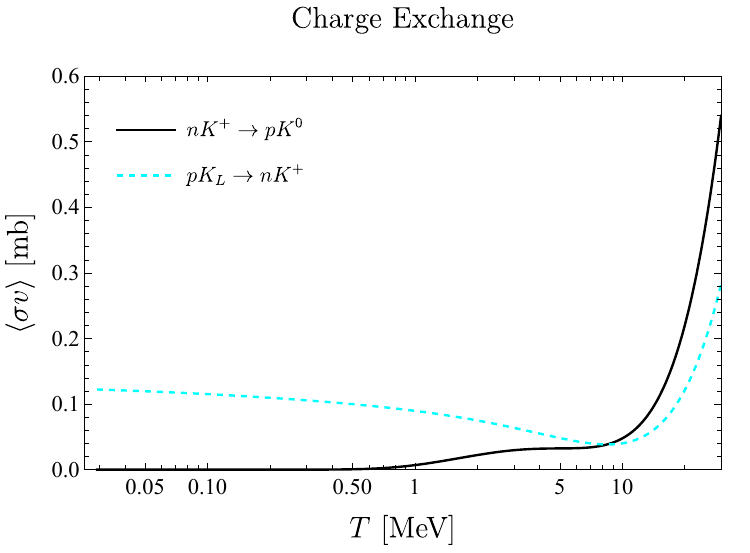}
    \caption{Cross sections for $p\,K_L \leftrightarrow n\,K^+$ as a function of $k_{\rm cm} $ (left) and the thermally averaged cross sections (right). Thermally averaged cross section for $p K_L \to p K^+$ is not used, as explained in the caption of Fig.\,\ref{Fig:nKL_Y}. The correct cross sections are shown in Fig.\,\ref{Fig:KLp_averaged}. 
    \label{Fig:pKL_Kp} }   
    \includegraphics[width=0.45\textwidth]{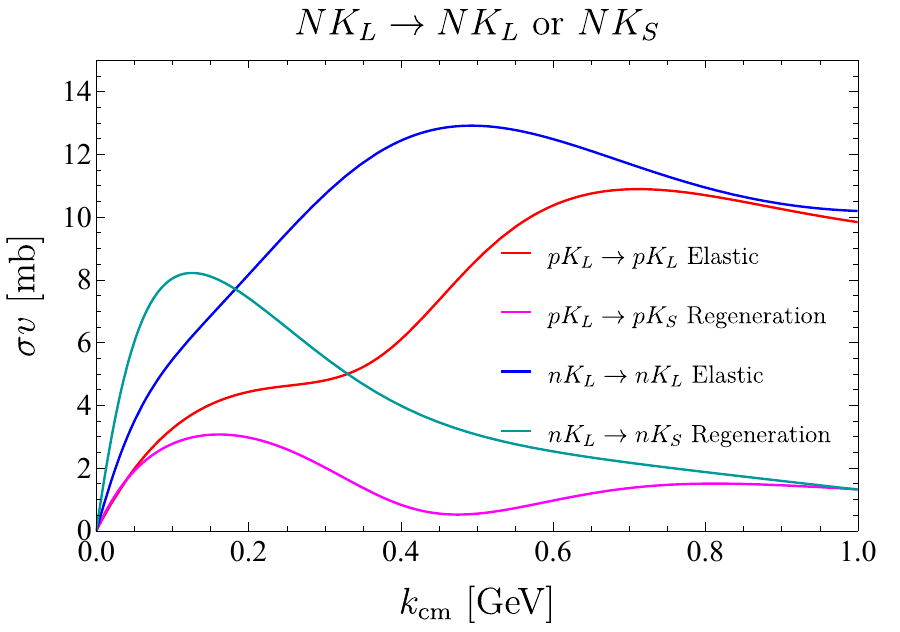}

    \caption{Cross sections for elastic processes $N K_L \to N K_L$ and regeneration processes $N K_L \to N K_S$ as a function of $k_{\rm cm}$. $N$ can be a proton or a neutron.}
    \label{fig:NKL_El_Rg}
\end{figure*}

The charge exchange processes, $p \, K^- \leftrightarrow n \, \bar K^0$,  are given by 
\bal
&\sigma (p \, K^- \to n \, \bar K^0)
= \frac{ \pi k_0 F(1, v)}{k}
\bigg(\left|  \frac{A_{1,s}^{\overline{K} N}-A_{0,s}^{\overline{K}  N}}{D}
\right|^2 \nn
\\ 
&\quad\quad+ 3(1+\eta^2)\left|T_p(A_{1,p}^{\overline{K}  N}) -T_p(A_{0,p}^{\overline{K}  N}) \right|^2 \bigg),
\\
&\sigma (n \, \bar K^0 \to p \, K^-)
=\frac{ \pi k F(1, v)}{k_0}
\bigg(\left|  \frac{A_{1,s}^{\overline{K} N}-A_{0,s}^{\overline{K}  N}}{D}
\right|^2 \nn
\\ 
&\quad\quad + 3(1+\eta^2)\left|T_p(A_{1,p}^{\overline{K}  N}) -T_p(A_{0,p}^{\overline{K}  N}) \right|^2 \bigg),
\eal
with $v$ being the relative velocity in the $p K^-$ system. $\bar K^0$ in the final state gives both $K_L$ and $K_S$ as $\sigma (p \, K^- \to n \, \bar K_{L(S)})=\frac{1}{2}\sigma (p \, K^- \to n \, \bar K^0)$.
The cross sections of $p K^- \leftrightarrow n \bar K^0$ scattering in the center-of-mass frame as well as the thermally averaged cross sections are shown in Fig.\,\ref{Fig:pKm_K0}.

\begin{figure*}
    \includegraphics[width=0.4\textwidth]{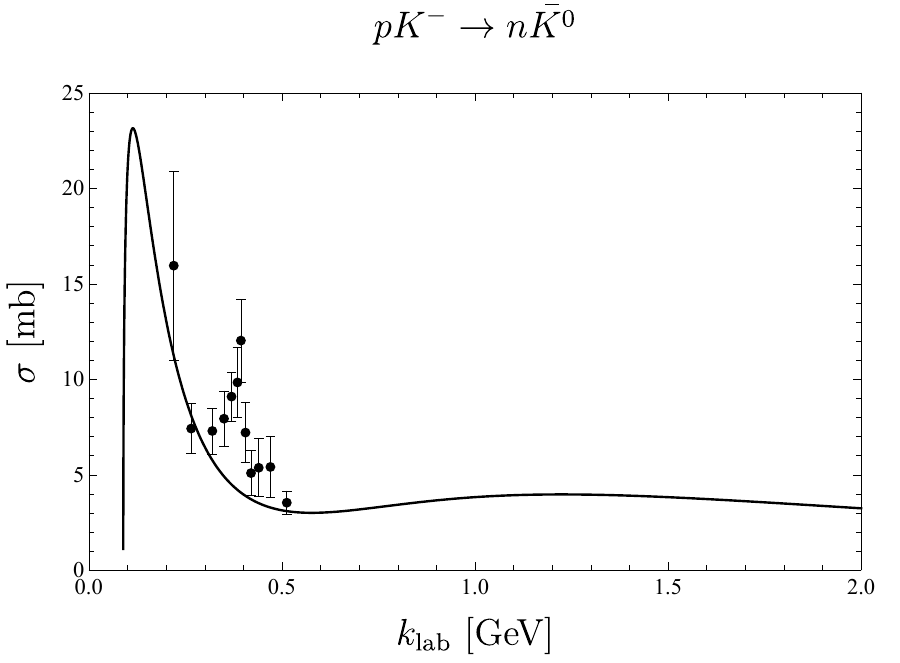}
    \includegraphics[width=0.4\textwidth]{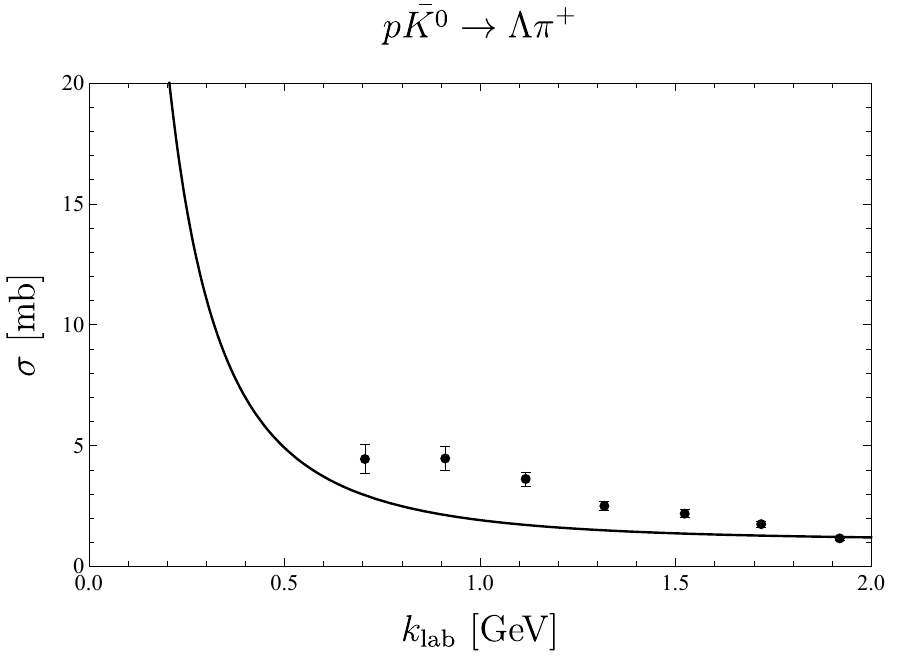}
    \\
    \includegraphics[width=0.4\textwidth]{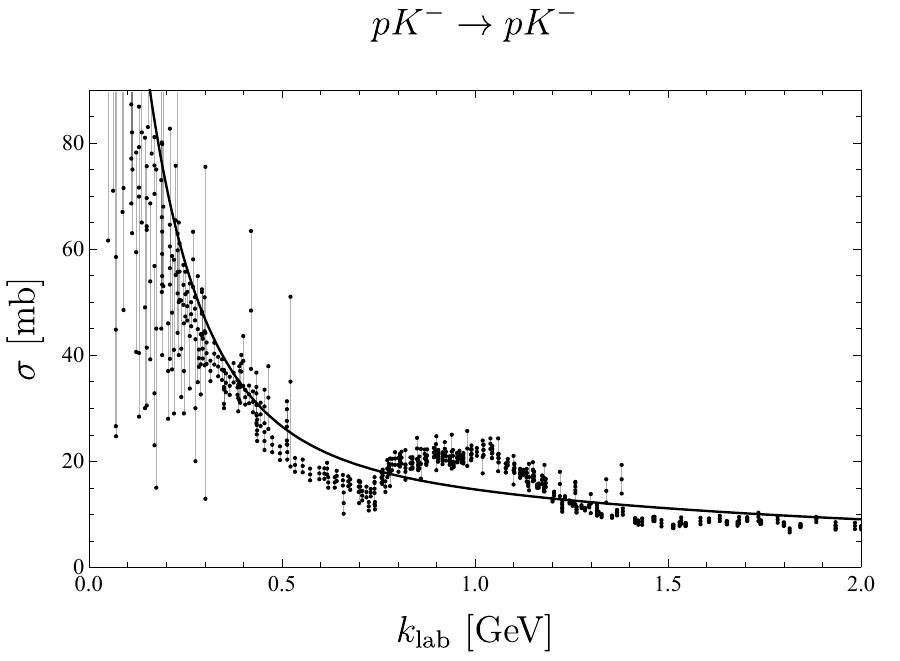}
    \includegraphics[width=0.4\textwidth]{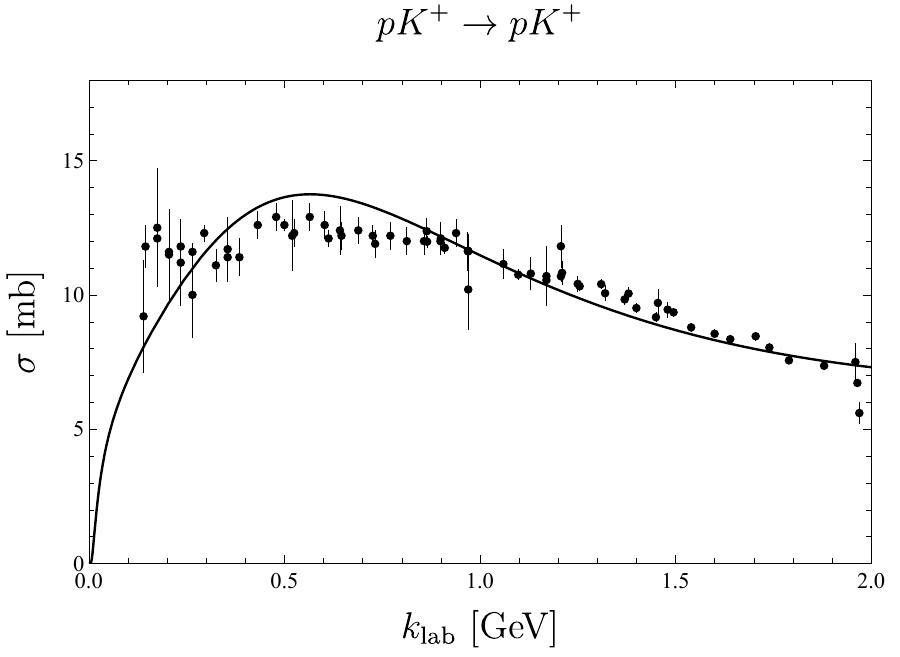}
    \includegraphics[width=0.4\textwidth]{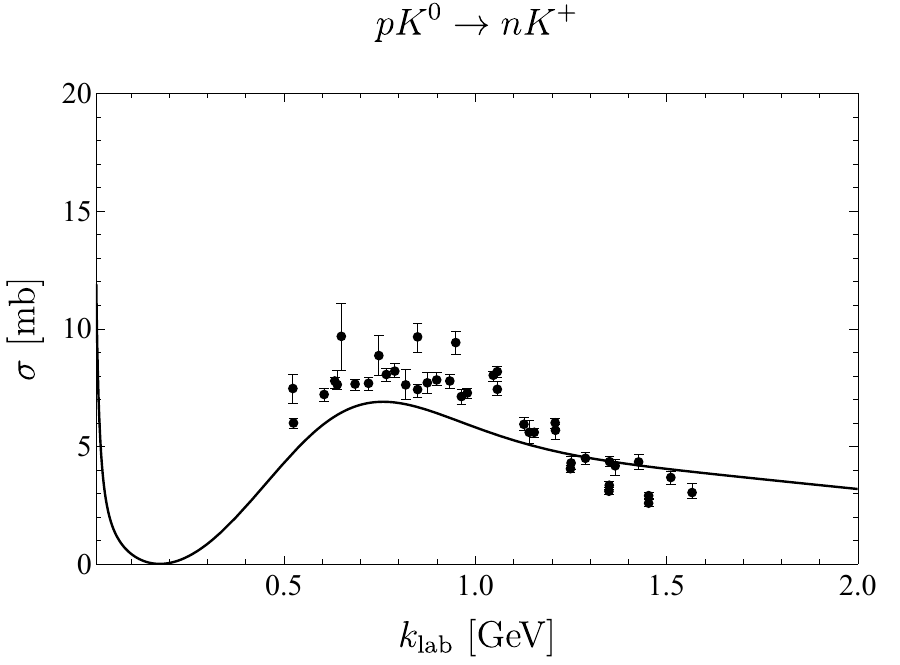}
    \includegraphics[width=0.4\textwidth]{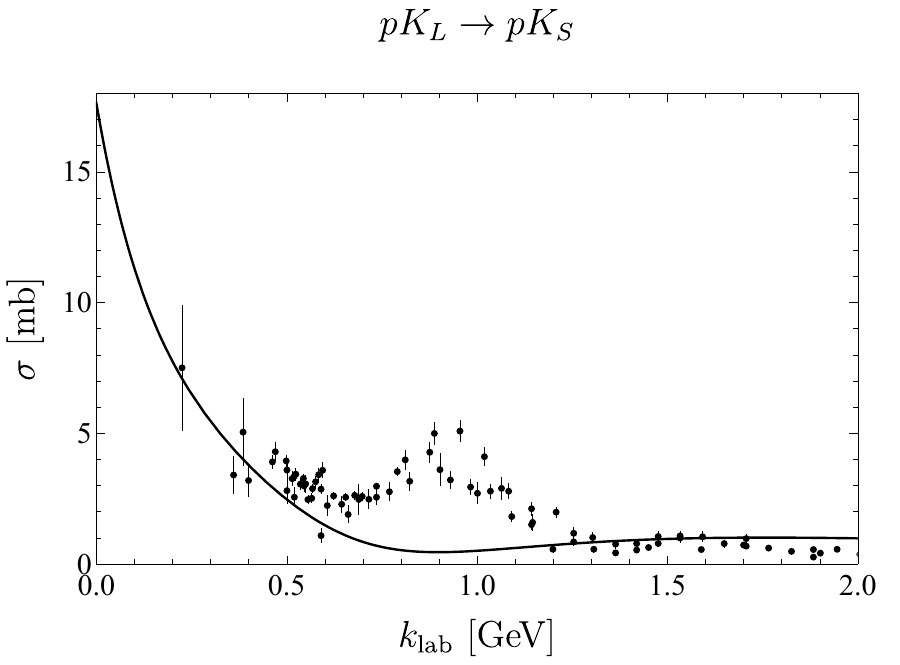}
    \includegraphics[width=0.4\textwidth]{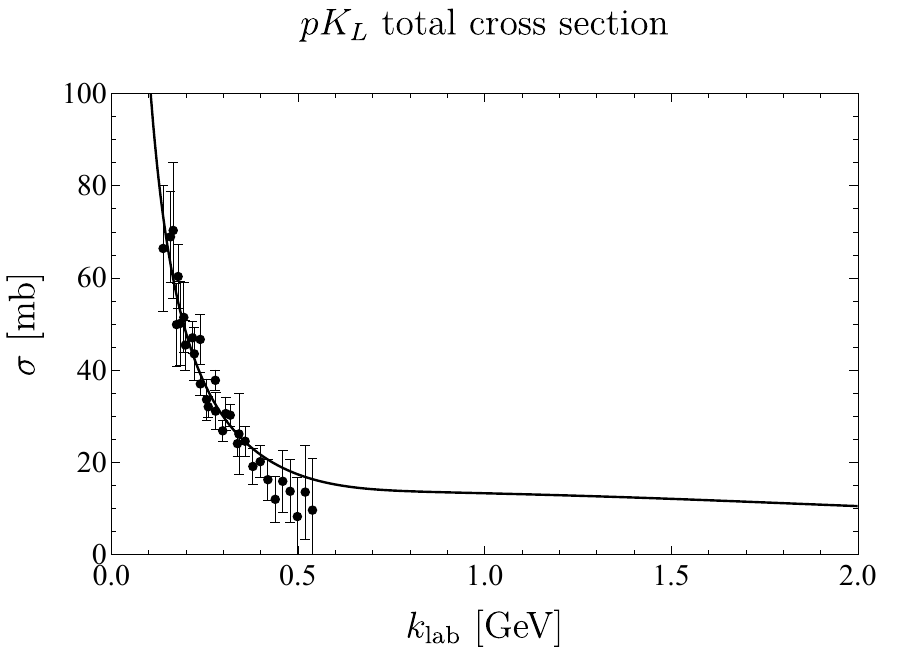}
    \caption{Ten parameters of our scattering lengths are determined by fitting seven different datasets, as indicated in each panel. The solid line shows our fit, and the data points with error bars represent the experimental measurements. See the main text for further details. Since we focus on $2\to 2$ scattering processes of $NK$ or $N\bar K$, the momentum range is restricted to $k_{\rm lab}<2\,\GeV$. }
    \label{fig:7_fitted_data}
\end{figure*}

The elastic scattering of $pK^-$ is also induced by the $\bar K N$ amplitude, 
\bal
&
\nn
\sigma( p K^- \to p K^-)=\pi F(1,v)^2 \big\{\left| T_s(A_{1,s}^{\overline{K}N})+T_s(A_{0,s}^{\overline{K}N}) \right|^2
\nn \\
&\quad\quad\quad +3(1+\eta^2)^2\left| T_p(A_{1,p}^{\overline{K}N})+T_p(A_{0,p}^{\overline{K}N}) \right|^2  \big\}. 
\eal
The cross section is shown in Fig.\,\ref{fig:7_fitted_data} (second row, left).

%



\vspace{0.2cm}
\noindent 
\underline{
\bf $K N$ cross sections}:
Unlike the $N \bar K$ system, hyperon production processes are forbidden, and therefore only four states are possible: $p K^+$, $p K^0$, $n K^+$, and $n K^0$ ($K^0$ is written with respect to $K_L$ and $K_S$).
Due to the absence of the hyperon channels, we consider both $s$ and $p$ waves with real scattering lengths. 
\bal
\sigma( p K^+ \to p K^+)&=4\pi F(-1,v)^2 \big\{\left| T_s(a_{1,s}^{KN}) \right|^2
\nn\\
&\quad\quad
+3(1+\eta^2)^2\left| T_p(a_{1,p}^{KN}) \right|^2  \big\}, 
\\
\sigma( p K^0 \to n K^+)&=\pi \frac{\kfcm}{\kicm} \big\{
\left| T_s(a_{1,s}^{KN}) -T_s(a_{0,s}^{KN}) \right|^2 
\nn \\ 
&\quad 
+3
\left| T_p(a_{1,p}^{KN}) -T_p(a_{0,p}^{KN}) \right|^2 \big\},
\\
\sigma(n K^+\to p K^0 )&=
\pi \frac{\kfcm}{\kicm} \big\{
\left| T_s(a_{1,s}^{KN}) -T_s(a_{0,s}^{KN}) \right|^2 
\nn \\ 
&\quad
+3
\left| T_p(a_{1,p}^{KN}) -T_p(a_{0,p}^{KN}) \right|^2 \big\},
\eal
where $\kfcm$ is the outgoing momentum in the center-of-mass frame. 
As before, when $K^0$ is replaced with $K_{L,S}$, the cross section is multiplied by 1/2. 
For the $s$-wave, we introduce $k$-dependence as $a_s=a_s^{(0)}(1+\kicm a_s^{(1)})$ to describe the $pK^-$ elastic scattering better. 

The cross sections of $p K^0 \leftrightarrow n \bar K^+$ scattering in the center-of-mass frame as well as the thermally averaged cross sections are shown in Fig.\,\ref{Fig:pKL_Kp}. The cross section of $pK^+$ elastic scattering is shown in Fig.\,\ref{fig:7_fitted_data} (second row, right). 

%

\begin{figure*}
    \includegraphics[width=0.58\textwidth]{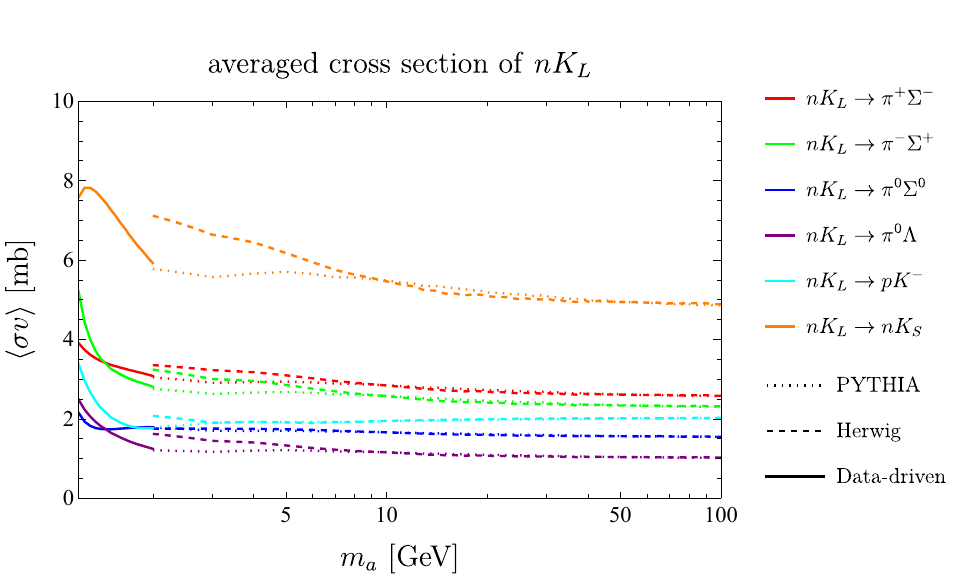}
    
    \caption{$nK_L$ cross sections averaged by using the kinematics from the axion decay. For $m_a < 2\,\rm GeV$, the data-driven method is used. For $m_a > 2\,\rm GeV$, two generators PYTHIA and Herwig are used. PYTHIA's result agrees better with the data-driven method at $2\,\rm GeV$.
    \label{Fig:KLn_averaged}}
    \includegraphics[width=0.58\textwidth]{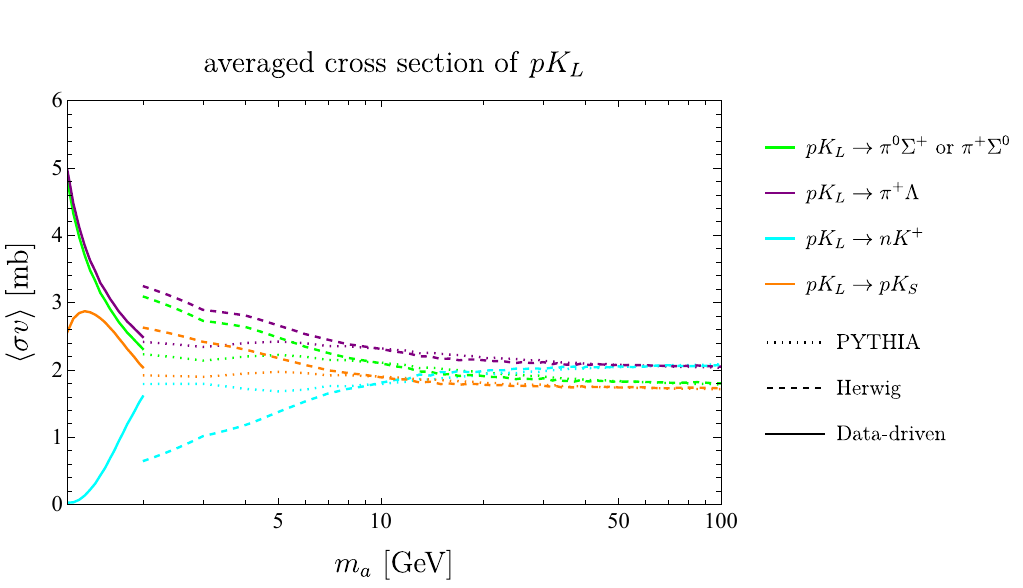}
    
    \caption{$pK_L$ cross sections averaged by using the kinematics from the axion decay. For $m_a < 2\,\rm GeV$, the data-driven method is used. For $m_a > 2\,\rm GeV$, two generators PYTHIA and Herwig are used. PYTHIA's result agrees better with the data-driven method at $2\,\rm GeV$.
    \label{Fig:KLp_averaged}}
\end{figure*}
~

\begin{figure*}[t]
    \centering
    \includegraphics[width=0.47\textwidth]{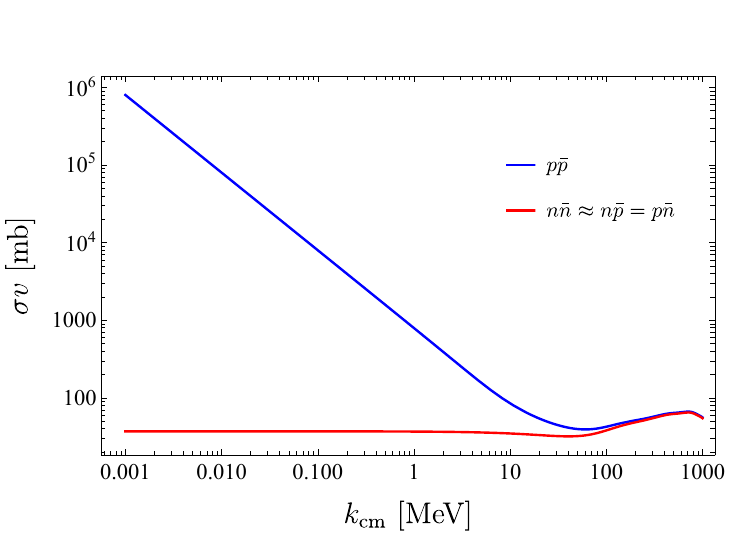}
    \includegraphics[width=0.47\textwidth]{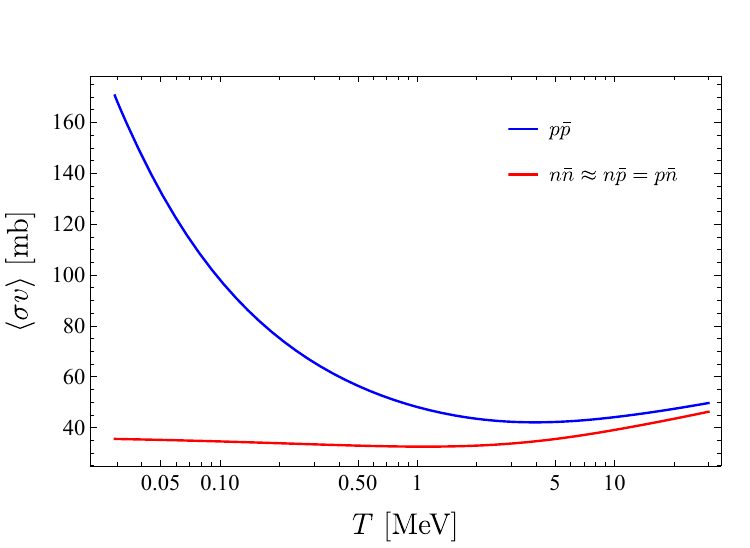}
    \caption{Cross sections for $N \bar N'$ annihilation. We make an estimation that $n \bar n$, $n \bar p$ and $p \bar n$ have the same cross sections. 
    \label{Fig:NNbar}}
\end{figure*}

\vspace{0.2cm}
\noindent 
\underline{
\bf $\bar K N$ and $ K N$ processes}:
Both $\bar K N$ and $K N$ amplitudes are necessary for the $K_L$ elastic scattering processes and the regeneration processes. 
\bal
    &\sigma( p K_L \to p K_S)
\nn
\\
&=
\! \pi 
\left| 
\frac{1}{2} \! \left(  T_s(a_{1,s}^{KN}) \! + \! T_s(a_{0,s}^{KN}) 
\right)
\! - \! T_s (A_{1,s}^{\overline{K}  N}) \right|^2\, \!\!\! 
\nn\\
&+3 
\pi
\left| 
\frac{1}{2} \! \left(  T_p(a_{1,p}^{KN}) \! + \! T_p(a_{0,p}^{KN}) 
\right)
\! - \! T_p (A_{1,p}^{\overline{K}  N})  \right|^2\, \!\!\! ,
\\
&\sigma( p K_L \to p K_L)
\nn
\\
&=
\! \pi 
\left| 
\frac{1}{2} \! \left(  T_s(a_{1,s}^{KN}) \! + \! T_s(a_{0,s}^{KN}) 
\right)
\! + \! T_s (A_{1,s}^{\overline{K}  N}) \right|^2\, \!\!\! 
\nn\\
&+3 
\pi
\left| 
\frac{1}{2} \! \left(  T_p(a_{1,p}^{KN}) \! + \! T_p(a_{0,p}^{KN}) 
\right)
\! + \! T_p (A_{1,p}^{\overline{K}  N})  \right|^2\, \!\!\! ,
\\
&\sigma( n K_L \to n K_S)
\nn
\\
&=
\! \pi 
\left|  
\!  \frac{1}{2} \! \left(T_s(A_{0,s}^{\overline{K} N})
\! + \! 
T_s(A_{1,s}^{\overline{K} N}) 
\right) -T_s(a_{1,s}^{KN})
\right|^2 \!\!\! 
\nn\\
&\! +3\pi 
\left| 
\!  \frac{1}{2} \! \left(T_p(A_{0,p}^{\overline{K} N})
\! + \! 
T_p(A_{1,p}^{\overline{K} N}) -T_p(a_{1,p}^{KN}) 
\right)
\right|^2 \!\!\! ,
\\
&\sigma( n K_L \to n K_L)
\nn
\\
&=
\! \pi 
\left|  
\!  \frac{1}{2} \! \left(T_s(A_{0,s}^{\overline{K} N})
\! + \! 
T_s(A_{1,s}^{\overline{K} N}) 
\right) +T_s(a_{1,s}^{KN})
\right|^2 \!\!\! 
\nn\\
&\! +3\pi 
\left| 
\!  \frac{1}{2} \! \left(T_p(A_{0,p}^{\overline{K} N})
\! + \! 
T_p(A_{1,p}^{\overline{K} N}) +T_p(a_{1,p}^{KN}) 
\right)
\right|^2 \!\!\! .
\eal
Including $p$-wave contribution is necessary to capture the features of charge exchange and regeneration processes. 
The cross sections of those processes are given in Fig.~\ref{fig:NKL_El_Rg}

As a consistency check, we verify unitarity for the $pK^-$ and $nK_L$ initial states in the limit where electromagnetic effects and mass differences are neglected: for each initial state, the sum of the cross sections over all open channels reproduces the total cross section given by the optical theorem,
$(2\pi/k)\,{\rm Im}\,T_{\rm elastic}$.


\vspace{0.2cm}
\noindent 
\underline{
\bf Fitting method and parameters}:
In order to determine ten parameters, we utilize seven datasets and evaluate the combined $\chi^2$. 

For the pure $\overline{K} N $ reactions, the measured processes are 
\bal
p K^- \to p K^-, 
\quad
&p \, K^- \to n \, \overline K^0, 
\quad
p \, \overline{K}^0 \to \pi^+ \Lambda,  
\quad
\eal
and the corresponding datasets are from the 2022 edition of PDG\,\cite{ParticleDataGroup:2022pth}, 
Ref.\,\cite{Ferro-Luzzi:1962jcn}, 
Fig.\,30 of Ref.\,\cite{GlueX:2017hgs} (originally Ref.\,\cite{Yamartino:1974sm})
respectively. 

In addition to Ref.~\cite{Ferro-Luzzi:1962jcn}, precise $pK^-\to n\bar K^0$ data are available in Ref.~\cite{Alston-Garnjost:1977zfn}, but we do not include them in our fit.
The peak structure observed in the data is from an excited hyperon resonance (see Fig.11 bottom of \cite{Alston-Garnjost:1977zfn}), which lies outside our framework of $2\to2$ scattering among $KN$, $\bar KN$, and $\pi Y$. Including them would significantly spoil the fit in the low-momentum region, which is most relevant for BBN. 

For the pure $K N $ reactions, we can use
\bal
p K^+ \to p K^+, \quad 
&p K^0 \to n K^+ , 
\eal
The corresponding datasets are from 2022 edition of PDG\,\cite{ParticleDataGroup:2022pth} and Fig.\,35 of Ref.\,\cite{GlueX:2017hgs} (Ref.\,\cite{Armitage:1977ms}),
respectively. 

Two $p K_L$  scattering processes are measured: 
\bal
&p K_L \to p K_S,
\nn \\
&{\rm Inclusive \ } p K_L \ (k_{\rm lab}<0.3\,\GeV).
\eal
The corresponding datasets are from Fig.\,27 of Ref.\,\cite{GlueX:2017hgs} (Ref.\,\cite{Capiluppi:1982fj}) and Fig.\,4 of Ref.\,\cite{Cleland:1975ex} (which includes data from Ref.\,\cite{Sayer:1968zz}), respectively. 

To perform the fit, we consider both statistical and systematic (if reported) uncertainties, and construct $\chi^2$ summing over the data points with $k_{\lab }<2\,\GeV$. 

Generating 200,000  initial seed points, we look for the local minimum of $\chi^2$, and pick the parameter set which gives the smallest $\chi^2$, that is, 
\bal
&a_{0,s}^{(0),KN} = 0.48\,{\fm},\quad  a_{1,s}^{(0),KN} = 0.23\, {\fm},
\nn \\
&a_{0,s}^{(1),KN}=-0.69\, {\fm},\quad  a_{1,s}^{(1),KN}=0.49\, {\fm},
\nn \\
&a_{0,p}^{KN}=0.28\, {\fm},
\quad a_{1,p}^{KN}=0.15\, {\fm}, 
\nn \\
&(A_{0,p}^{\overline{K}N})^3= (-0.012+0.053\,i) \, {\fm}^3,
\nn \\
&(A_{1,p}^{\overline{K}N})^3= (0.0017+0.00050\,i) \, {\fm}^3. 
\eal
The fitted curves and the data with error bars are shown in Fig.\,\ref{fig:7_fitted_data}. 
Recall that four parameters are from Eqs.\eqref{Eq:A0} and \eqref{Eq:A1}.  
 
These data are fitted fairly well, as the data points with errors are shown if a measurement exists.  We do not evaluate the uncertainties because this is beyond the scope of this paper.  
 


\subsection{Injected baryons}\label{sec:xsec-baryon}
Since the axion has neither a baryon number nor an electric charge, the net proton number and neutron number from one axion decay must be zero.
This implies that $X_n$ is not modified if an injected anti-proton (or anti-neutron) annihilates with a proton (or neutron) while $X_n$ receives an effective modification only when an anti-proton (or anti-neutron) annihilates with a neutron (or proton).
So, we need to know four possible annihilation channels: $p \bar p$, $n \bar p$, $p \bar n$, and $n \bar n$ into $\pi\pi$ (annihilation into $\gamma\gamma$ is QED-suppressed).

For this, we follow the analysis presented in Ref.\,\cite{Lee:2015hma};
\bal
\sigma_{\rm ann} =
\frac{\pi}{k^2}
\sum_{L=0}^{L_{\rm max}}
(2L+1) T_L(k) G_L(k),
\eal
where
\bal
&T_L= \frac{4 s_L K R}{\Delta_L^2+(s_L+KR)^2}
\eal
for $K=\sqrt{k^2+2\mu V_0}$ and
\bal
&s_L = R \left( \frac{g_L (df_L/dr)-f_L(dg_L/dr)}{g_L^2+f_L^2} \right)_{r=R}
\\
&\Delta_L=R \left( \frac{g_L (df_L/dr)+f_L(dg_L/dr)}{g_L^2+f_L^2} \right)_{r=R},
\\
&f_L(r) = \left(\frac{\pi k r}{2}\right)^{1/2} J_{L+1/2}(kr),
\\
&g_L(r)=-\left(\frac{\pi k r}{2}\right)^{1/2} Y_{L+1/2}(kr).
\eal
Here, $J_n(kr)$ and $Y_n(kr)$ are respectively Bessel and Neumann functions.
The Coulomb correction is included in $G_L$ ($G_L=1$ for $n\bar n$, $p \bar n$ and $n \bar p$),
\bal
G_L(k)
=
\frac{(L^2+\xi^2)((L-1)^2+\xi^2) \cdots (1+\xi^2)}{(L!)^2}
F(Z,v).
\eal
$R$ and $V_0$ are the parameters of a square well potential, $V(r)=-V_0 \Theta(R-r)$, whose physical meanings can be interpreted as nuclear contact radius and the strong interaction potential depth.
As analyzed in Ref.\,\cite{Lee:2015hma}, these parametrizations agree with experimental data well using $R = 0.97\,\fm$ and $V_0=85\,\MeV$.


\section{Reshaping $K_L$ distribution \label{sec:KLreshape}}

Based on our assessment of various $K_L N$ scattering cross sections, we find that elastic scattering is significant in the high-momentum region, $1\,\GeV \lesssim E_{K_L} \lesssim 2\,\GeV$. Although elastic scattering does not contribute to the Boltzmann equations as a number-changing process, it does modify the $K_L$ energy spectrum relative to the distribution originating from axion decay. In this appendix, we present prescriptions for incorporating elastic scattering effects and obtaining the reshaped $K_L$ spectrum, which is relevant for subsequent scattering processes in which $K_L$ is depleted. 

Firstly, we ignore both the Hubble expansion and $K_L$ decay, which are good approximations since the timescale of $K_L$ scattering with a nucleon is quite short.
Then, a $K_L$ with energy $E$ has two branches: a fraction $r(E)$ redistributes to different energies due to elastic scattering, and the remaining fraction, $1 - r(E)$, disappears due to number-changing scattering processes. The redistributed spectrum undergoes another iteration, and eventually, the remaining component becomes negligible after enough iterations.

A specific algorithm is as follows. We bin the $K_L$ energy spectrum from threshold to $E_{K_L} = 2.023\,\GeV$
(equivalent to $k_{\lab} = 2\,\GeV$), with a bin size of 50\,MeV (25 MeV for $m_a\leq 2\,\GeV$). Let $n_i$ denote the number of $K_L$ particles in the $i$-th bin, such that the total $\sum_i n_i$ is $N_{a \to K_L}$. Then, $n_i$ either migrates to other bins $n_{j \leq i}$ through elastic scattering, or is added to $n_i^{\rm dis}$, which is initially zero:
\begin{align}
\!\!&\text{Elastic scattering:} &&\!\!n_j = \!\!\sum_{i=1}^{N_{\rm bins}}\!\! n_i r(E_i) f(E_i \to E_j), \\
\!\!&\text{Disappearance:} &&\!\!n_i^{\rm dis} +\!\!= n_i (1 - r(E_i)).
\end{align}
Here, $f(E_i \to E_j)$ is the probability distribution for elastic scattering (assuming isotropic in the center-of-mass frame), where the $K_L$ energy changes from $E_i$ to $E_j$. The function $r(E_i)$ depends on the cross-section ratios and also on $X_n$, 
\begin{align}
r(E) = \frac{X_n\sigma^{\rm elastic}_{nK_L} + (1-X_n)\sigma^{\rm elastic}_{pK_L}}{X_n\sigma^{\rm tot}_{nK_L} + (1-X_n)\sigma^{\rm tot}_{pK_L}}.
\end{align}
 Fortunately, $r(E)$ is less sensitive to $X_n$ because the cross section ratios are accidentally similar, $\sigma^{\rm elastic}_{nK_L}/\sigma^{\rm tot}_{nK_L}\simeq\sigma^{\rm elastic}_{pK_L}/\sigma^{\rm tot}_{pK_L}$. We choose $X_n = 0.5$ as a representative value.
The elastically scattered component, $n_j$, is recycled in the next migration step, and $n_i^{\rm dis}$ is incrementally built up in each iteration.

After about six iterations, the $K_L$ disappears, and the reshaping procedure is complete. The modified spectrum used for computing averaged cross sections is stored in $n_i^{\rm dis}$, where $\sum_i n_i^{\rm dis} \approx N_{a \to K_L}$. Note that the updated $K_L$ spectrum is still very different from the thermal distribution. 


\end{appendix}

\bibliographystyle{utphys}
\bibliography{ref}

\end{document}